\documentclass[onecolumn,10pt,nofootinbib]{revtex4}
\usepackage{graphicx,amsmath,amsthm,hyperref,verbatim,amsfonts,amssymb,mathtools,bm}
\usepackage{caption,subcaption}
\usepackage{float,subfloat}
\usepackage[capitalize]{cleveref}
\usepackage{physics}
\usepackage{orcidlink}
\usepackage{lmodern,pdfpages}
\usepackage[a4paper,margin=0.5in]{geometry}
\usepackage{tabu,ulem,tabularx,array,booktabs,float}
\usepackage[explicit]{titlesec}
\usepackage[toc]{appendix}

\hypersetup{colorlinks=true,urlcolor=cyan,citecolor=blue,linkcolor=red,linktocpage=true}
\numberwithin{equation}{section}

\newcommand{\cA}{{\mathcal{A}}}
\newcommand{\cC}{{\mathcal{C}}}
\newcommand{\cD}{{\mathcal{D}}}

\newcommand{\cI}{{\mathcal{I}}}

\newcommand{\cM}{{\mathcal{M}}}
\newcommand{\cN}{{\mathcal{N}}}
\newcommand{\cO}{{\mathcal{O}}}
\newcommand{\cP}{{\mathcal{P}}}
\newcommand{\cQ}{{\mathcal{Q}}}

\newcommand{\cV}{{\mathcal{V}}}

\newcommand{\cZ}{{\mathcal{Z}}}

\newcommand{\sF}{{\mathsf{F}}}

\newcommand{\sM}{{\mathsf{M}}}

\newcommand{\fM}{{\mathfrak{M}}}
\newcommand{\fS}{{\mathfrak{S}}}
\newcommand{\fone}{{\mathfrak{1}}}
\newcommand{\ftwo}{{\mathfrak{2}}}

\begin{document}
\renewcommand{\baselinestretch}{1.2}
\parskip=.5em
\title{Quiver superconformal index and giant gravitons: asymptotics and expansions}
\author{Souradeep Purkayastha$^{a,b}$\orcidlink{0000-0001-8284-5063}}
\email[Email: ]{spurkayastha@fuw.edu.pl}
\author{Zishen Qu$^{c,d}$}
\email[Email: ]{zishenq2@illinois.edu}
\author{Ali Zahabi$^{a}$\orcidlink{0000-0001-6596-6816}}
\email[Email: ]{zahabi.ali@gmail.com}
\affiliation{$^{a}$Institut de Math\'ematiques de Bourgogne, Universit\'e de Bourgogne Franche-Comt\'e, Dijon 21000, France}
\affiliation{$^{b}$Faculty of Physics, University of Warsaw, Warsaw 02-093, Poland}
\affiliation{$^{c}$Department of Combinatorics and Optimization, University of Waterloo, Waterloo N2L 3G1, Canada}
\affiliation{$^{d}$ Department of Mathematics, University of Illinois Urbana-Champaign, IL 61801, USA}
\begin{abstract}
\section*{\large{Abstract}}
\noindent
We study asymptotics of the $d=4$, $\cN=1$ superconformal index for toric quiver gauge theories. 
Using graph-theoretic and algebraic factorization techniques, we obtain a cycle expansion for the large$-N$ index in terms of the $R$-charge-weighted adjacency matrix.
Applying saddle-point techniques at the on-shell $R$-charges, we determine the asymptotic degeneracy in the univariate specialization for $\hat{A}_{m}$, and along the main diagonal for the bivariate index for $\cN=4$ and  $\hat{A}_{3}$. In these cases we find $\ln |c_n|\sim \gamma n^{\frac{1}{2}}+ \beta \ln n+ \alpha$ (Hardy--Ramanujan type).
We also identify polynomial growth for $dP3$, $Y^{3,3}$ and $Y^{p,0}$, and give numerical evidence for $\gamma$ in further $Y^{p,p}$ examples.
Finally, we generalize Murthy’s giant graviton expansion via the Hubbard–-Stratonovich transformation and Borodin–-Okounkov formula to multi-matrix models relevant for quivers.
\end{abstract}
\maketitle
\tableofcontents
\newpage

\section{Introduction}
\label{section:intro}

One of the fundamental advances in late 20th century theoretical physics was the discovery of a number of dualities between a string theory or quantum field theory on a manifold with boundary, and another one in lower dimension living on the boundary.  
The holographic principle in string theory, first noted in~\cite{tHooft:1993dmi,susskind1995}, conjectures such a correspondence between observables of a gravitational theory to analogous ones of a boundary conformal field theory (CFT).
The principle is formulated in the language of partition and correlation functions.
The AdS/CFT correspondence~\cite{maldacenaoriginal}, one of its most effective realizations, may be used to obtain information about the entropy of gravitational systems in the bulk Anti--de Sitter (AdS) space from the holographically-dual counting on the boundary CFT.
See~\cite{natsuume} and references therein for more details on the correspondence.
Since the development of the Bekenstein--Hawking formula~\cite{hawkingentropyoriginal} for black hole entropy, and an explicit micro-state counting of black holes via D-branes~\cite{stromingervafa1996}, thermodynamics has been one of the primary topics of study in pursuit of a quantum theory of black holes and, eventually, of gravity. 

CFTs in this correspondence are typically supersymmetric, and the most well-studied such \textit{superconformal} field theory (SCFT) is $\cN=4$ Super--Yang--Mills (SYM) in four dimensions, first introduced in~\cite{maldacenaoriginal}. 
This is the simplest example of a \textit{quiver gauge theory}, an SCFT that arises from the compactification of $d=10$ type IIB superstring theory~\cite{douglas1996dbranes}. 
Quiver gauge theories are the low energy limit of stacks of $D3$-branes probing orbifolds of the form $\mathbb{C}^{3}/\Gamma$, where $\Gamma$ is taken to be a discrete group of symmetries so that $\mathbb{C}^{3}/\Gamma$ is Calabi--Yau; see~\cite{he2004lectures} and references therein for a detailed review. 
They can be described by directed graphs called \textit{quivers}. 
Each vertex of the quiver represents a gauge group and a vector multiplet in its adjoint representation, and each directed edge represents a chiral multiplet in a bifundamental representation of the source (fundamental) and target (antifundamental) gauge groups. 
Each constituent chiral multiplet has an assigned set of $R$-charges; vector multiplets by definition carry no $R$-charge. 
In general there can be additional global flavour symmetries, but in this work we switch them off.

In any supersymmetric theory one can define a~\textit{supersymmetric index}, a weighted partition function over the states of the theory~\cite{WITTEN1982253}; if the theory exhibits conformal invariance, we term it the \textit{superconformal index}.
It is a topological invariant, and may be generalized to include various switched-on fugacities representing internal symmetries. 
The superconformal index was first introduced in the context of $\cN=4$ SYM~\cite{shirazall}, where it counts $\frac{1}{16}-$BPS states.
In this paper we concern ourselves with the $d=4$, $\cN=1$ superconformal index (simply called the \textit{index} for the remainder of this paper) for the specific class of toric quiver gauge theories, for which the compactified Calabi--Yau threefold is a toric variety, and work with the assignment of a $\mathrm{U}(N)$ gauge group at each vertex for a specific $N$. 
The \textit{large $N$ index} for toric quivers is the leading order of the index under the limit $N \rightarrow \infty$ of the common gauge group dimension, and can be expressed in terms of the determinant of the $R$-charge-weighted adjacency matrix of the quiver~\cite{Mattioli2015}. We shall call this matrix the \textit{large $N$ index matrix}. 
The index $\cI(p,q)$ we consider exhibits bivariate dependency on two switched-on fugacities $p$ and $q$ by construction, but properties of the $d=4$, $\cN=1$ superconformal algebra lead to the large $N$ index matrix $M(t)$ exhibiting a functional dependency on the single quantity $t=(pq)^{1/2}$, see~\cref{sec:review}. 

The study of the large $R$-charge limit of the gauge theory is of particular interest in the context of duality.
We examine the degeneracy of states in the large charge limit by expanding the large $N$ index as a formal power series $\sum_{m,n} c_{mn}p^{m}q^{n}$ and observing the behaviour of the weighted state degeneracy $c_{mn}$ as $m,n \rightarrow \infty$, and it is in this context that asymptotic analysis becomes relevant.
The asymptotic growth of $c_{mn}$ in the large $N$ limit is expected to estimate a state count of a dual theory of type $AdS_{5} \times SE_{5}$, for a Sasaki--Einstein manifold $SE_{5}$; see \cite{sasaki} and references therein for more details. 
The logarithm $\ln |c_{mn}|$ similarly estimates the entropy of the dual theory.
In this paper we study such asymptotics for various toric quivers.
Since the index is a weighted sum of the ground state degeneracy and vanishes due to parity over excited states, $|c_{mn}|$ can only be a lower bound of the true count. 
Work~\cite{shirazall,murthy2020} in this context for $\cN=4$ SYM indicates that the growth of these coefficients is expected to reproduce the entropy of a gas of gravitons at lower charges, but undergoes a Hawking-Page transition into the expression for entropy of an AdS black hole at higher charges.

The asymptotic analysis of the unrefined univariate index for $\cN=4$ SYM, with one fugacity variable $p=q=x$, was done in~\cite{murthy2020}; we extend this work by analyzing the asymptotics of the univariate and bivariate index for other classes of toric quivers.
In the bivariate case, we utilize techniques from multivariate analytic combinatorics. 
We investigate the asymptotics of coefficients of terms of the form $p^{nr}q^{ns}$ as $n$ grows, for fixed directions $(r,s)$, hence converting the bivariate problem into a univariate one which can be analyzed by classical saddle-point techniques.
Specifically we consider the \textit{main diagonal} $(1,1)$ -- we
study a new generating function obtained from the bivariate index by trimming out the diagonal terms of of the form $(pq)^{n}$.
See e.g.~\cite{Pemantle_Wilson_Melczer_2024} for a review.
The framework used for such main-diagonal analysis has connections to partition functions for integers, and Lie algebras~\cite{macdonald,nekrasovokounkovmain}. 

Our approach to asymptotics is made possible via an interesting factorization of the determinant of the large $N$ index matrix into an infinite product reminiscent of generating functions for integer partitions. 
It was first seen in~\cite{holocheck}, and eventually proven in~\cite{eagerschmudetachikawa}, that for toric quivers $\det M(t)$ factorizes as a polynomial in $t$, and the factorization appears to hold true even with the switching-on of flavour symmetries which introduce their own fugacities. 
This factorization is valid for exact (physical) $R$-charges, but~\cite{zigzag} provides evidence of its validity even when the $R$-charge constraints are relaxed. 
In several cases for large quivers we find it convenient to arrive at $\det M(t)$ for exact $R$-charges by first obtaining it for inexact $R$-charges using the \textit{method of zig-zag paths} outlined in~\cite{zigzag}, where it has been shown to work correctly in several circumstances. 
We then pass to the exact $R$ charge regime by the principle of \textit{$a$-maximization}~\cite{intriligatormax}, where we find the configuration maximizing the $a$ central charge of the theory. 
While studying these factorizations we notice the appearance of various \textit{cycles}, i.e. non-trivial closed walks of the quiver in the form of sums of their $R$-charges, earlier observed, for example, in~\cite{Pasukonis_2013,chiralring}, and we outline the structure of these cycles for the examples we consider. 
Besides $\cN=4$ SYM, we consider the $Y^{p,q}$ and $\hat{A}_{m}$ infinite quiver families, and the $dP3$ quiver.
For the $\hat{A}_{m}$ family, the superpotential is not known in general, but we utilize an ansatz that correctly outputs $\det M(t)$ on-shell.

Having arrived at generating function equations, we derive asymptotics using saddle-point techniques dating back to the late 19th century~\cite{bruijn1958}, utilized in the classic \textit{circle method} derivation of integer partition asymptotics by Hardy and Ramanujan~\cite{hardyramanujan}, and independently by Uspensky~\cite{uspensky_asymptotics}, and refined in the century since; see~\cite{vaughan_1997} for a review.
In particular, the infinite product structure of the large $N$ index is similar to partition generating functions, having a dense ring of singularities on the unit circle, typically with a finite set of them providing dominating contributions to asymptotics. 
We derive asymptotics for some of the quivers under consideration, including all members of the family $\hat{A}_{m}$, by a study of their generating functions near the singularities $\pm1$. 
When this approach succeeds, we obtain Hardy--Ramanujan asymptotics of the form $\ln |c_{n}| \sim \gamma n^{\frac{1}{2}} +\beta \ln n+\alpha$ for numbers $\alpha \ne 0,\beta,\gamma$.
The $\cO(n^{\frac{1}{2}})$ leading order behaviour observed is characteristic of $d=4$ BPS state count~\cite{Sen_2010} (also see~\cite{Sen_2014} for a broad review).
Through $\gamma$ we can introduce the notion of an effective central charge for the theory, as has been recently done in a $d=3$ context~\cite{ceffgukov,adams2025crmeffresurgencestokes,harichurn2025ctexteffsurgerymodularity}.  
At the sub-leading order, the appearance of the logarithmic correction term $\beta \ln n$ too is expected~\cite{Solodukhin_1995,Kaul_2000,Das:2001ic,Sen_2010,2012asen,Sen_2013} for microstate counts of black-hole entropy, and it has been explicitly calculated in the dual picture for $\cN=4$ SYM~\cite{lezcano2020paper,Gonz_lez_Lezcano_2021}. 
Our asymptotic results are supported by verification using the SageMath and Mathematica computer algebra systems. 

Our principal mathematical result is~\eqref{finalAmresult}, the asymptotic growth, up to sign, of the coefficients of the $\hat{A}_{m}$ univariate generating function $g(z)$ given by~\eqref{Amgenfunc}; $m=1$ corresponds to $\cN=4$ SYM, and $m=2$ to $Y^{1,1}$. 
This result arises from the dominant contribution to the contour integral from the singularity at $z=-1$.
We are able to generalize our analysis to a more general function $G(z)$ given by~\eqref{eq_remark_generalGF}, for which we find the asymptotic growth formulae, up to sign,~\eqref{eq_remark_generalGF_z=1} and~\eqref{eq_remark_generalGF_z=-1} respectively, in the cases that the singularities contributing dominantly to the contour integral are $z=1$ and $z=-1$ respectively.
\eqref{eq_remark_generalGF_z=1} generalizes special cases previously discussed in the literature.
Intuitively speaking, the function~\eqref{eq_remark_generalGF} may be viewed as a generator of coloured integer partitions with both `bosonic' and `fermionic' colours.
In the bivariate case, we consider two cases of bivariate generating functions, for the $\cN=4$ SYM and the $\hat{A}_{3}$ quivers, which we analyze in a univariate setup along the main diagonal.
We also find a few quivers whose asymptotics do not display Hardy--Ramanujan asymptotic behaviour, but grow polynomially.
In addition, we are able to numerically conjecture $\gamma$ for a few cases which are not treatable by our asymptotic techniques, but appear to exhibit Hardy--Ramanujan growth in the exponential sector.
See~\cref{subchap:asymp_summary} for a more detailed summary of our obtained results and numeric checks on them.

The finite $N$ index is not mathematically as wieldy as the large $N$ infinite product in general, but nonetheless is of interest and importance.
For example, it has been discovered in the last few years that various unitary matrix models admit a so-called \textit{giant graviton expansion}~\cite{gaiotto_giantgraviton,murthy_giantgravitons,deddo2024giantgravitonexpansionbubbling,imamura2022analyticcontinuationgiantgravitons,imamura2024giantgravitonexpansiongeneral,Fujiwara:2023bdc}, where the integral can be written as an infinite series of terms which, on the gravity side, appear to hold the interpretation of expressing the contribution from collections of giant gravitons.
On the gauge theory side, this expansion can be viewed as an iterative correction on the large $N$ index to output the finite $N$ index.
For $\cN=4$ SYM in particular, this has been studied for the $\frac{1}{16}-$, $\frac{1}{8}-$ (Schur), modified Schur and $\frac{1}{2}-$BPS indices~\cite{murthy_giantgravitons,Liu_2023,Ezroura:2024wmp,Eleftheriou_2024,hatsuda2025deformedschurindicesmacdonald,Lee2024,Gautason:2024nru}, and in our paper we take a brief look at the expansion for quiver gauge theories with possibly more than one vertex, i.e. a \textit{matrix-coupling} generalization.

The rest of the paper is organized as follows.
In \cref{sec:review} we review the $\cN=1$ superconformal index for toric quivers and its large $N$ limit, and the factorization and zig-zag path techniques used to factorize $\det M(t)$ and the $a$-maximization procedure used to obtain the exact \textit{on-shell} $R$-charges. 
In \cref{section:factorizations} we obtain the factorization of $\det M(t)$ for the $Y^{p,q}$ and $\hat{A}_{m}$ families of quivers, along with the $dP3$ quiver. 
The obtained factorizations are listed out in \cref{subsec:summary}. 
In \cref{sec:analysis} we obtain asymptotics for the aforementioned classes of quivers, first in the univariate regime and then in the bivariate regime, along the main diagonal.
Finally, in~\cref{section:matrixmodelandgiantgrav} we discuss the giant graviton expansion for the generalized matrix-coupling model of which our index is a special case, concluding with a look at a few examples of the expansion parameters for toric quivers.
Appendices~\ref{appendix:jacobitripleproductandmacdonaldidentities} and~\ref{appendix:misc} provide supplementary information and evidence.

\section{Toric quiver superconformal index}
\label{sec:review}

We now provide a brief review of the $\cN=1$ superconformal index, and its large $N$ limit, for toric quivers. 
Further details on the index can be found in~\cite{gadde,holocheck} and references therein, and in~\cref{section:matrixmodelandgiantgrav} we briefly review the underlying matrix model. 
The index for radially-quantized SCFTs on $S^{3}\times \mathbb{R}$ was first investigated in~\cite{shirazall}. 
The $\cN=1$ index may be written in its \textit{left-handed} form with a particular selection of charges\footnote{This form is common in the literature~\cite{holocheck,gadde}. The index may be equivalently defined in a \textit{right-handed} fashion, or with other combinations of charges that commute with the regulator $\Xi$.} as
\begin{equation}
\cI= \tr (-1)^{F}e^{-\beta(D-2J_{1}+\frac{3}{2}R)}p^{\frac{1}{3}(D+J_{1})+J_{2}}q^{\frac{1}{3}(D+J_{1})-J_{2}}, \label{defsuperconformalindexnew}
\end{equation}  
where $F$ grades the supersymmetric Hilbert space by returning an eigenvalue of $0$ for bosonic states and $1$ for fermionic states, $\Xi=D-2J_{1}+\frac{3}{2}R$ is the modified Hamiltonian\footnote{It is necessary to modify the usual Hamiltonian of a radially quantized superconformal field theory, the dilation operator $D$, in order to meaningfully define the index~\cite{romelsberger}.}, and $\cM_{1,2}=\frac{1}{3}(D+J_{1})\pm J_{2}$ are ordinary charges with corresponding fugacity variables $p$ and $q$.

The $d=4$, $\cN=1$ superconformal algebra relates $\Xi$ to a particular selection of supercharges $Q,Q^{\dag}$ as $\Xi=\{Q,Q^{\dag}\}$. 
The set $\{D,J_{1},J_{2}\}$ generates the Cartan subalgebra of the $d=4$ conformal algebra, which along with the $R$-charge sector generated by $R$ constitutes the bosonic part of the superalgebra. 
The operators are selected so that $[\cM_{1,2},\Xi]=0$ and that $\Xi$ vanishes on the supersymmetric ground states, allowing the index \eqref{defsuperconformalindexnew} to quantify the anisomorphism between bosonic and fermionic ground state spaces. 
For these states, the vanishing of $\Xi$ imparts a functional interdependency between $D$ and $R$; hence the fugacity exponents $\cM_{1,2}$ in the index directly quantify the $R$-charge.
The state-operator correspondence of the theory lets us count these states in terms of primary operators generating them out of the vacuum in the infinite past; see e.g.~\cite{rychkov} and references therein for a review. 

Note that it is a characteristic feature of SCFTs that the superalgebra for a general degree of supersymmetry $\cN$ intertwines the $R$-charges with the other constituent generators~\cite{eberhardt}. 
Also, in general the system may possess extended $\cN>1$ supersymmetry, giving rise to an extended superconformal algebra, but as mentioned in~\cref{section:intro}, we will always consider the $\cN=1$ index. 
If there is extended supersymmetry, we condense the $R$-charge sector into a reduced set of $\cN=1$ $R$-charges. 
Reviews of the general $d=4$ superconformal algebra may be found in~\cite{eberhardt,shirazall}.

\subsection*{Quiver gauge theories}

A quiver gauge theory can be formally described by a set of integer-labelled vertices $\cV=\{ k : 1 \le k \le V \}$ and directed edges $\varepsilon:i\rightarrow j$ between these vertices encoded by the $V\times V$ adjacency matrix $A$ of the graph. 
Associated with a quiver, there may also be a superpotential, whose data can be graphically encoded as a subset of cycles on the quiver.
Associated to each vertex $i$ is a gauge group $G_{i}$; the full gauge symmetry of the theory is the product group $\bigtimes_{i} G_{i}$. 
Each edge $\varepsilon:i \rightarrow j$ corresponds to a bifundamental chiral multiplet to which we assign an $R$-charge $r(\varepsilon)$, and which transforms under the fundamental (respectively, antifundamental) representation of $G_{i}$ (respectively, $G_{j}$). 
Each vertex $i$ further corresponds to an $R$-charge-less vector multiplet in the adjoint representation of $G_{i}$. 

To evaluate the index~\eqref{defsuperconformalindexnew}, we consider the action of the traced operator on each constituent multiplet, and then multiply the individual indices together. 
We incorporate gauge degrees of freedom in~\eqref{defsuperconformalindexnew} by appending characters in the representations of the various fields in the multiplets; these are then integrated out over the full gauge group.
It is convenient to first evaluate the \textit{single-letter} index $\cI_{\mathrm{s.l.}}$, which is the restricted sum over all primary operators and their derivatives, and then apply the operation of plethystic exponentiation to arrive at the full index~\cite{gadde, holocheck}. 
If $f(x_{i})$ is a function of $x_{1},\ldots,x_{N}$, the plethystic exponential of $f$ at $(x_{i})$ is defined to be 
\begin{equation}
\operatorname{PE}[f(x_{i})]= \exp \left( \sum_{n=1}^{\infty} \frac{1}{n}f(x_{i}^{n})\right). \label{defplethystic}
\end{equation}

For a toric quiver, each gauge group has the same degree, and in this work we assume all gauge groups are $G_{i}=\mathrm{U}(N)$, meaning a $\mathrm{U}(N)$ gauge freedom is associated with each vertex. 
In the case of $\mathrm{U}(N)$ the number $N$ is the rank of the group.
Incorporating the gauge symmetry, the single-letter and full indices are, respectively
\begin{subequations}
\begin{equation}
\cI^{s}_{N}(p,q,U_{i},U^{\dag}_{i})=\underbrace{\sum_{i=1}^{V} \frac{-p-q+2pq}{(1-p)(1-q)} \tr(U_{i}) \tr(U^{\dag}_{i})}_{\textrm{Contribution from vector multiplets}}+\underbrace{\sum_{i,j=1}^{V}\sum_{\varepsilon:i\rightarrow j}\frac{(pq)^{\frac{1}{2}r(\varepsilon)}\tr(U^{\dag}_{i})\tr(U_{j})-(pq)^{1-\frac{1}{2}r(\varepsilon)} \tr(U_{i})\tr(U^{\dag}_{j})}{(1-p)(1-q)}}_{\textrm{Contribution from chiral multiplets}}, \label{quiversingleletterindexoriginal}
\end{equation}
and
\begin{equation}
\cI_{N}(p,q)=\int_{\bigtimes_{i=1}^{V}\mathrm{U}(N)} \prod_{i=1}^{V}\dd U_{i} \, \operatorname{PE}\left[\cI^{s}_{N}(p,q,U_{i},U^{\dag}_{i})\right]. \label{defindextoricquiver}
\end{equation}
\end{subequations}
The integral \eqref{defindextoricquiver} is a multi-dimensional matrix model over $V$ copies of $\mathrm{U}(N)$ -- a so-called `double-trace' model, because of the quadratic adjoint and bifundamental character terms.
In order to express it in the eigenvalue representation we employ the machinery of the Weyl integral formula, which allows for an integral over a compact Lie group to be expressed as one over a maximal torus of the group. 
For a single compact gauge group $G$ with a maximal torus $T$ and Weyl group $W$, the formula is
\begin{equation}
\int_{G} \dd g \, f(g) = \frac{1}{|W|} \int_{T} \dd t \left[ \det \left(I_{G/T}-\mathrm{Ad}_{G/T}(t^{-1}) \right) \int_{G} \dd g \, f(gtg^{-1}) \right], \label{weylintegrationformulamain}
\end{equation}    
where $g$ and $t$ parametrize $G$ and $T$, respectively, and the measure transformation factor $J(t)=\det \left(I_{G/T}-\mathrm{Ad}_{G/T}(t^{-1}) \right)$ is the Vandermonde determinant, expressible in terms of the maximal toral parameters (see~\cite{fultonharris,Adams1983-vh} for details). 
The Weyl group and maximal torus for $\mathrm{U}(N)$ are the permutation group $S_{N}$ and the standard $N$-torus $\mathbb{T}^{N}$, respectively, and we assume that the Haar measures on $G$ and $T$ are appropriately normalized. 
The complex-valued function $f$ appearing in \eqref{weylintegrationformulamain}, when specialized to our setting involving characters, becomes a class function. 

With variables $z_{ik}$ for $1\le k \le N$ parametrizing the $i$th maximal torus\footnote{We perform an inversion $z \mapsto 1/z$ of fugacity variables to transpose and be consistent with the usual depictions of the index and its large $N$ limit in the literature~\cite{holocheck,zigzag}. Equivalently, this is the invariance under inversion of the Haar measures on the $\mathrm{U}(N)$, meaning $\dd U_{i}=\dd U_{i}^{\dag}$ in \eqref{defindextoricquiver}.}, we are able to evaluate the full index at finite $N$ as

\begin{subequations}
\begin{eqnarray}
\cI_{N}(p,q) &=& \frac{1}{(N!)^{V}} \prod_{i=1}^{V}\prod_{j=1}^{N} \oint \frac{\dd z_{ij}}{2\pi i z_{ij}} \underbrace{\left[ \prod_{i=1}^{V} \prod_{\substack{k,l=1 \\ k \ne l}}^{N} \left|1-z_{ik}/z_{il} \right| \right]}_{\textrm{Vandermonde determinant}} \operatorname{PE} \left[ \sum_{i=1}^{V} \left( \frac{pq-1}{(1-p)(1-q)} +1\right) \sum_{k,l=1}^{N} z_{ik}/z_{il} \right. \nonumber \\ & & \left. +\sum_{i,j=1}^{V} \sum_{\varepsilon:i \rightarrow j} \sum_{r,s=1}^{N} \frac{(pq)^{\frac{1}{2}r(\varepsilon)}z_{ir}z^{-1}_{js}-(pq)^{1-\frac{1}{2}r(\varepsilon)}z^{-1}_{ir}z_{js}}{(1-p)(1-q)} \right] \label{formforlargeN} \\[+2mm]
&=& \frac{\prod_{k=1}^{\infty} \left(1-p^{k}\right)^{VN}\left(1-q^{k} \right)^{VN}}{(N!)^{V}} \prod_{i=1}^{V}\prod_{j=1}^{N} \oint \frac{\dd z_{ij}}{2\pi i z_{ij}} \prod_{i=1}^{V} \prod_{\substack{k,l=1 \\ k \ne l}}^{N} \left[ \frac{1}{\Gamma(z_{ik}z_{il}^{-1})} \right] \prod_{i,j=1}^{V} \prod_{\varepsilon :i \rightarrow j} \prod_{r,s=1}^{N} \Gamma \left((pq)^{\frac{1}{2}r(\varepsilon)}z_{ir}z^{-1}_{js} \right), \;\;\;\;\;\;\;\;\; \label{quiverfullindex}
\end{eqnarray}  
\end{subequations}
which is written in terms of the elliptic gamma function $\Gamma(z;p,q)=\prod_{i,j=0}^{\infty} \left( 1-p^{i+1}q^{j+1}/z \right)/\left(1-p^{i}q^{j}z \right)$; we suppress writing the dependence on $p$ and $q$ for brevity.
Note that for $N=1$ the quantities $ \frac{1}{\Gamma(z_{ik}z_{il}^{-1})}$ are identically taken as $1$.
The finite $N$ index, and in general such a double-trace matrix model, may be equivalently expressed in terms of characters of the symmetric group $S_{N}$, which are indexed by integer partitions; this has been performed for scalar coupling constants in~\cite{murthy_giantgravitons}, but we do not pursue this direction in this paper. 
In~\cref{section:matrixmodelandgiantgrav}, utilizing the Hubbard--Stratonovich transformation \cite{stratonovich,hubbard} to express this model in terms a generalized Gross--Witten--Wadia model~\cite{gwworiginal1,wadia1980cp,wadia2012study}, we proceed on to the giant graviton expansion of the index.

\subsection{Large \texorpdfstring{$N$}{N} Index}

Using standard Coulomb gas techniques ~\cite{Dolan_2009,shirazall,Benvenuti_2007,holocheck,hagedorn} (also see \cref{section:matrixmodelandgiantgrav}), the large $N$ index, meaning the large $N$ limit at leading order of the matrix model \eqref{formforlargeN}, may be derived as
\begin{equation}
\cI_{\infty}(p,q)=\prod_{k=1}^{\infty}\frac{\left(1-p^{k}\right)^{V}\left(1-q^{k}\right)^{V}}{\det M\left(p^{k},q^{k}\right)} \equiv \prod_{k=1}^{\infty}\frac{\left(1-p^{k}\right)^{V}\left(1-q^{k}\right)^{V}}{\det M\left(t^{k}\right)}, \label{deflargeNindex}
\end{equation}
with the normalization $\cI_{\infty}(0,0)=1$, and $M(p,q) \equiv M(t)$ is the $V\times V$ \textit{large $N$ index matrix} with univariate dependence on $t=(pq)^{1/2}$. 
The large $N$ index matrix is defined by the equation
\begin{equation}
M(t)=(1-t^{2})I-\cM(t)+t^{2}\cM^{T}(t^{-1}) \label{mainmatrix}
\end{equation}  
with $\cM(t)$ being the $V\times V$ $R$-charge weighted adjacency matrix of the quiver, having entries
\begin{equation}
\cM_{ij}(t)=\sum_{\varepsilon:i \rightarrow j}t^{r(\varepsilon)}, \label{weightedadjacencymatrix} 
\end{equation}
so that the usual adjacency matrix for the underlying graph of the quiver is then $A=\cM(1)$. 
Despite emerging out of the large $N$ limit of the full index \eqref{quiverfullindex}, $M(t)$ can be expressed purely in terms of the contents of the single-letter index plethystically exponentiated in \eqref{formforlargeN}. 
The factorization of $\det M(t)$, discussed in \cref{subsec:factorization}, plays an important role in our eventual asymptotic analysis in \cref{sec:analysis}, where $\cI(p,q)$ has the interpretation of a generating function. 

\subsubsection*{Bivariate and univariate series expansion}
The index \eqref{deflargeNindex} has a series expansion
\begin{equation}
\cI_{\infty}(p,q)= \sum_{m,n}c_{mn}p^{m}q^{n}, \label{indexasgeneratingfunction}
\end{equation}
with $|c_{mn}|$ providing a bound on the state degeneracy for for eigenvalues $m$ and $n$ of the operators $\cM_{1}$ and $\cM_2$ respectively. 
The allowed eigenvalues of the dilation operator $D$ (the scaling dimension) are always positive integers or half-integers and form an infinite set bounded from below\footnote{This follows from the state-operator correspondence and unitarity bounds on the conformal dimension; see~\cite{gadde,rychkov,shirazsolo} for details.}, and the eigenvalues of the angular momentum operators $J_{1,2}$ are $\pm \frac{1}{2}$. 
Hence, the indices $m$ and $n$ in \eqref{indexasgeneratingfunction} are bounded from below. 

We may convert the bivariate index~\eqref{deflargeNindex} into a \textit{univariate} index by removing a fugacity degree of freedom.
For example, we can set $p=q=t$ to obtain the univariate index
\begin{equation}
\cI_{\infty}(t)= \prod_{k=1}^{\infty}\frac{\left(1-t^{k}\right)^{2V}}{\det M\left(t^{k}\right)}= \sum_{n}c_{n}t^{n} \label{indexasgeneratingfunctionunivariate}
\end{equation}
in one fugacity variable $t$, and in this interpretation $|c_{n}|$ bounds the degeneracy at eigenvalue $n$ for the operator $\cM_{1}+\cM_{2}$. 
This index has been analyzed in~\cite{murthy2020} for $\cN=4$ SYM.
Alternatively, we may take a subsequence of~\eqref{indexasgeneratingfunction} along a specific direction $(r,s)$,
\begin{equation}
    \cI_{\infty}^{(r,s)}(p^{r}q^{s})=\sum_{n} c_{nr,ns}p^{nr}q^{ns}. \label{indexsubsequence}
\end{equation}
Such an index is also univariate in the variable $p^{r}q^{s}$, but it is possibly more effective than~\eqref{indexasgeneratingfunctionunivariate} at communicating information about the dual gravity theory; see~\cref{subsection:bivariate} for further details, where we consider the main diagonal $(1,1)$.

\subsection{Factorization of the Index}
\label{subsec:factorization}

We now discuss the techniques we use to factorize the determinant of the large $N$ index matrix \eqref{mainmatrix} of a toric quiver, beginning with the relatively easy example of $\cN=4$ SYM. 
As observed in~\cite{holocheck}, and later proven in~\cite{eagerschmudetachikawa} for exact $R$-charges, the factorization of the large $N$ index matrix \eqref{mainmatrix} has the generic form
\begin{equation}
\det M(t)= \prod_{i=1}^{Z} \left(1-t^{f_{i}(r(\varepsilon))} \right) \label{formoffactorizationofdetMtintro}
\end{equation}
for some positive integer $Z$ and functions $f_{i}$ of the $R$-charges $r(\varepsilon)$ of the quiver. 
This factorization also seems to hold true when the $R$ charges are relaxed by parametrizing them with some free variables~\cite{zigzag}. 
The factorization may be performed explicitly by hand or with a computer algebra system in the simpler cases of low rank or in the presence of symmetry. 
Unfortunately, this can become complicated for larger quivers. To work around this difficulty, an alternative graphical method has been demonstrated to work for several classes of toric quivers. 
In the examples we will study, cycle structures in the factorization appear only upon the imposition on certain constraints on the $R$-charges; it appears unlikely for a completely unconstrained set of $R$-charges. 

The strongest set of constraints would be to fix the $R$-charges exactly through the process of $a$-maximization~\cite{intriligatormax, eagerschmudetachikawa}, reviewed in \cref{subsec_a_maximization}, but we shall see that a weaker set of constraints is sufficient. 
The first condition\footnote{This follows from $F$-terms in the superfield formalism.} is that all terms of the superpotential have a combined $R$-charge of $2$.
The second relates to the vanishing of the NSVZ beta function at each node~\cite{zigzag}, i.e.
\begin{equation}
\sum_{i=1}^{n_{k}} (1-r_{i})=2, \label{mainconstraint}   
\end{equation}
for any node $k$, with the sum containing the $R$-charges $r_{i}$ of all $n_{k}$ bifundamental fields charged under the $k$th gauge group.
Note that this has the consequence that cycles are taken twice. 

If explicit factorization is not feasible by direct computation, we may use the \textit{method of zig-zag paths} on the dual graph of the toric embedding of the quiver, which is a \textit{dimer model}. 
This technique, implemented graphically for toric quivers with a superpotential, is conjectural for inexact $R$-charges. A detailed description of dimer models and zig-zag paths may be found in~\cite{dimermain,kenyon,zigzag}, and we give a brief summary here. 
For toric quivers, each field occurs exactly twice in the superpotential in terms of alternating sign. This property allows us to `open up' the quiver, which may consist of multiple edges connecting two nodes, into a planar quiver, whose opposite edges are identified allowing an embedding\footnote{This is possible because the Euler characteristic of the graph vanishes~\cite{branedimers}.} on the torus $S^{1} \times S^{1}$ as a periodic bipartite tiling.

We then can define certain zig-zag paths on the dual dimer that weave and cut across the edges. 
This dual dimer (henceforth just called `dimer') is also toric, periodic and bipartite and has an isoradial embedding, i.e. each face can be inscribed in a circle whose centre lies inside the face.
Under this embedding, the zig-zag paths are non-intersecting simple closed curves on the torus, and the entire dimer is spanned by a unique set of the paths such that each edge has exactly two such paths going through it in opposite directions.

Since the edges of the dimer represent the chiral multiplets, we can associate a set of $R$-charges to any zig-zag path corresponding to the edges it passes through. 
Let $Z$ be the number of zig-zag paths, and for $i = 1,2,\ldots,Z$ let $\{Z_{i}\}$ be the set of edges of the path $Z_{i}$. 
Then we have 
\begin{equation}
\det M(t)=\prod_{i=1}^{Z} \left(1-t^{\sum_{j \in \{ Z_{i}\}}1-r_{j}^{(i)}}\right), \label{deffactorzigzag}
\end{equation}
where $r_{j}^{(i)}$ is the $R$-charge associated with edge $j$ of the $i$th path. 
Following~\cite{zigzag}, we adopt the terminology of calling the situation of exactly fixed $R$-charges as \textit{on-shell}, and that of partial fixing with the superpotential and beta function conditions~\eqref{mainconstraint} as \textit{off-shell}. 
The authors of~\cite{zigzag} have checked \eqref{deffactorzigzag} to be true off-shell for several examples of quivers, some of which we revisit in this paper. 
We observe that under off-shell conditions, the factorization \eqref{deffactorzigzag} appears to correspond exactly to a factorization over the cycles of the quiver, with a one-to-one correspondence between the cycles and the zig-zag paths.
This corroborates the more general observations in~\cite{Pasukonis_2013} --
specifically in our treatment with flavour symmetries switched off, it seems that \eqref{deffactorzigzag} can generally be rewritten as 
\begin{equation}
\det M(t)=\prod_{i=1}^{Z} \left(1-t^{\sum_{j \in \{L_{i}\}}r_{j}^{(i)}}\right), \label{deffactorcycles}    
\end{equation}
where $L_{i}$ is the $i$th cycle consisting of a set of edges $\{L_{i}\}$, and $r_{j}^{(i)}$ is the $R$-charge associated with edge $j \in \{L_{i}\}$. 
In \cref{section:factorizations} we detail the factorization of $\det M(t)$ for a few examples of quivers, and we investigate whether going off-shell is necessary or sufficient for the emergence of the cycle structure \eqref{deffactorcycles}. 
In other words, we test whether the charges can be unconstrained, need to be off-shell, or must be exactly specified on-shell. 
We verify our obtained results by explicit factorization for each example, or for a few selected quivers from each family of examples. 

\begin{figure}[H]
\centering
\includegraphics[width=5 cm]{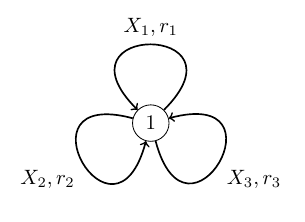}
\caption{The clover quiver for $\cN=4$ SYM.}
\label{clover}
\end{figure}

For instance, we demonstrate the factorization for the clover quiver describing $\cN=4$ SYM and depicted by~\cref{clover}. 
This quiver exhibits $\cN=4$ supersymmetry, but we work with the $\cN=1$ index. 
The dimer model corresponding to this is described in~\cite{dimermain}, and the zig-zag path calculations are done in~\cite{zigzag}; we reproduce these here. 
For this quiver there are three chiral multiplets $X_{1,2,3}$ with respective $R$-charges $r_{1,2,3}$ corresponding to the three cycles. 
$M(t)$ in this case is just a number, so from \eqref{mainmatrix} we have 
\begin{equation}
\det M_{\mathrm{clover}}(t)=1-t^{2}+\sum_{i=1}^{3}\left(t^{2-r_{i}}-t^{r_{i}}\right). \label{detclovermain}
\end{equation}

\begin{figure}[H]
\centering
\subfloat[The planar quiver.]
{
\includegraphics[width=5 cm]{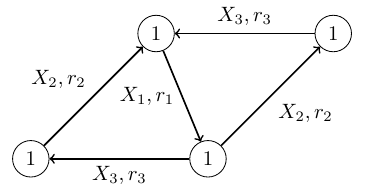}
\label{clover2}
}
\qquad 
\subfloat[The dimer model.]
{
\includegraphics[width=6 cm]{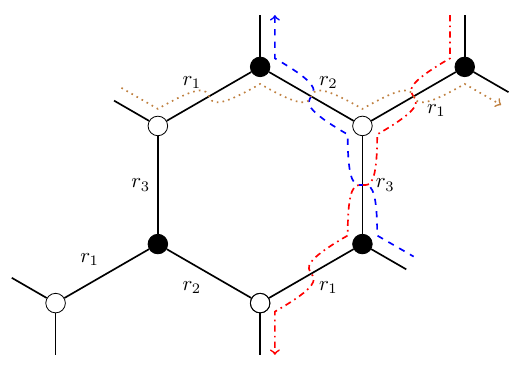}
\label{clover3}
}
\caption{The periodic tiling and the corresponding dimer with three zig-zag paths for $\cN=4$ SYM.}
\label{clover23}
\end{figure}

A cycle structure is not immediately obvious in \eqref{detclovermain}, and to get it we need to go off-shell. 
The superpotential condition \eqref{mainconstraint} for the sole node gives $\sum_{i=1}^{3}2(1-r_{i})=2$, or $r_{1}+r_{2}+r_{3}=2$. 
The superpotential for this quiver has the form $X_{1}X_{2}X_{3}-X_{1}X_{3}X_{2}$~\cite{dimermain}, and we note that this gives back the same condition. 
Plugging this into \eqref{detclovermain} gives
\begin{equation}
\det M_{\mathrm{clover}}(t)= \left(1-t^{r_{1}}\right)\left(1-t^{r_{2}}\right)\left(1-t^{r_{3}}\right), \label{detclover1}
\end{equation}
where we are able to easily identify the cycles as the three directed edges that form the clover. 
We now calculate this factorization using the method of zig-zag paths to demonstrate the dimer model technique, opening up the quiver selecting the $X_{1}$ edge as common for the cycles describing the superpotential terms. 
The toric dimer is then constructed as a tiling of hexagons on the plane. 
\cref{clover23} shows the planar quiver and the dimer with the three zig-zag paths, denoted by dashed blue, dash dot red and dotted brown, with their associated $R$-charges.
Only one iteration of each path is shown for brevity.
White and black nodes demarcate two different perfect matchings for the dimer, and opposite edges and nodes are identified in both cases. 
Identifying the $R$-charges associated with these paths using~\eqref{deffactorzigzag}, we are able to reproduce \eqref{detclover1} by going off-shell,
\begin{equation}
\det M_{\mathrm{clover}}(t)=\left(1-t^{2-r_{2}-r_{3}}\right)\left(1-t^{2-r_{3}-r_{1}}\right)\left(1-t^{2-r_{1}-r_{2}}\right)=\left(1-t^{r_{1}} \right)\left(1-t^{r_{2}} \right)\left(1-t^{r_{3}} \right). \label{detcloveragain}    
\end{equation} 

\subsection{\texorpdfstring{$a$}{a}-maximization}
\label{subsec_a_maximization}

The off-shell $R$-charges in any quiver gauge theory are typically constrained to be expressed in terms of a set of parameters, less than or equal to the number of $R$-charges. 
To further constrain the $R$-charges to a unique set of values, the authors of~\cite{intriligatormax} devised a method, briefly revisited here, involving locally maximizing the $a$ central charge.
For any~$d=4$ CFT, we can define two central charges, $a$ and $c$~\cite{Anselmi_1998}, in terms of the $R$-charge operator $R$, 
\begin{equation}
a(R)=\frac{3}{32} \left(3 \tr R^{3}-\tr R \right), \; c(R)=\frac{1}{32} \left(9 \tr R^{3}-5\tr R \right). \label{amaximizationmain}
\end{equation}
These quantify 't Hooft anomalies~\cite{sasaki,intriligatormax}.
The traces over $R$ or $R^{3}$ are fermionic traces: they extract out the $R$-charges of the individual species of fermions in the chiral and vector multiplets. 
For the theories we consider, each vector multiplet has one fermion, a gaugino, with $R$-charge $1$, and for any chiral multiplet represented by an edge $\varepsilon$ the corresponding fermion has $R$-charge $r(\varepsilon)-1$. 
In our notation, for off-shell quivers we then have
\begin{equation}
\tr R= V+ \sum_{i,j=1}^{V} \sum_{\varepsilon:i \rightarrow j} (r(\varepsilon)-1), \; \tr R^{3}= V+ \sum_{i,j=1}^{V} \sum_{\varepsilon:i \rightarrow j} (r(\varepsilon)-1)^{3}. \label{traceofR_R3}
\end{equation}
It has been proven that for superconformal quivers, $\tr R$ vanishes~\cite{newresults}, and this is straightforward to check for the examples presented in \cref{section:factorizations}.
This has the consequence that $a=c$.\footnote{This is more generally true for theories with $AdS_{5}$ gravity duals at large $N$~\cite{Henningson_1998}.} 
Ignoring multiplicative and additive terms, the function to maximize becomes
\begin{equation}
a(R)= \sum_{i,j=1}^{V} \sum_{\varepsilon:i \rightarrow j} (r(\varepsilon)-1)^{3}. \label{amaximizationnext}
\end{equation}

Let us assume the off-shell $R$-charges are parametrized by variables $x_{1},x_{2},\ldots,x_{s}$ with $1 \le s \le V$, so that we have the functional form $a(R) \equiv a(x_{1},x_{2},\ldots,x_{s})$. 
The standard way to find maxima of $a(R)$ is then to solve the system of equations defined by the vanishing of the partial derivatives $\partial a/\partial x_i$, using the eigenvalues of the Hessian matrix with entries $H_{ij}=\partial^{2}a/\partial x_{i} \partial x_{j}$ to characterize when a point is a local maxima. 
In particular, the points for which the Hessian is negative-definite (all eigenvalues are negative) are local maxima, however we may be unable to conclude anything about those points where the Hessian has zero as an eigenvalue, in which case we need to analyze higher derivatives. 
In \cref{section:factorizations} we find or conjecture the local maxima of \eqref{amaximizationnext} for all considered classes of toric quivers. 

As an example, it is straightforward to reproduce the exact $R$-charges for the $\cN=4$ SYM quiver by $a$-maximization to match the values taken for the $\cN=4$ SYM Lagrangian~\cite{shirazall,murthybizet}.~\eqref{amaximizationnext} is parametrized by two variables and has the form
\begin{equation}
a(r_{1},r_{2})=(r_{1}-1)^{3}+(r_{2}-1)^{3}+(1-r_{1}-r_{2})^{3}, \label{afunctionclover}
\end{equation}
hence at the extrema or saddle points we have 
\begin{equation}
(r_{1}-1)^{2}-(1-r_{1}-r_{2})^{2}=(r_{2}-1)^{2}-(1-r_{1}-r_{2})^{2}=0, \label{dervanclover}
\end{equation}
implying either $r_{1}=r_{2}$ or $r_{1}+r_{2}=2$. 
The case $r_{1}=r_{2}$ gives $(r_{1},r_{2},r_{3})=(\frac{2}{3},\frac{2}{3},\frac{2}{3})$ or $(0,0,2)$, while the other case gives $(r_{1},r_{2},r_{3})=(2,0,0)$ or $(0,2,0)$. 
The Hessian matrix is 
\begin{equation}
H= 6\begin{pmatrix}
-r_{2} & 1-r_{1}-r_{2} \\ 1-r_{1}-r_{2} & -r_{1} \label{hessianclover}
\end{pmatrix},
\end{equation}   
and it may be verified that only the solution $(\frac{2}{3},\frac{2}{3},\frac{2}{3})$ is a local maximum, resulting in all negative eigenvalues. 
This is the on-shell $R$-charge assignment for the clover quiver and is consistent with descriptions of $\cN=4$ SYM in the literature. 

\section{Factorization and cycle structure}
\label{section:factorizations}
In this section we work out techniques to factor $\det M(t)$ for a collection of toric quivers. We also derive, conjecture or mention the exact (on-shell) $R$-charges in each case;  typically, we need to go off-shell to arrive at the cycle structure. 
The cycles in our examples are sometimes non-trivial to identify in their respective quivers, and require an algorithmic description.
We get the impression that the cycles come in families, but at this stage are unable to provide a full categorization of cycles for a general toric quiver or a scheme for classification of toric quivers based on the kinds of cycles they contain. 

\subsection{Summary of factorizations}
\label{subsec:summary}

Here we summarize the factorization results obtained in this section; the reader wishing to skip our derivations may proceed directly to~\cref{sec:analysis} after surveying these results. We take up the $Y^{p,q}$ and $\hat{A}_{m}$ infinite families of quivers, and the $dP3$ quiver, obtaining a factorization of $\det M(t)$ off-shell in each case. For the cases not previously determined in the literature, we calculate or indicate the probable on-shell $R$-charges associated with these quivers using $a$-maximization. 

For completeness we reproduce here the factorization for $Y^{p,q}$~\cite{holocheck,zigzag}. 
The cycle expansion for $Y^{p,q}$, and the factorization, $a$-maximization and cycle expansions for other two cases are, to the best of our knowledge, new.
\begin{itemize}
\item[(a)] For the $Y^{p,q}$ family of quivers, we reproduce the dimer model described in~\cite{zigzag,branedimers} to obtain the off-shell cycle factorization
\begin{equation}
\det M_{Y^{p,q}}(t)= \left(1-t^{pr_{U}+qr_{V}+(p-q)r_{Z}} \right)^{2}\left(1-t^{(p-q)r_{U}+pr_{Y}}\right)^{2}. \label{detMpqsummary}
\end{equation}
Note that \eqref{detMpqsummary} corresponds off-shell to result (A.2) in~\cite{zigzag}, and is consistent on-shell with result (5.16) of~\cite{holocheck} for the special cases of $q=p$ and $q=0$. In our analysis we use the superpotential described in~\cite{holocheck,sasaki}. For $p>q \ne 0$ we corroborate the results of~\cite{zigzag} and for the special cases of $q=p$ and $q=0$ we are able to verify \eqref{detMpqsummary} by hand due to the block-circulant nature of $M(t)$. We also verify a few cases of $p>q \ne 0$ using computer algebra. The off-shell $R$-charges are parametrized by two variables $x$ and $y$~\cite{holocheck,sasaki} via the equations
\begin{equation}
r_{U}=1-\frac{1}{2}(x+y),\;\;\; r_{V}=1+\frac{1}{2}(x-y), \;\;\; r_{Y}=y, \;\;\; r_{Z}=x, \label{partlyfixingRcharges}
\end{equation}
and the on-shell values for $q>0$ are given~\cite{holocheck,sasaki} by the substitutions 
\begin{equation}
y= \frac{1}{3q^{2}} \left(-4p^{2}+2pq+3q^{2}+(2p-q)\sqrt{4p^{2}-3q^{2}}  \right), \;
x= \frac{1}{3q^{2}} \left(-4p^{2}-2pq+3q^{2}+(2p+q)\sqrt{4p^{2}-3q^{2}}  \right). \label{pqforalphamaximization}
\end{equation} 
For $q=0$ we instead obtain 
\begin{equation}
x=y=\frac{1}{2} \label{xy_q=0}
\end{equation}
in~\eqref{partlyfixingRcharges}, with $r_{V}$ set to $0$, which is consistent with~\cite{holocheck}.

\item[(b)] For the $\hat{A}_{m}$ family of quivers we fix a superpotential and use the dimer and zig-zag path techniques to obtain the cycle factorization of $\det M(t)$ off-shell. 
We obtain
\begin{equation}
\det M_{\hat{A}_{m}}(t)= \left( 1-t^{\sum_{k=1}^{n}r_{V_{k}}}\right)\left( 1-t^{\sum_{k=1}^{n}r_{U_{k}}}\right)\prod_{k=1}^{n}\left( 1-t^{r_{Y_{k}}}\right), \label{detMAnsummary} 
\end{equation}
where the $r_{Y_{k}}$ are equal. 
For the special off-shell case of equal $r_{U_{k}}$ and equal $r_{V_{k}}$, which covers our claimed on-shell values, we are able to verify this calculation by hand due to the circulant nature of $M(t)$, and we also verify some general off-shell cases using computer algebra. We calculate that the exact $R$-charges are
\begin{equation}
r_{U_{k}}=r_{V_{k}}=r_{Y_{k}}=\frac{2}{3}. \label{exactRchargesAn}
\end{equation}

\item[c)] 
For the $dP3$ quiver we use the dimer model and superpotential described in~\cite{branedimers,delpezzo3} to obtain the off-shell cycle factorization
\begin{equation}
\det M_{dP3}(t)= \prod_{k=1}^{6} \left(1-t^{r_{k-2,k}+r_{k,k+2}+r_{k+2,k+3}+r_{k+3,k-2}} \right), \label{detMdP3summary}
\end{equation}
which is verified with computer algebra. The exact $R$-charges in this setting are
\begin{equation}
r_{k,k+1}=\frac{1}{3} \text{ and } r_{k,k+2}=\frac{2}{3}. \label{exactRchargespP3}
\end{equation}
\end{itemize}

\subsection{The \texorpdfstring{$Y^{p,q}$}{Ypq} family}
\label{summaryYpq}
We first investigate the infinite class of toric quivers $Y^{p,q}$, whose vertices can be arranged in a $2p$-gon. 
Index calculations for this class have previously been discussed in~\cite{zigzag,holocheck}. 
The $Y^{p,q}$ quivers consist of $2p$ $U$ fields, $2q$ $V$ fields, $p+q$ $Y$ fields and $p-q$ $Z$ fields. 
Note that these variables $p$ and $q$ are not to be confused with the fugacity factors.
We sketch the construction of the $Y^{p,p}$ quivers first, from which the $Y^{p,q}$ quivers can be built iteratively. 
A complete description of these constructions can be found in~\cite{sasaki,holocheck}.

\begin{figure}[H]
\centering
\subfloat[Removing $Y:8 \rightarrow 1,\, Y:4 \rightarrow 5$.]
{
\includegraphics[width=5 cm]{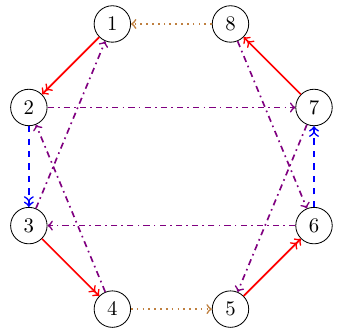}
\label{Y42a}}
\qquad 
\subfloat[Removing $Y:8 \rightarrow 1, \, Y:6 \rightarrow 7$.]
{
\includegraphics[width=5 cm]{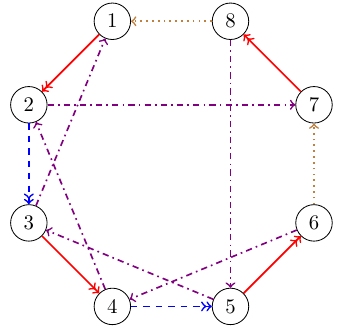}
\label{Y42b}}
\caption{An example of toric duality in the $Y^{4,2}$ quiver.}
\label{y42}
\end{figure}

For the $Y^{p,p}$ quivers there are $2p$ fields of the form $U_{i}^{\alpha}: 2i-1 \twoheadrightarrow 2i$, $2p$ fields $V_{i}^{\alpha}: 2i \twoheadrightarrow 2i+1$ and $2p$ fields $Y: j \rightarrow j-2$, where $\alpha\in\{1,2\}$ with $1\le i \le p$ and $1\le j \le 2p$ and the quantities are taken modulo $2p$. The superpotential for a $Y^{p,p}$ quiver has the form
\begin{equation}
W_{Y^{p,p}}=\sum_{k=1}^{p} \epsilon_{\alpha \beta} \left(U_{k}^{\alpha}V_{k}^{\beta}Y_{2k+1}+V_{k}^{\alpha}U_{k+1}^{\beta} Y_{2k+2} \right), \label{superpotentialypp}
\end{equation} 
which may be interpreted as a sum over triangular $U-V-Y$ cycles in the quiver. 
Let us consider four consecutive nodes $\{2i-1,2i,2i+1,2i+2\}$, with $1 \le i \le 2p$ and all indices modulo $2p$ as usual. 
To get a $Y^{p,p-1}$ quiver from a $Y^{p,p}$ quiver we remove $Y_{2i+1}$ and $Y_{2i+2}$ and introduce $Y_{2i+2}:2i+2 \rightarrow 2i-1$, then we replace the two $V^{\alpha}_{i}$ by one $Z_{i}: 2i \rightarrow 2i+1$. 
This process, continued $p-q$ times, yields a $Y^{p,q}$ quiver, with superpotential given by
\begin{equation}
W_{Y^{p,q}}=\sum_{k} \epsilon_{\alpha \beta} \left(U_{k}^{\alpha}V_{k}^{\beta}Y_{2k+1}+V_{k}^{\alpha}U_{k+1}^{\beta} Y_{2k+2} \right)+\sum_{l \ne k} \epsilon_{\alpha \beta} Z_{l}U_{l+1}^{\alpha}Y_{2l+2}U_{l}^{\beta}, \label{superpotentialypq}
\end{equation}
where $k$ takes entries from a finite subset of $ \{1,2,\ldots,p\}$ with cardinality $q$, and $l$ takes the remaining $p-q$ entries.
The freedom of choice in selecting $k$ and $l$ manifests physically as the toric duality of such quivers~\cite{Feng_2001,holocheck,romelsberger}. 
It has been worked out in several cases that the exact superconformal index for finite $N$ is identical for dual quivers\footnote{This has interesting connections to identities of hypergeometric functions as the finite $N$ index is written in terms of elliptic gamma functions, see~\cite{spiridonov,SPIRIDONOV2010192} and also~\cite{holocheck}.}~\cite{Dolan_2009}, so we expect the same to hold true at large $N$.
Borrowing from~\cite{holocheck}, in~\cref{y42} we show two different representations of the $Y^{4,2}$ quiver, constructed from the $Y^{4,4}$ quiver by removing different $Y$ fields.
For both of these quivers, we will shortly present the cycle structure.
Following~\cite{holocheck}, we use solid red for $U$ fields, dashed blue for $V$ fields, dash-dot violet for $Y$ fields and dotted brown for $Z$ fields.

\subsubsection*{General formulation using zig-zag paths}

\begin{figure}[H]
    \centering
    \subfloat[The repetitive unit in the planar quiver for $Y^{p,p}$.]
    {
    \includegraphics[width=8 cm]{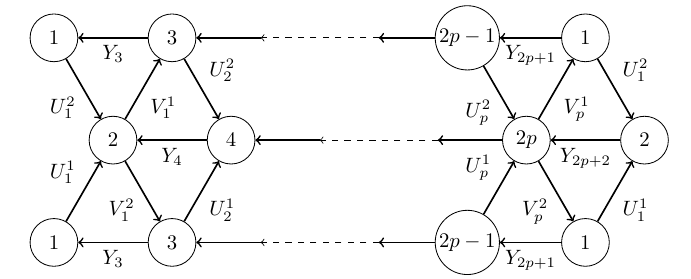}
    \label{Yppplanar}
    }
    \subfloat[The substitution carried out at the $m$th block. The periodic structure completes the two superpotential cycles.]
    {
     \includegraphics[width=8 cm]{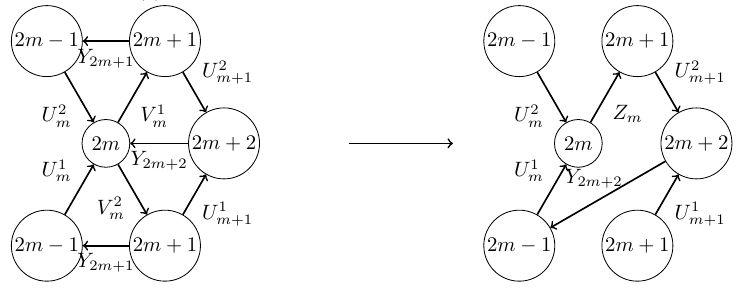}
     \label{Ypqsubstitution}
     }
    \caption{Construction of the planar quiver for $Y^{p,q}$.}
    \label{Ypqplanar}
\end{figure}

It is instructive to find $\det M(t)$ for the $Y^{p,q}$ quivers using the method of zig-zag paths, and then check our results by hand and computer algebra for various cases.
Imposing off-shell conditions, we reproduce the dimer described in~\cite{holocheck,branedimers}. 
First, we construct the planar quiver corresponding to the $Y^{p,p}$ quiver in terms of the repetitive unit shown in~\cref{Yppplanar}, which has $p$ blocks and whose opposite edges and nodes are identified. 
The substitution mechanism we have discussed corresponds to changing $p-q$ of these blocks in the manner shown in~\cref{Ypqsubstitution}, again with toric identification of edges and nodes.

\begin{figure}[H]
\centering
\subfloat[Unsubstituted blocks.]
{
\includegraphics[width=4 cm]{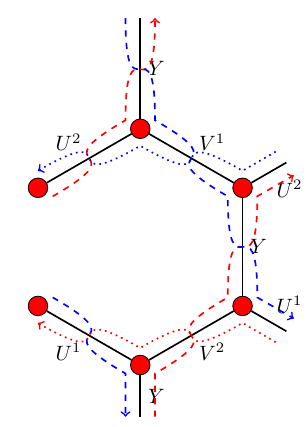}
\label{Ypqdimer1}
}
\qquad
\subfloat[$Z$-substituted blocks.]
{
\includegraphics[width=4 cm]{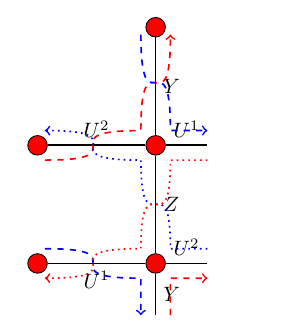}
\label{Ypqdimer2}
}
\caption{The four zig-zag paths for the $Y^{p,q}$ dimer, presented in dotted red and blue, and dashed red and blue, respectively.}
\label{Ypqdimer}
\end{figure}

The dimer corresponding to the quiver is then a set of hexagons and deformed hexagons, which can also be interpreted in terms of blocks. These may be found from each of the two types of blocks in \cref{Ypqsubstitution}, and are associated with exactly four zig-zag paths that traverse the dimer, as shown in~\cref{Ypqdimer}.
For simplicity we do not show a more extended picture of the embedding and the perfect matchings.
We indicate the spinor index of the $U$ and $V$ fields to indicate the warping structure for each block.

We now use these zig-zag paths to evaluate $\det M(t)$. 
Since the doublet $U$ and $V$ fields have same $R$-charge, a priori there are $p+q+p+q+p-q=3p+q$ independent $R$-charges~\cite{sasaki}.
Note that the dotted paths are equivalent to each other in the type of fields they weave across, and similarly for the dashed paths. In the $Y^{p,q}$ quiver, the unsubstituted blocks indicated by \cref{Ypqdimer1} will appear $q$ times, and the $Z$-substituted blocks indicated by \cref{Ypqdimer2} will appear $p-q$ times. Hence from \eqref{deffactorzigzag}, and with the same notation of \eqref{superpotentialypq}, we achieve the factorization
\begin{equation}
\begin{multlined}
\det M_{Y^{p,q}}(t)= \underbrace{\left(1-t^{\sum_{k} \left(4-r_{V_{k}}-r_{U_{k}}-r_{Y_{2k+1}}-r_{Y_{2k+2}} \right)+\sum_{l \ne k} \left(2-r_{U_{l}}-r_{Y_{2l+2}} \right)} \right)^{2}}_{\textrm{from dashed paths}} \underbrace{\left(1-t^{\sum_{k}(2-r_{U_{k}}-r_{V_{k}})+\sum_{l\ne k}(2-r_{U_{l}}-r_{Z_{l}})}\right)^{2}}_{\textrm{from dotted paths}} \\
= \left(1-t^{2(p+q)-\sum_{k} \left(r_{V_{k}}+r_{U_{k}}+r_{Y_{2k+1}}+r_{Y_{2k+2}} \right)-\sum_{l \ne k} \left(r_{U_{l}}+r_{Y_{2l+2}} \right)} \right)^{2} \left(1-t^{2p-\sum_{k}(r_{U_{k}}+r_{V_{k}})-\sum_{l\ne k}(r_{U_{l}}+r_{Z_{l}})}\right)^{2}.
\label{detYpqunconstrained}
\end{multlined}
\end{equation}
We need to invoke the off-shell constraints to be able to identify cycles in this factorization. 
For any $Y^{p,q}$ quiver, the off-shell assignment of $R$-charges leads to exactly two unconstrained variables~\cite{sasaki,holocheck}, which may be parametrized by~\eqref{partlyfixingRcharges}, where
the $R$-charges for fields of the same type are equal. 
Using mathematical identities derived from~\eqref{partlyfixingRcharges} in the first relation of \eqref{detYpqunconstrained}, we get
\begin{eqnarray}
\det M_{Y^{p,q}}(t) &=& \left(1-t^{q(4-r_{U}-r_{V}-2r_{Y})+(p-q)(2-r_{U}-r_{Y})}\right)^{2} \left(1-t^{q(2-r_{U}-r_{V})+(p-q)(2-r_{U}-r_{Z})}\right)^{2} \nonumber \\
&=& \left(1-t^{q(r_{U}+r_{V})+(p-q)(r_{U}+r_{Z})} \right)^{2}  \left(1-t^{qr_{Y}+(p-q)(r_{U}+r_{Y})}\right)^{2} \nonumber \\ &=& \left(1-t^{pr_{U}+qr_{V}+(p-q)r_{Z}} \right)^{2}\left(1-t^{(p-q)r_{U}+pr_{Y}}\right)^{2}. \label{detMpqoffshell}
\end{eqnarray}  
We can now identify two types of cycles in the quiver. 
The two cycles corresponding to the dashed zig-zag paths, i.e. the first factor in \eqref{detMpqoffshell}, are obvious: they run around the whole $2p$-gon cyclically, passing through $p$ $U$-edges, $q$ $V$-edges and $(p-q)$ $Z$-edges;
we name them \textit{ring cycles}.
In general, the other two cycles corresponding to the dotted zig-zag paths, i.e. the second factor in \eqref{detMpqoffshell}, are less obvious: they involve twists due to the `corruption' introduced by the substitution procedure; 
we name them \textit{$U-Y$ cycles}.
For the special case of $q=p$, they may be identified as `rings' over the $p$ odd and $p$ even $Y$-edges respectively. 
For the special case of $q=0$, they can be identified as a pair of symmetric `twists' that advance one node through $U$ and go back three nodes through $Y$, and do so $p$ times. 

\subsubsection*{Constructing the $U-Y$ cycles for $0<q \le p$.}

We now consider the $U-Y$ cycles for general $0<q \le p$.
Since the $Z$-substitution is done $p-q$ times, there are $2p-2(p-q)=2q$ \textit{old-type} $Y$-edges that warp back two nodes, and $p-q$ \textit{new-type} ones that warp back three nodes. 
The old-type edges can be further classed according to parity of the source and target vertices: $p$ of them, the `odd' ones, link odd vertices, and the other $p$ `even' ones link even vertices.
This splitting is possible because the substitution procedure always removes one even old-type $Y$-edge and one odd old-type $Y$-edge.
Traversing each new-type $Y$-edge results in a change of parity: all new-type $Y$-edges source from even vertices and target to odd vertices.
Furthermore, we note that every even vertex is the source of exactly one $Y$-edge and, similarly, every odd vertex is the target of exactly one $Y$-edge.
If the resultant parity upon traversal is even, a $U$-edge starts from the target vertex, whereas if it is odd, a $U$-edge leads into the target vertex. 

The preceding observations allow us to construct an algorithm to identify two classes of $U-Y$ cycles in a $Y^{p,q}$ quiver for $0<q\le p$. 
The algorithm distinguishes between the `even' and the `odd' cycles, which comprise only even (respectively, odd) $Y$-type edges.
For the even class of cycles, we proceed as follows.

\begin{itemize}
    \item[(a)] Start from an even vertex as source and follow the $Y$-edge leading out of it.
    \item[(b)] If the target is even, repeat step (a); if not, follow a $U$-edge to an adjoining even vertex. This decreases the vertex count by $2$ modulo $2p$.
    \item[(c)] Continue the traversal until all $q$ even old-type $Y$-edges, $p-q$ new-type $Y$-edges and $p$ adjoining $U$-edges are covered.
    This completes an even cycle as all the even vertices are traversed.
\end{itemize}

For the odd class of cycles, we proceed by reversing the direction of all the arrows in the quiver.
Now each odd vertex is the source of exactly one $Y$-edge.
All new-type reversed $Y$-edges now source from odd vertices and target to even vertices; upon traversing one of them, further traversing an adjoining `reversed' $U$-edge enables us to restore odd parity.
Now we may execute the algorithm outlined for the even cycle, but this time starting from an odd vertex. This is possible because the relevant properties of the mirrored quiver are formally equivalent. 
We obtain an odd cycle as we traverse all the odd vertices (all $q$ odd old-type $Y$-edges, $p-q$ new-type $Y$-edges and $p$ adjoining $U$-edges) but the cycle obtained is reversed, so we reverse the edges once again to restore the original quiver and the correct cycle.

Hence, the two classes of $U-Y$ cycles seem to be structured according to a split of the $2q$ old-type $Y$-edges into odd and even categories; each has $q$ old-type $Y$-edges, $p-q$ new-type $Y$-edges and $p$ $U$-edges. 
For $q=0$, as there are no old-type $Y$-edges, the two classes degenerate into a single class of cycle.

\subsubsection*{Cycle structures for $Y^{4,2}$}

As an example we explicitly describe the cycle structure for the two toric modes for $Y^{4,2}$ depicted by \cref{y42}, using standard cycle notation based on the vertex labels.
For both modes, the two ring cycles may be given by $(1,2,3,4,5,6,7,8,1)$, running around the periphery of the graph. 
The $U-Y$ cycles differ for the two modes. 
For the first mode described by \cref{Y42a}, they are given by $(2,7,8,6,3,4,2)$ (even) and $(2,7,5,6,3,1,2)$ (odd), while for the second mode described by \cref{Y42b} they are given by $(2,7,8,5,6,4,2)$ (even) and $(2,7,8,5,3,1,2)$ (odd).

\subsubsection*{Comparison with literature}

We have agreement of \eqref{detMpqoffshell} with previously published off-shell and on-shell results: \eqref{detMpqoffshell} matches result (A.2) in~\cite{zigzag} for off-shell $Y^{p,q}$ (we independently reproduce the same dimer model), and is also consistent with result (5.16) in~\cite{holocheck} for on-shell $Y^{p,p}$ and $Y^{p,0}$ quivers. 
In an on-shell assignment for $q \ne 0$, $a$-maximization is used~\cite{holocheck,sasaki} to further constrain \eqref{partlyfixingRcharges} to the exact values~\eqref{pqforalphamaximization}, giving $r_{U}=r_{V}=r_{Y}=\frac{2}{3}$ for $Y^{p,p}$, and for $Y^{p,0}$ we have the on-shell assignment $r_{U}=r_{Z}=r_{Y}=\frac{1}{2}$, as will be verified at the end of this subsection. For $p>q \ne 0$ we fail to corroborate the results of~\cite{holocheck}. 

\subsubsection*{Analytic verification for $q=p$ and $q=0$}

We demonstrate the analytic off-shell calculations for $Y^{p,p}$ and $Y^{p,0}$ to corroborate \eqref{detMpqoffshell} for $q=p$ and $q=0$, respectively. 
Let us first try the $Y^{p,p}$ case. 
Before explicitly writing down $M(t)$ note that~\eqref{mainmatrix} indicates the contributions to the entries $M_{ij}(t)$ come in several blocks which may repeat modulo $2p$ with respect to the matrix indices. 
These blocks, described below, can be identified as elements originating out of $I$, $\cM(t)$ or $\cM^{T}(t^{-1})$, and correspond to the multiplets in the quiver.

\begin{equation}
\begin{split}
\;\;\; D&=1-t^{2} \in M_{jj}(t), \\
A&=-2t^{r_{U}} \in M_{2i-1,2i}(t), \;\;\; A'=2t^{2-r_{U}} \in M_{2i,2i-1}(t), \\
B&=-2t^{r_{V}} \in M_{2i,2i+1}(t), \;\;\; B'=2t^{2-r_{V}} \in M_{2i+1,2i}(t), \\
C&=-t^{r_{Y}} \in M_{j+2,j}(t), \;\;\;\;\;\;\; C'=t^{2-r_{Y}} \in M_{j,j+2}(t). \;\;\;\;\;\;\;\; \label{ABCD}
\end{split}
\end{equation}  

Here $1 \le i \le p$ and $1 \le j \le 2p$, and all indices are identified modulo $2p$ as usual. For general $p$, we have $10p$ contributions from the blocks in \eqref{ABCD} to the $4p^{2}$ entries of $M_{p,p}(t)$. 
Since $4p^{2}>10p$ only when $p \ge 3$, for $p=1,2$ there are several `wrap-arounds' leading to $M_{Y^{1,1}}(t)$ and $M_{Y^{2,2}}(t)$ having entries with multiple contributions.
For $p\geq3$ the matrices $M_{Y^{p,p}}(t)$ begin to `empty up'. Explicitly, the first two matrices are

\begin{equation}
M_{Y^{1,1}}(t)=\begin{pmatrix}
D+C+C' & A+B' \\ A'+B & D+C+C'
\end{pmatrix}, 
\;\;
M_{Y^{2,2}}(t)=\begin{pmatrix} 
D & A & C+C' & B' \\ A' & D & B & C+C' \\
C+C' & B' & D & A \\
B & C+C' & A' & D
\end{pmatrix}, \label{Y11andY22matrices}
\end{equation}
and we note they are circulant in $1 \times 1$ and $2\times 2$ blocks respectively. By direct expansion, or using a computer algebra system, we can directly calculate $\det M(t)$ in these cases achieve the factorization~\eqref{detMpqoffshell}. 
For $p\geq3$, we obtain a $2p \times 2p$ matrix circulant in $2 \times 2$ blocks,

\begin{equation}
M_{Y^{p,p}}(t)= \begin{pmatrix}
D & A & C' & \mathbf{0} & C & B' \\
A'& D & B & C' & \mathbf{0} & C \\
C & B' & D & A & C' & \mathbf{0} \\
\vdots & \vdots & \vdots & \vdots & \vdots & \vdots \\
\mathbf{0} & C & A' & D & B & C' \\
C' & \mathbf{0} & C & B' & D & A \\
B & C' & \mathbf{0} & C & A' & D
\end{pmatrix}, \label{Mpp}
\end{equation}
where $\mathbf{0}$ represents a row of $2p-5$ consecutive zeroes, and the $2p-6$ rows with vertical dots in each entry have the unbroken strings $C-B'-D-A-C'$ or $C-A'-D-B-C'$ with, zeros on either end.
The general matrix \eqref{Mpp} may be thought of as being generated by the $2 \times 2$-block vector\footnote{We consider the rows for convenience.}
\begin{equation}
v= \left(\begin{pmatrix}
D & A \\ A' & D
\end{pmatrix}, \begin{pmatrix}
C' & 0 \\ B & C'
\end{pmatrix}, \underbrace{\begin{pmatrix}
0 & 0 \\ 0 & 0
\end{pmatrix} \ldots,\begin{pmatrix}
0 & 0 \\ 0 & 0
\end{pmatrix}}_{p-3}, \begin{pmatrix}
C & B' \\ 0 & C
\end{pmatrix} \right). \label{Mppblockvector}
\end{equation}
These blocks also appear in the~\eqref{Y11andY22matrices}, but some or all are coincident, in which case their entries add up.

The eigenvalues of a $b$-block-circulant $nb \times nb$ matrix, for $b \ge 1$, are expressible in terms of the $n$th roots of unity~\cite{circulant,blockcirculant} (note that $b=1$-block-circulant matrices are termed \textit{circulant}).
With $\omega=\exp \left(2\pi i/p \right)$, the eigenvalues of \eqref{Mpp} are given by the $2p$ eigenvalues of the $p$ matrices
\begin{equation}
H_{j}=\begin{pmatrix}
D & A \\ A' & D
\end{pmatrix}+\begin{pmatrix}
C' & 0 \\ B & C'
\end{pmatrix}\omega^{j}+\begin{pmatrix}
C & B' \\ 0 & C
\end{pmatrix} \omega^{-j}, \label{Mppblockmatrix}
\end{equation}
for $0 \le j \le p-1$. 
The product of eigenvalues of the terms generated by~\eqref{Mppblockmatrix} may be found and factored: using \eqref{partlyfixingRcharges}, at the $j$th level we have
\begin{align}
\lambda_{j} \lambda'_{j} &= \left(D+C'\omega^{j}+C \omega^{-j} \right)^{2}-\left(AA'+BB'+AB \omega^{j}+A'B' \omega^{-j} \right) \nonumber
\\ &= \left(1-t^{2}+t^{2-y}\omega^{j}-t^{y}\omega^{-j} \right)^{2}+4t^{2} \left(2-t^{-y}\omega^{j}-t^{y}\omega^{-j} \right) = \left(1-t^{y}\omega^{-j}\right)^{2}\left(1-t^{2-y}\omega^{j}\right)^{2}. \label{jthleveldetMpp}
\end{align}
Hence using properties of the $p$th roots of unity and~\eqref{partlyfixingRcharges}, the
determinant of \eqref{Mpp} is
\begin{align}
\det M_{Y^{p,p}}(t)= \prod_{j=0}^{p-1} \lambda_{j} \lambda'_{j} 
&= \left(1-t^{py}\right)^{2} \left(1-t^{p(2-y)}\right)^{2} = \left(1-t^{pr_{Y}}\right)^{2} \left(1-t^{p(r_{U}+r_{V})}\right)^{2}, \label{detMppdirectly}
\end{align}
consistent with \eqref{detMpqoffshell} for $q=p$.

To generalize this technique to evaluate $\det M_{Y^{p,q}}(t)$ with $p\geq3$, we need to perform the block substitutions
\begin{equation}
\begin{pmatrix}
C' & 0 \\ B & C'
\end{pmatrix} \rightarrow \begin{pmatrix}
0 & C' \\ E & 0
\end{pmatrix}, \qquad
\begin{pmatrix}
C & B' \\ 0 & C
\end{pmatrix} \rightarrow \begin{pmatrix}
0 & E' \\ C & 0
\end{pmatrix} \label{substitutionsinMpp}
\end{equation}
in the matrices~\eqref{Mpp} $p-q$ times, replacing $B$ by $E=-t^{r_{Z}}$ and $B'$ by $E'=t^{2-r_{Z}}$.
These substitutions (which may be analogously performed in the~\eqref{Y11andY22matrices}, where these two blocks coincide) are such that each substituted block containing $C'$ terms is anti-diagonal to each block containing $C$ terms.
For general $p$ and $q$, we do not get back a block-circulant matrix, but we do get one for $q=0$.
When $q=0$ and $p\geq3$ we have a $2p \times 2p$ block-circulant matrix generated by the $2 \times 2$-block vector\footnote{Again, we consider the rows for convenience.}
\begin{equation}
\tilde{v}= \left(\begin{pmatrix}
D & A \\ A' & D
\end{pmatrix}, \begin{pmatrix}
0 & C' \\ E & 0
\end{pmatrix}, \underbrace{\begin{pmatrix}
0 & 0 \\ 0 & 0
\end{pmatrix} \ldots,\begin{pmatrix}
0 & 0 \\ 0 & 0
\end{pmatrix}}_{p-3}, \begin{pmatrix}
0 & E' \\ C & 0
\end{pmatrix} \right), \label{Mp0blockvector}
\end{equation}
and can proceed similarly to the $q=p$ case. 
The cases when $p=1,2$ are also block-circulant, and may be checked by hand or with a computer algebra system. 
Analogous to our treatment of the matrices \eqref{Mppblockmatrix}, we need to find the $2p$ eigenvalues of the matrices
\begin{equation}
H_{j}=\begin{pmatrix}
D & A \\ A' & D
\end{pmatrix}+\begin{pmatrix}
0 & C' \\ E & 0
\end{pmatrix}\omega^{j}+\begin{pmatrix}
0 & E' \\ C & 0
\end{pmatrix} \omega^{-j}, \label{Mp0blockmatrix}
\end{equation}
for $0 \le j \le p-1$, where once again $\omega=\exp \left(2\pi i/p \right)$. 
Using \eqref{partlyfixingRcharges}, the product of the eigenvalues of \eqref{Mp0blockmatrix} at the $j$th level is given by
\begin{eqnarray}
\tilde{\lambda}_{j} \tilde{\lambda'}_{j} &=& D^{2}-\left(A+C'\omega^{j}+E'\omega^{-j} \right)\left(A'+E \omega^{j}+C\omega^{-j} \right) \nonumber
\\ &=& \left(1-t^{2} \right)^{2}-AA'-CC'-EE'-(AE+A'C')\omega^{j}-C'E\omega^{2j}-(A'E'+AC)\omega^{-j}-CE'\omega^{-2j} \nonumber \\
&=& \left(1-t^{1-\frac{1}{2}(x-y)}\omega^{-j}\right)^{2}\left(1-t^{1+\frac{1}{2}(x-y)}\omega^{j}\right)^{2}. \label{jthleveldetMp0}
\end{eqnarray}
Hence, the determinant of the large $N$ index matrix is 
\begin{align}
\det M_{Y^{p,0}}(t) = \prod_{j=0}^{p-1} \tilde{\lambda}_{j} \tilde{\lambda'}_{j}  &= \prod_{j=0}^{p-1} \left(\omega^{j}-t^{1-\frac{1}{2}(x-y)}\right)^{2}\left(\omega^{-j}-t^{1+\frac{1}{2}(x-y)}\right)^{2} \nonumber \\
&= \left(1-t^{p\left(1-\frac{1}{2}(x-y) \right)}\right)^{2}\left(1-t^{p\left(1+\frac{1}{2}(x-y)\right)}\right)^{2} = \left(1-t^{p(r_{U}+r_{Y})}\right)^{2} \left(1-t^{p(r_{U}+r_{Z})}\right)^{2}, \label{detMp0directly}
\end{align}
consistent with \eqref{detMpqoffshell} for $q=0$. 
We have also verified this result by computer algebra for various examples of $p>q \ne 0$.

\subsubsection*{$a$-maximization}

The local maxima of \eqref{amaximizationnext} in the case of $Y^{p,q}$ quivers for $q \ne 0$ was computed in~\cite{sasaki} to be \eqref{pqforalphamaximization}. 
In this case, the extremization was carried out over two variables. 
When $q=0$, however, \eqref{pqforalphamaximization} blows up, so we need to treat this case differently. 
For the $Y^{p,0}$ quivers we have the two-variable parametrization
\begin{equation}
a(x,y)=-\frac{p}{4}(x+y)^{3}+p(x-1)^{3}+p(y-1)^{3} \label{afunctionYp0}
\end{equation}
from \eqref{pqforalphamaximization}, and need to solve
\begin{equation}
(x+y)^{2}-4(x-1)^{2}=(x+y)^{2}-4(y-1)^{2}=0. \label{dervanYp0}
\end{equation}
A solution to~\eqref{dervanYp0} must satisfy $x=y$, in which case $(x,y)=(\frac{1}{2},\frac{1}{2})$, or $x+y=2$, in which case $(x,y)=(2,0)$ or $(x,y)=(0,2)$. Only the first case is a local extremum, as can be verified from the eigenvalues of the Hessian
\begin{equation}
H=\frac{3p}{2} \begin{pmatrix}
3x-y-4 & -(x+y) \\ -(x+y) & 3y-x-4 \label{hessianYp0}
\end{pmatrix}.
\end{equation}

Hence the on-shell $R$-charge assignment is $r_{U}=r_{Y}=r_{Z}=\frac{1}{2}$, matching the values used in~\cite{holocheck}.
Note that the matrices~\eqref{Y11andY22matrices} and~\eqref{Mpp} are circulant on-shell, but the latter's fully-substituted version generated by~\eqref{Mp0blockvector} is not.

\subsection{The \texorpdfstring{$\hat{A}_{m}$}{An} family}

The $\hat{A}_{m}$ quivers are another infinite family of quivers, part of the so-called ADE sub-class of toric quivers named as they may be constructed out of the extended Dynkin diagrams of the $A_{m}$, $D_{m}$ and $E_{m}$ Lie algebras~\cite{mckay1}.
In each case, we get the $\hat{A}_{m}$, $\hat{D}_{m}$ and $\hat{E}_{m}$ quivers and their reduced forms $\tilde{A}_{m}$, $\tilde{D}_{m}$ and $\tilde{E}_{m}$, respectively. 
The $A_{m}$ and $D_{m}$ Lie algebras are infinite, and we have just three unique $E_{m}$ members $E_{6}$, $E_{7}$ and $E_{8}$. 
In this subsection we work out $\det M(t)$ for the full $\hat{A}_{m}$ quivers.
They are described by $m$-gons~\cite{chiralring}; with vertices labelled by $1\le k \le m$, the fields constituting the quiver structure into cycles at each vertex and two concentric cycles circumscribing the $m$-gon in opposite directions\footnote{See e.g.~\cite{Mukhi:2002ck} for details; note that this family of quivers in fact describes $\cN=2$ theories.}. 
We may define the fields as $U_{k}:k \rightarrow k+1$, $V_{k}:k \rightarrow k-1$ and $Y_{k}:k \rightarrow k$, where the indices are taken modulo $m$. 
To reduce the $\hat{A}_{m}$ quiver to the $A_{m}$ quiver (thereby generating another infinite family) we remove the $Y$-fields.

\begin{figure}[H]
\centering
\includegraphics[width = 6 cm]{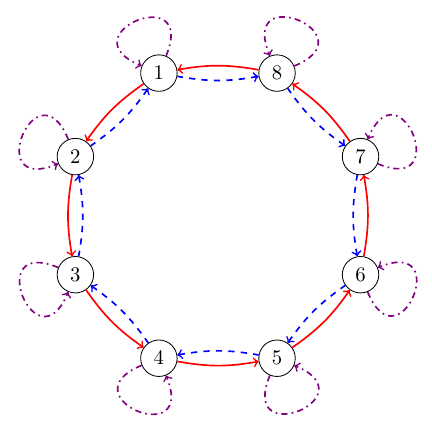}
\caption{The $\hat{A}_{8}$ quiver. We have three types of fields, the $U$ fields represented by solid red, the $V$ fields by dashed blue and the $Y$ fields by dash-dot violet.}
\label{A8}
\end{figure}

\begin{figure}[H]
\centering
\subfloat[The repeating unit in the planar quiver.]
{
\includegraphics[width=5 cm]{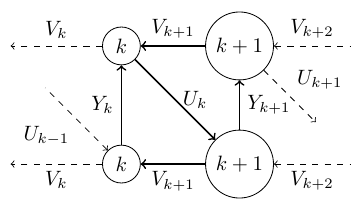}
\label{An1}
}
\qquad
\subfloat[One block of the dimer with its three zig-zag paths, presented in dashed red, dashed blue and dotted black.]
{
\includegraphics[width=5 cm]{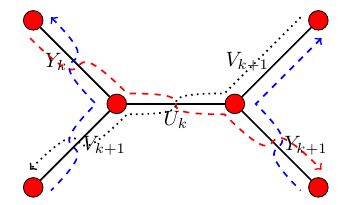}
\label{An2}
}
\caption{The planar quiver and dimer for the $\hat{A}_{m}$ quiver is composed of $m$ repeating blocks of the form described in \cref{An1}. The $m$ dotted black paths are unique to their respective blocks, while the dashed paths run around the whole quiver.}
\label{An}
\end{figure}

For the $\hat{A}_{m}$ quiver, exact forms of the superpotential appear not to be determined.
In accordance with the properties of superpotentials for toric quivers~\cite{zigzag,sasaki,dimermain}, we assume the form\footnote{There may be more than one candidate superpotential for a quiver, and the exact superpotential is in general non-trivial to determine.}
\begin{equation}
W=\sum_{k=1}^{m}\left(U_{k}Y_{k+1}V_{k+1}-Y_{k}U_{k}V_{k+1}\right), \label{superpotentialan}
\end{equation}
and shortly we shall see that this candidate superpotential correctly reproduces $\det M$ on-shell.
From the structure of \eqref{superpotentialan} the planar quiver and dimer may be constructed as shown in~\cref{An}.
We are able to identify $m+2$ zig-zag paths.   

Analogous to the analysis for the $Y^{p,q}$ quivers, assuming off-shell conditions we can now evaluate $\det M(t)$ as
\begin{equation}
\det M_{\hat{A}_{m}}(t)= \left( 1-t^{\sum_{k=1}^{m}\left(2-r_{U_{k}}-r_{Y_{k}}\right)}\right)\left( 1-t^{\sum_{k=1}^{m}\left(2-r_{V_{k+1}}-r_{Y_{k+1}}\right)}\right)\prod_{k=1}^{m}\left( 1-t^{\left(2-r_{V_{k+1}}-r_{U_{k}}\right)}\right). \label{detMAn1}
\end{equation}
To identify the cycle structure we use the off-shell constraints arising from the superpotential terms,
\begin{equation}
r_{U_{K}}+r_{V_{K+1}}+r_{Y_{k}}=r_{U_{K}}+r_{V_{K+1}}+r_{Y_{k+1}}=2. \label{Anoffshell1}
\end{equation}
This is the sole condition as the beta function vanishing condition~\eqref{mainconstraint} is just the sum of two such terms in \eqref{Anoffshell1}. 
We note that \eqref{Anoffshell1} forces the equality of the $r_{Y_{k}}$ to a common $R$-charge $r_{Y}$ for the $Y$-fields. 
Using \eqref{Anoffshell1} in \eqref{detMAn1} and shifting dummy indices, we get
\begin{eqnarray}
\det M_{\hat{A}_{m}}(t) &=& \left( 1-t^{\sum_{k=1}^{m}r_{V_{k}}}\right)\left( 1-t^{\sum_{k=1}^{m}r_{U_{k}}}\right)\prod_{k=1}^{m}\left( 1-t^{r_{Y_{k}}}\right) \nonumber \\
&=& \left( 1-t^{\sum_{k=1}^{m}r_{V_{k}}}\right)\left( 1-t^{\sum_{k=1}^{m}r_{U_{k}}}\right)\left( 1-t^{r_{Y}}\right)^{m}. \label{detMAn}
\end{eqnarray} 
We identify the cycles as two full $U$- and $V$-cycles in opposite directions around the whole quiver, and $m$ $Y$-cycles at each vertex.  

We have checked~\eqref{detMAn} for several values of $m$ using computer algebra.
In addition, it can be verified by hand for the special off-shell case of equal $r_{U_{k}}$ and equal $r_{V_{k}}$, which includes our predicted on-shell condition of $r_{U}=r_{V}=r_{Y}=\frac{2}{3}$ -- which we will derive shortly -- by using circulant matrices. 
In this case \eqref{Anoffshell1} gives $r_{U}+r_{V}+r_{Y}=2$. 
Proceeding as we did in~\cref{summaryYpq}, $M_{\hat{A}_{m}}(t)$ is ($1$-block) circulant and is generated for $m \ge 3$ by 
\begin{equation}
(1-t^{2}-t^{r_{Y}}+t^{2-r_{Y}},t^{2-r_{V}}-t^{r_{U}},\underbrace{0,\ldots,0}_{m-3},t^{2-r_{U}}-t^{r_{V}})^{T}. \label{generator_vector_Ahatm}
\end{equation}
The cases $m=1,2$ may be checked by hand; in fact they correspond to the $\cN=4$ SYM and $Y^{1,1}$ quivers respectively on-shell. 
For $m \ge2$, with $\omega=\exp(2 \pi i/m)$, we have 
\begin{align}
\det M_{\hat{A}_{m}}(t) &= \prod_{j=0}^{m-1} \left(1-t^{2}-t^{r_{Y}}+t^{2-r_{Y}}+(t^{2-r_{V}}-t^{r_{U}})\omega^{j}+(t^{2-r_{U}}-t^{r_{V}})\omega^{-j} \right) \nonumber \\
&= \prod_{j=0}^{m-1} \left(1-t^{r_{U}+r_{V}+r_{Y}}-t^{r_{Y}}+t^{r_{U}+r_{V}}+(t^{r_{U}+r_{Y}}-t^{r_{U}})\omega^{j}+(t^{r_{V}+r_{Y}}-t^{r_{V}})\omega^{-j} \right) \nonumber \\
&= \prod_{j=0}^{m-1} \left(1-t^{r_{Y}}\right)\left(1-t^{r_{U}}\omega^{j} \right)\left(1-t^{r_{V}}\omega^{-j} \right)= \left(1-t^{r_{Y}}\right)^{m} \left(1-t^{mr_{U}} \right)\left(1-t^{mr_{V}} \right). \label{detMAn2}
\end{align}
This is consistent with \eqref{detMAn} for equal $r_{U_{k}}$ and equal $r_{V_{k}}$.

To obtain the exact $R$-charges for this quiver family we have 
$m+1$ parameters arising out of the constraints \eqref{Anoffshell1}, which we  
set to be $y=r_{Y_{k}}$ and $x_{k}=r_{U_{k}}$ for $1 \le k \le m$. 
Thus, we seek the local maxima of
\begin{equation}
a(y,x_{1},x_{2},\ldots,x_{n})=m(y-1)^{3}+\sum_{k=1}^{m} \left((x_{k}-1)^{3}+(1-x_{k}-y)^{3}\right). \label{afunctionAn}
\end{equation}

The extremization conditions give $m+1$ equations
\begin{subequations} \label{dervanAn}
\begin{eqnarray} 
m(y-1)^{2} &=& \sum_{l=1}^{m}(1-x_{l}-y)^{2}, \label{dervanAnfirst} \\
(x_{k}-1)^{2} &=& (1-x_{k}-y)^{2} \qquad (1 \leq k \leq m), \label{dervanAnsecond}
\end{eqnarray}
\end{subequations}
and, out of all solutions, we find by trial and error that the case 
with $y-1=x_{k}-1=1-x_{k}-y$ for all $k$ gives a local maximum. 
This case corresponds to $x_{k}=y=\frac{2}{3}$. 
We can analytically verify that the Hessian at these coordinates,
\begin{equation}
H= 2 \begin{pmatrix}
-2m & (-1)_{1 \times m} \\
(-1)_{m \times 1} & -2I_{m \times m}
\end{pmatrix}_{(m+1) \times (m+1)}, \label{hessianAnmaximum}
\end{equation} 
has eigenvalues $2\left(-m-1\pm \sqrt{m^{2}-m+1}\right)$, with respective eigenvectors $(m-1\mp\sqrt{m^{2}-m+1},\textbf{1})^{T}$, and eigenvalue $-4$ with multiplicity $m-1$ and a set of eigenvectors $\left(1,\textbf{0},1,\textbf{0}-1 \right)^{T}$ which have an entry of $1$ in the $k$th position for $1\le k \le m-1$ between the first and last entries, the rest being $0$. 
All these are negative, hence this is a local maximum. It remains to check the other solutions of \eqref{dervanAn} do not yield local maxima. 
First, assume that $x_{k}-1=1-x_{k}-y$ holds true for all $k$ from \eqref{dervanAnsecond}, but that from \eqref{dervanAnfirst} we instead have $y-1=y-x_{k}-1$ for all $k$. This forces $y=2$ and $x_{k}=0$, and the Hessian at these coordinates,
\begin{equation}
H= 6 \begin{pmatrix}
0 & (-1)_{1 \times m} \\
(-1)_{m \times 1} & -2I_{m \times m}
\end{pmatrix}_{(m+1) \times (m+1)}, \label{hessianAnmaximum2nd}
\end{equation}
has $6(\sqrt{m+1}-1)$ as a positive eigenvalue with eigenvector $(-\sqrt{m+1}-1,1,\ldots,1)^{T}$, meaning we do not have a local maxima. 
For the remaining cases, assume that for a particular index $k_{0}$ equation~\eqref{dervanAnsecond} does not yield $x_{k_{0}}-1=1-x_{k_{0}}-y$. 
This forces $y=0$. 
However for $y=0$ we always get the Hessian
\begin{equation}
H=-6 \begin{pmatrix}
\sum_{k=1}^{m} x_{k} & \textbf{x}^{T}_{1 \times m} \\ \mathbf{x}_{m \times 1} & \mathbf{0}_{m \times m}
\end{pmatrix}_{(m+1) \times (m+1)}, \label{hessianAnothercases} 
\end{equation}
where $\mathbf{x}=\left(1-x_{1},\ldots,1-x_{m} \right)^{T}$, which has vanishing determinant and a zero eigenvalue. 
Thus, we get the sole local maximum $x_{k}=y=\frac{2}{3}$ corresponding to $r_{U_{k}}=r_{V_{k}}=r_{Y_{k}}=\frac{2}{3}$.

\subsection{\texorpdfstring{$dP3$}{dP3}}
\begin{figure}[h!]
\centering
\subfloat[The quiver with one highlighted cycle.]
{
\includegraphics[width=5 cm]{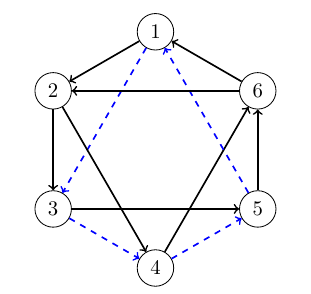}
\label{dP3a}
}
\qquad 
\subfloat[The repeating unit in the dual dimer with the zig-zag paths.]
{
\includegraphics[width=5 cm]{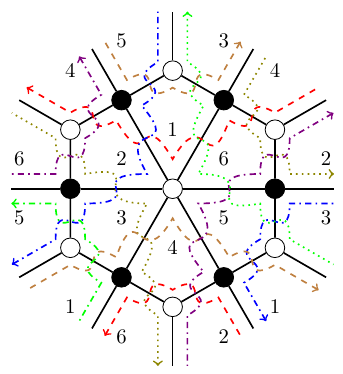}
\label{dP3b}
}
\caption{The $dP3$ quiver and its six zig-zag paths.}
\label{dP3}
\end{figure}

The $dPn$ quivers are associated to the $n$th del Pezzo surfaces. 
For $n=1,2,3$ these quivers are well-studied~\cite{Bo_Feng_2002,zigzag,delpezzo2,branedimers,delpezzo3}.
Here we take up $dP3$, which may be described in terms of a planar hexagon or an octahedron in three dimensions. We use the model of $dP3$ with the associated dimer from~\cite{branedimers,delpezzo3}, and attempt to calculate $\det M(t)$ using the method of zig-zag paths.
The quiver has the twelve fields $X_{i,i+1}:i \rightarrow i+1$, $X_{i,i+2}:i \rightarrow i+2$, and the superpotential is given by
\begin{equation}
W= X_{12}X_{23}X_{34}X_{45}X_{56}X_{61}+X_{13}X_{35}X_{51}+X_{24}X_{46}X_{62}-X_{12}X_{24}X_{45}X_{51}-X_{34}X_{46}X_{61}X_{13}-X_{56}X_{62}X_{23}X_{35}, \label{superpotentialdP3}
\end{equation} 
resulting in a hexagonal cycle, two triangular cycles and three rectangular cycles. 
Note a superficial similarity to the $Y^{3,3}$ quiver except that the $i \rightarrow i+2$ arrows are `backwards', and the superpotential is more complicated. 
The construction of the dimer model for $dP3$ is detailed in~\cite{branedimers,delpezzo3}; basing on which we just depict the dual and the zig-zag paths on it alongside the quiver diagram in~\cref{dP3}.
One of the six kite-shaped cycles is outlined in dashed blue in~\cref{dP3a} (the field labelling is canonical).
The six zig-zag paths are denoted by dashed red, dash dot blue, dotted olive, dashed brown, dash dot violet and dotted green respectively.

From \cref{dP3b} it is straightforward to obtain $\det M(t)$ in a cyclic notation with $r_{ij}$ denoting the $R$-charge of the field $X_{ij}$,
\begin{equation}
\det M_{dP3}(t)= \prod_{k=1}^{6} \left(1-t^{4-r_{k-1,k}-r_{k,k+1}-r_{k+1,k+3}-r_{k+3,k+5}} \right). \label{detMdP31}
\end{equation}
Equation~\eqref{detMdP31} may be simplified with the off-shell conditions. We have the constraints~\eqref{mainconstraint} from the beta function vanishing,
\begin{equation}
r_{k-2,k}+r_{k,k+2}+r_{k-1,k}+r_{k,k+1}=2, \label{offshelldP3beta}
\end{equation}
and using \eqref{offshelldP3beta} twice in \eqref{detMdP31}, the second time with the dummy substitution $k \rightarrow k+3$, gives
\begin{equation}
\det M_{dP3}(t)= \prod_{k=1}^{6} \left(1-t^{r_{k-2,k}+r_{k,k+2}+r_{k+2,k+3}+r_{k+3,k-2}} \right). \label{detMdP3}
\end{equation}
The six kite-shaped cycles in \eqref{detMdP3} can be identified from \cref{dP3a}. 
In cyclic notation, the cycles in are given by $(1345)$, $(2456)$ and so forth. 
We note that we only used the beta function vanishing constraints~\eqref{mainconstraint} and did not require those from the superpotential.
To find the exact $R$-charges, aside the beta function vanishing conditions \eqref{offshelldP3beta} we also have the conditions from the superpotential,
\begin{equation}
\sum_{k} r_{k,k+1} = \sum_{k} r_{k,k+2} = r_{k,k+1}+r_{k+1,k+3}+r_{k+3,k-2}+r_{k-2,k}=2. \label{offshelldP3suppot}
\end{equation}
The first sum in \eqref{offshelldP3suppot} is over all possible $k$, and the second over all possible $k$ of the same parity. The third equation can be considered for $1 \le k \le 3$. There are in total twelve apparent constraints.
From \eqref{offshelldP3beta} and the second set of constraints in \eqref{offshelldP3suppot}, respectively, we have
\begin{equation}
r_{k,k+2}=2-r_{k+2,k+4}-r_{k+4,k}=r_{k+3,k+4}+r_{k+4,k+5}. \label{constraintmaindP3}
\end{equation}
Computer algebra shows that~\eqref{offshelldP3beta} and~\eqref{offshelldP3suppot} together equate to seven independent constraints. 
Hence we can parametrize in five variables $x_{k}$ for $1 \le k \le 5$, which we take such that $r_{k,k+1}=x_{k}$, and seek to locally maximize
\begin{equation}
a(x_{1},x_{2},x_{3},x_{4},x_{5})=\sum_{k=1}^{5} \left(x_{k}-1\right)^{3}+\left(1-\sum_{k=1}^{5}x_{k} \right)^{3}+\sum_{k=1}^{4} \left(x_{k}+x_{k+1}-1 \right)^{3}+\left(1-\sum_{k=1}^{4}x_{k}\right)^{3}+\left(1-\sum_{k=2}^{5}x_{k}\right)^{3}. \label{afunctiondP3}
\end{equation}
Selecting $x_{k}=\frac{1}{3}$ by symmetry considerations, we find it is a local maximum using computer algebra.
To prove that this is the global maximum, we argue that no other local maxima exist. 
For if there were another local maximum $P'=(x'_{1},x'_{2},x'_{3},x'_{4},x'_{5})$, consider the restriction of $a$ to the line joining $P=(\frac{1}{3},\frac{1}{3},\frac{1}{3},\frac{1}{3},\frac{1}{3})$ to $P'$. 
Then this restriction is at most a cubic function of one real coordinate, and it cannot have more than one maximum, contradicting our supposition.
Hence the on-shell $R$-charges are $r_{k,k+1}=\frac{1}{3},r_{k,k+2}=\frac{2}{3}$. 

We have verified~\eqref{detMdP3} off-shell by computer in the variables $r_{k,k+1}$, obeying the first condition in~\eqref{offshelldP3suppot}.

\section{Index asymptotics}
\label{sec:analysis}

In this section we develop asymptotic formulae for some of the generating functions appearing above. 
We begin with a summary of the generating functions and our obtained results.

\subsection{Summary of obtained results}
\label{subchap:asymp_summary}

Consider first the on-shell univariate index generating functions \eqref{indexasgeneratingfunctionunivariate}, with the exact $R$-charges listed in \cref{subsec_a_maximization} and \cref{subsec:summary}. 
We find univariate asymptotics for the $\hat{A}_m$ family using the saddle-point method described below, including the $\cN=4$ SYM case (equal to $\hat{A}_1$) and the $Y^{1,1}$ case (equal to $\hat{A}_2$). 
We make the variable substitution $t=z^3$ in our generating functions to work with power series.

\begin{table}[H]
\begin{center}
\begin{tabular}{|>{$}c<{$}|>{$}c<{$}|>{$}c<{$}|}
\hline
\textrm{Quiver} & \; \textrm{Generating function }g(z)=\sum_{n \ge 0}c_{n}z^{n} \; & \; \textrm{Asymptotics }\cA[c_{n}]\textrm{ of }c_{n} \;
\\ \hline \hline && \\[-1em]
\cN=4 \mathrm{ SYM} = \hat{A}_1 & \prod_{k=1}^{\infty} \dfrac{(1-z^{3k})^{2}}{(1-z^{2k})^{3}} & (-1)^{n}\frac{1}{9 \sqrt{2}}n^{-1} \exp\left(\frac{2 \sqrt{2} \pi}{3}n^{\frac{1}{2}} \right) \\
\hline && \\[-1em] && \\[-1em]
Y^{1,1} = \hat{A}_2 & \prod_{k=1}^{\infty} \dfrac{\left(1-z^{3k}\right)^{4}}{\left(1-z^{4k}\right)^{2}\left(1-z^{2k}\right)^{2}} & (-1)^{n} \frac{7^{\frac{1}{4}}}{9\sqrt{6}} n^{-\frac{3}{4}} \exp\left(\frac{\pi\sqrt{7}}{3}n^{\frac{1}{2}} \right) \\
\hline && \\[-1em] && \\[-1em]
\hat{A}_{m} & \prod_{k=1}^{\infty} \dfrac{ \left(1-z^{3k}\right)^{2m}}{(1-z^{2km})^{2}(1-z^{2k})^{m}} & \substack{ (-1)^{n}2^{\frac{m-7}{4}}3^{\frac{-3m-3}{4}}m \left(\frac{m}{3}+\frac{1}{m} \right)^{\frac{3-m}{4}} \\ n^{\frac{m-5}{4}} \exp \left( \sqrt{\frac{2}{3}\left(\frac{m}{3}+\frac{1}{m} \right)}\pi n^{\frac{1}{2}} \right)} \\
\hline
\end{tabular}
\caption{Univariate asymptotics from the saddle-point method.}
\label{tab:saddleasm}
\end{center}
\end{table}

\begin{figure}[H]
\centering
\includegraphics[width=10cm]{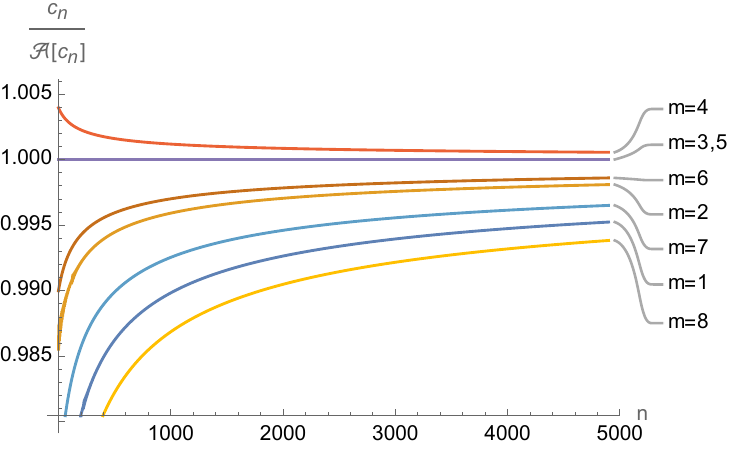}
\caption{The fit of the coefficient $c_{n}$ to the corresponding obtained asymptotic formula $\cA[c_{n}]$ for the $\hat{A}_{m}$ quivers.}
\label{fig_univar}
\end{figure}

The asymptotic behaviour listed in~\cref{tab:saddleasm} is typical of the saddle-point method, i.e. of the form $\alpha n^{\beta}\exp \left(\gamma n^{\frac{1}{2}}\right)$ for $\gamma \ne 0$.
In all these cases, the dominant contribution to the asymptotics is from the singularity at $z=-1$.
Presented in~\cref{fig_univar} is a computer check of these formulae, for  $1 \le m \le 8$.
For simplicity just the relevant sector of the interpolating curves is shown, for $100 \le n \le 5000$. Interestingly, for $m=3,5$ the coefficients rapidly approach the asymptotic formula, and the interpolation curves are effectively indistinguishable.

\begin{figure}[H]
\centering
\includegraphics[width=10cm]{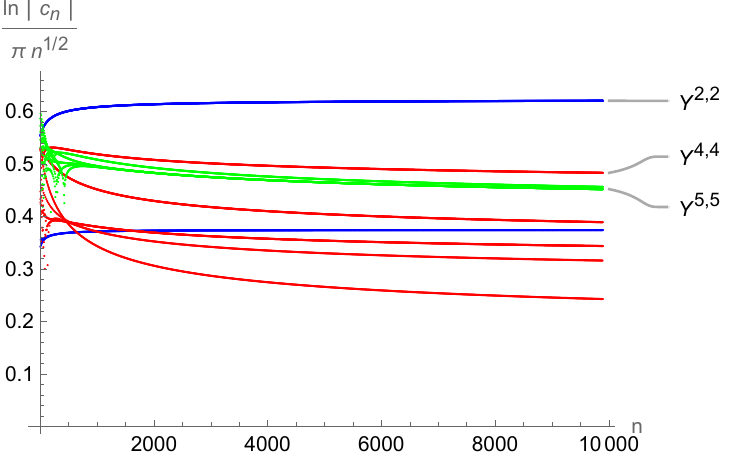}
\caption{
The growth of the exponential sector $\ln |c_{n}|$ of the $Y^{p,p}$ quivers.}
\label{fig_Y_asymptotics}
\end{figure}

The general $Y^{p,q}$ quiver admits a generating function of the form 
\begin{equation} 
g(z)=\prod_{k=1}^{\infty} \frac{\left(1-z^{3k}\right)^{4p}}{\left(1-z^{3k\left[p+q+\frac{p-q}{2}x-\frac{p+q}{2}y \right]}\right)^{2}\left(1-z^{3k\left[p-q- \frac{p-q}{2}x+\frac{p+q}{2}y \right]}\right)^{2}}, \label{gf_Ypq_q_ne_0}
\end{equation}
with the variable substitution $t=z^{3}$ and on-shell $x$ and $y$ given by~\eqref{pqforalphamaximization} for $q \ne 0$.
Of note is the special case $q=p$, for which this is
\begin{equation}
g(z)=\prod_{k=1}^{\infty} \frac{(1-z^{3k})^{4p}}{(1-z^{4kp})^{2}(1-z^{2kp})^{2}}. \label{gf_Ypp}
\end{equation}
For $q=0$ the generating function instead becomes
\begin{equation}
g(z)=\prod_{k=1}^{\infty} \frac{\left(1-z^{k}\right)^{4p}}{\left(1-z^{kp}\right)^{4}}, \label{gf_Yp0}
\end{equation}
with the variable substitution $t=z$.
We leave a full saddle-point analysis of the univariate asymptotics for this family to future work.
However, from numerical investigation we have found that the exponential sector of the asymptotics may be of the Hardy--Ramanujan form in at
least a few cases. 
For example, see~\cref{fig_Y_asymptotics} for some computer checks,~\cref{tab:Ypp} for two explicit conjectures based on these checks, and~\cref{tab:Y22} for supporting numeric evidence.
In~\cref{fig_Y_asymptotics} we plot just the points for $100 \le n \le 10000$ without interpolation for $p=2,4,5$.
The three quivers are demarcated by colour: blue, red and green respectively. 
The growth appears roughly to be along multiple subsequences in each case; note that $Y^{5,5}$ exhibits two closely-spaced growth curves that converge to the same value. 
The number of such subsequences appears to increase with $p$.
There exist subsequences for which $c_{n}$ vanishes, which we ignore when plotting.

\begin{table}[H]
\begin{center}
\begin{tabular}{|>{$}c<{$}|>{$}c<{$}|>{$}c<{$}|}
\hline
\textrm{Quiver} & \; \textrm{Generating function }g(z)=\sum_{n \ge 0}c_{n}z^{n} \; & \; \textrm{Asymptotics of }\ln |c_{n}|  \;
\\ \hline \hline && \\[-1em]
Y^{2,2} & \prod_{k=1}^{\infty} \dfrac{(1-z^{3k})^{8}}{(1-z^{8k})^{2}(1-z^{4k})^2} & \frac{3}{8} \pi n^{\frac{1}{2}}, \; n \equiv 3 \pmod 4; \; \frac{5}{8} \pi n^{\frac{1}{2}}, \; n \equiv 0 \pmod 2 \\
\hline
\end{tabular}
\caption{Conjectured asymptotics for the exponential sector of the $Y^{2,2}$ quiver.}
\label{tab:Ypp}
\end{center}
\end{table}
For the $Y^{2,2}$ generating function in~\cref{tab:Ypp}, the coefficients $c_{n}$ vanish for all $n \equiv 1 \pmod 4$; this is because such terms must get contributions from $\prod_{k=1}^{\infty}(1-z^{3k})^{8}$, and for the generating series $\prod_{k=1}^{\infty}(1-z^{k})^{8}=\sum_{n \ge 0}c_{n}z^{n}$ one can prove that $c_{n}=0$ for $n \equiv 3 \pmod 4$ (see~\cref{appendix:jacobitripleproductandmacdonaldidentities}).

\begin{figure}[H]
\centering
\begin{subfigure}[b]{.5\textwidth}
\centering
\includegraphics[width=8cm]{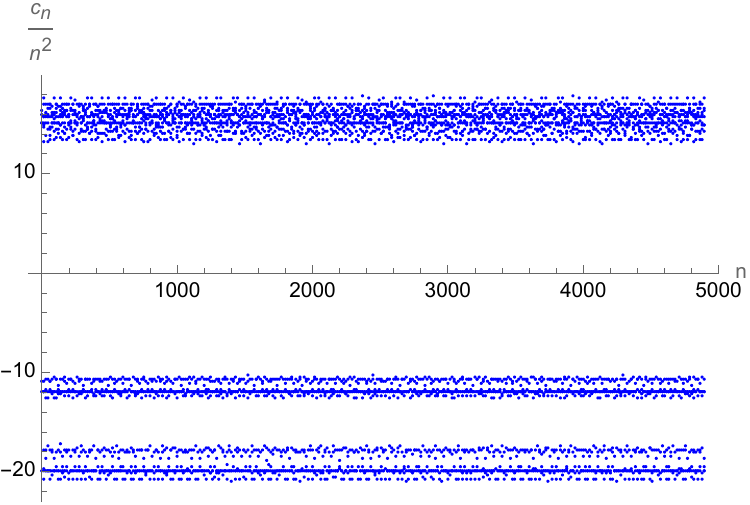}
\caption{$dP3$}
\label{fig_dp3}
\end{subfigure}%
\begin{subfigure}[b]{.5\textwidth}
\centering
\includegraphics[width=8cm]{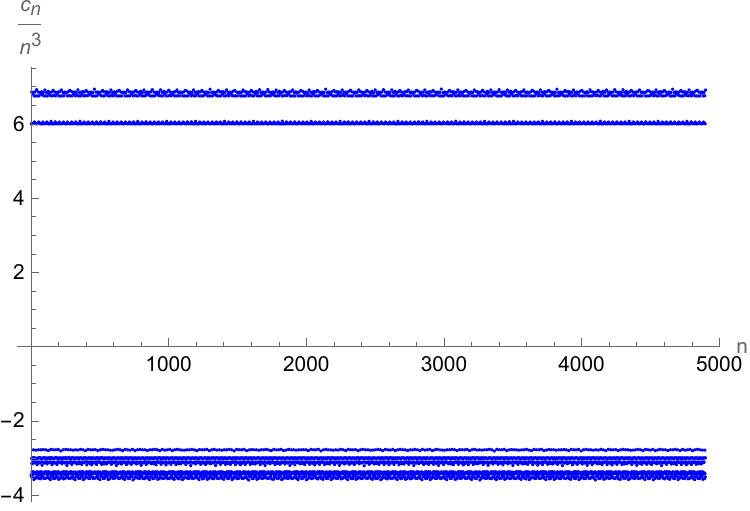}
\caption{$Y^{3,0}$}
\label{fig_y30}
\end{subfigure}
\caption{Polynomially growing asymptotics of the $dP3$ and $Y^{3,0}$ univariate index.}
\label{fig_polynomial_asymptotics}
\end{figure}

We also find, as detailed in~\cref{sec:polyasm}, that for some univariate quivers we get only \textit{polynomial growth}.
In some instances, we are able to prove the growth using the Jacobi triple product identity and other results. 
Presented in~\cref{tab:polyasm} is a summary of all polynomial growth cases we obtain and conjecture.
The variable specializations, sum of squares function $r_{s}(n)$ and $\Theta$ notation are defined in~\cref{sec:polyasm}.
One may identify distinct coefficient subsequences; see~\cref{fig_polynomial_asymptotics} for a few examples, where we plot just the points for $100 \le n \le 5000$ without interpolation.

\begin{table}[H]
\begin{center}
\begin{tabular}{|>{$}c<{$}|>{$}c<{$}|>{$}c<{$}|}
\hline
\textrm{Quiver} & \; \textrm{Generating function }g(z)=\sum_{n \ge 0}c_{n}z^{n} \; & \; \textrm{Behaviour of }c_{n} \;
\\ \hline \hline && \\[-1em]
dP3 &  \prod_{k=1}^{\infty} \dfrac{(1-z^{k})^{12}}{(1-z^{2k})^{6}} & (-1)^{n}r_{6}(n) = \Theta\left(n^{2}\right) \\
\hline && \\[-1em] && \\[-1em]
Y^{3,3} &  \prod_{k=1}^{\infty} \dfrac{\left(1-z^{3k}\right)^{12}}{\left(1-z^{12k}\right)^{2}\left(1-z^{6k}\right)^{2}} & \Theta \left(n^{3}\right) \\
\hline && \\[-1em] && \\[-1em]
Y^{2,0} &  \prod_{k=1}^{\infty} \dfrac{\left(1-z^{k}\right)^{8}}{\left(1-z^{2k}\right)^{4}} & (-1)^{n}r_{4}(n)= \Theta \left(n\right) \\
\hline && \\[-1em] && \\[-1em]
Y^{p,0}, \, p \ge 3 & \prod_{k=1}^{\infty} \dfrac{\left(1-z^{k}\right)^{4p}}{\left(1-z^{kp}\right)^{4}} & \Theta \left(n^{2p-3}\right) \\
\hline 
\end{tabular}
\caption{The univariate quivers exhibiting polynomial growth asymptotics.}
\label{tab:polyasm}
\end{center}
\end{table}

\begin{figure}[H]
\centering
\includegraphics[width=10cm]{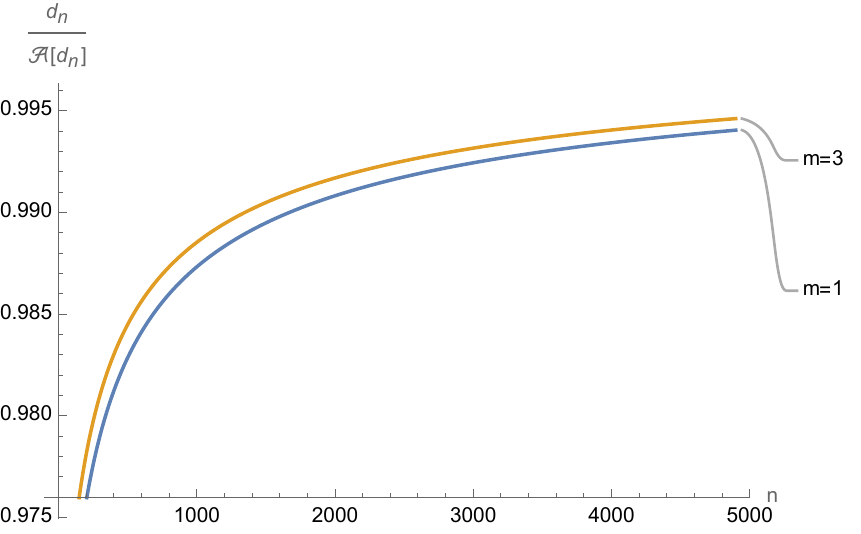}
\caption{The fit of the coefficient $d_{n}$ to the corresponding obtained asymptotic formula $\cA[d_{n}]$ for the $\hat{A}_{1}$ and $\hat{A}_{3}$ quiver bivariate generating functions along the main diagonal.}
\label{fig_bivar}
\end{figure}

Finally, to complement our univariate results, in~\cref{subsection:bivariate} we also derive asymptotics for the main diagonal coefficients $(pq)^{n}$ of the bivariate index~\eqref{deflargeNindex} in two cases; see~\cref{tab:biasym}.
In both these cases, the dominant contribution to the asymptotics is from the singularity at $z=1$, and of Hardy--Ramanujan type.
\begin{table}[H]
\begin{center}
\begin{tabular}{|>{$}c<{$}|>{$}c<{$}|>{$}c<{$}|}
\hline
\textrm{Quiver} & \; \textrm{Generating function }g(z)=\sum_{n \ge 0}d_{n}z^{n} \; & \; \textrm{Asymptotics }\cA[d_{n}]\textrm{ of }d_{n} \;
\\ \hline \hline && \\[-1em]
\cN=4 \mathrm{ SYM} = \hat{A}_{1} & \prod_{k=1}^{\infty} \dfrac{(1-z^{9k})}{(1-z^{3k}+z^{6k})(1-z^{k})^3} & \frac{2^{\frac{1}{4}}}{24} n^{-\frac{5}{4}}\exp \left(\sqrt{2}\pi n^{\frac{1}{2}}\right) \\
\hline && \\[-1em] && \\[-1em]
\hat{A}_{3} & \left[1+ 8 \sum_{k=1}^{\infty} \dfrac{kz^{3k}(1-3z^{3k})}{1-z^{6k}} \right] \prod_{k=1}^{\infty} \dfrac{ \left(1-z^{3k}\right)\left(1+z^{3k}\right)^{2}}{(1-z^{3k})^{2}(1-z^{k})^{3}} & \frac{11^{\frac{3}{4}}2^{\frac{1}{4}}}{72 \pi} n^{-\frac{5}{4}}\exp \left( \frac{\pi\sqrt{22}}{3} n^{\frac{1}{2}}\right) \\
\hline
\end{tabular}
\caption{Diagonal coefficients $d_n$ for $\hat{A}_{1}$ and $\hat{A}_{3}$}
\label{tab:biasym}
\end{center}
\end{table}
These asymptotics have been computationally verified; see~\cref{fig_bivar}, where for simplicity just the relevant sector of the interpolating curves is shown, and plotting is done for $100 \le n \le 5000$.
\subsection{Saddle-point asymptotics for the \texorpdfstring{$\hat{A}_{m}$}{A_m} generating function}
\label{subsec:asymototics_review}
Let $G(z)$ be an analytic function with power series expansion $G(z)=\sum_{n=0}^{\infty}c_{n}z^{n}$ converging in the unit disc centered at the origin;
we view $G(z)$ as the generating function encoding its coefficient sequence $(c_n)$. The \textit{Cauchy integral theorem} implies $c_n$ can be encoded by a complex contour integral 
\begin{equation}
c_{n}= \frac{1}{2\pi i} \oint_{\cC} G(z)z^{-n-1} \, \dd z , \label{contourdefinition}
\end{equation}
where $\cC$ is any circle centered at the origin with radius less than $1$. Making the substitution $z=e^{-\xi+i\theta}$, where $\xi>0$ and $\theta$ ranges over the interval $(-\pi,\pi)$, gives
\begin{equation}
c_{n}= \frac{1}{2\pi} \int_{-\pi}^{\pi} \exp\left(n\xi-ni \theta+\ln G\left(e^{-\xi+i\theta}\right) \right) \, \dd \theta. \label{Amgenfunc}
\end{equation}

The methods of analytic combinatorics use such integral representations to connect the behaviour of $G(z)$ near its singularities to the asymptotic behaviour of $c_n$. In the simplest cases, $G(z)$ admits a finite number of singularities, and asymptotics are deduced by a local study of $G(z)$ near each of them. Unfortunately, the generating functions of interest to us admit a dense subset of points of the unit circle as singularities. 
This complicates the analysis, but mirrors the behaviour of generating functions enumerating classes of integer partitions; the classical \textit{saddle-point method}~\cite[Ch.~5]{bruijn1958} shows how, under assumptions often holding in practice, the asymptotic behaviour of $c_n$ is still dictated by a study of $G(z)$ near a finite set of its singularities.
The key to the saddle-point method is identifying the point(s) on the unit circle where local behaviour dictates asymptotics. Consider the generating function
\begin{equation}
g(z) = \prod_{k=1}^{\infty}\frac{ \left(1-z^{3k}\right)^{2m}}{(1-z^{2km})^{2}(1-z^{2k})^{m}} 
\end{equation}
for the $\hat{A}_{m}$ quiver (we remind the reader that the cases $m=1,2$ are equivalent to the $\cN=4$ SYM and $Y^{1,1}$ quivers respectively). 
If $g$ were a rational function, then its poles on the unit circle with maximal order would be the points determining asymptotics. 
In this case $g$ is meromorphic, and all even roots of unity are roots of an infinite number of factors in its denominator, so talking about multiplicities does not make sense. 
Nonetheless, if the infinite product is truncated at any point then the resulting rational function has a pole at $z=-1$ of strictly higher order than any other pole (note that factors of $1-z$ in the denominator cancel with factors of $1-z$ in the numerator for all $k$, while factors of $1+z$ only appear in the numerator when $k$ is even). 
This suggests heuristically that the behaviour of $g(z)$ near $z=-1$ will determine the asymptotic behaviour of $c_n$, an assertion we will provide analytical justification for.

Because it is easier to work with points on the positive real line, we make the substitution 
\begin{equation}
G(z) = g(-z) = \prod_{k=1}^{\infty}\frac{ \left(1-z^{6k}\right)^{2m} \left(1+z^{3(2k-1)}\right)^{2m}}{(1-z^{2km})^{2}(1-z^{2k})^{m}}. \label{eq:GFtransformed}
\end{equation}
The rules of coefficient extraction for series imply that the coefficient $c_n$ in $G(z) = \sum_{n \geq 0}c_nz^n$ is $(-1)^n$ times the coefficient of $z^n$ in $g(z)$, so it is sufficient to find asymptotics for $c_n$.

To apply the saddle-point method to $G$ we first choose the parameter $\xi$, describing the radius of the circle of integration $e^{-\xi+i\theta}$, so that $G(z)$ has maximum magnitude near the positive real line that drops sharply as $\theta$ moves away from $0$. For a suitable choice of $\xi$, depending on $n$ and approaching $0$ as $n\rightarrow\infty$, we will replace the domain of integration $(-\pi,\pi)$ in~\eqref{Amgenfunc} with a small interval $(-\epsilon,\epsilon)$ for some $\epsilon>0$ also going to zero as $n\rightarrow\infty$.
The value of $\epsilon=\epsilon(n)$ is chosen carefully: it must be large enough so that the error introduced by this truncation of the domain of integration is asymptotically negligible, while also small enough so that the integrand is well-approximated by the initial terms in its power series expansion near $z=\xi$.

\subsubsection*{Finding the saddle-point}

In order for the modulus of the integrand in~\eqref{Amgenfunc} to rapidly decay away from the positive real line on the circle of radius $z=e^{-\xi}$ in the complex plane, we require the derivative with respect to $\xi$ of the \textit{phase function}
\begin{equation}
\phi(\xi)=n\xi+\ln G \left(e^{-\xi} \right) \label{phixi}
\end{equation}
to vanish (note that we have set $\theta=0$ in the integrand because our previous change of variables means we are interested in points near the positive real line). In other words, we seek a solution to the equation
\begin{align}
0=\phi'(\xi)
&= n + \left(2m\sum_{k \geq 1} \ln \left(1-e^{-6k\xi}\right)+2m\sum_{k \geq 1} \ln \left(1+e^{-3(2k-1)\xi}\right) - 2\sum_{k \geq 1} \ln \left(1-e^{-2km\xi}\right) - m\sum_{k \geq 1} \ln \left(1-e^{-2k\xi}\right)\right)' \nonumber \\[+2mm]
&= n-2m \sum_{k \geq 1} \frac{6k}{1-e^{6k\xi}} - 2m \sum_{k \geq 1} \frac{3(2k-1)}{1+e^{3(2k-1)\xi}} + 2 \sum_{k \geq 1} \frac{2km}{1-e^{2km\xi}} + m \sum_{k \geq 1} \frac{2k}{1-e^{2k\xi}}. \label{eq:saddlepoint}
\end{align}
Using the Euler-Maclaurin approximations~\cite[Sect.~3.6]{bruijn1958}
\begin{align*}
\sum_{k=1}^{\infty} \frac{sk}{1-e^{\xi sk}} &= \frac{1}{s \xi^{2}}\int_{0}^{\infty}\frac{y}{1-e^{y}} \, \dd y +\frac{s}{2\xi} \left[ \frac{y}{1-e^{sy}} \right]^{\infty}_{0} +\cO(1) =-\frac{\pi^{2}}{6s \xi^{2}}+\frac{1}{2\xi}+\cO(1), \\
\sum_{k=1}^{\infty} \frac{sk}{1+e^{\xi sk}} &= \frac{1}{s \xi^{2}} \int_{0}^{\infty} \frac{y}{1+e^{y}} \, \dd y+\frac{s}{2\xi} \left[ \frac{y}{1+e^{sy}} \right]_{0}^{\infty} +\cO(1) =\frac{\pi^{2}}{12s \xi^{2}}+\cO(1), \\
\sum_{k=1}^{\infty} \frac{s(2k-1)}{1+e^{\xi s(2k-1)}} &= \sum_{k=1}^{\infty} \frac{sk}{1+e^{\xi sk}}-\sum_{k=1}^{\infty} \frac{2sk}{1+e^{2\xi sk}}=\frac{\pi^{2}}{24s \xi^{2}}+\cO(1),
\end{align*}
for real $s,\xi>0$, we see that~\eqref{eq:saddlepoint} is satisfied by $\xi=\xi_{\mathrm{max}}$, where
\begin{equation} 
n = \frac{\pi^2}{\xi_{\mathrm{max}}^2}\left(\frac{m}{18} + \frac{1}{6m}\right) + \left(\frac{m-2}{2}\right) \frac{1}{\xi_{\mathrm{max}}} +\cO(1). \label{eq:n_in_terms_of_xi} 
\end{equation}
Inverting this equation gives
\begin{equation}
\xi_\mathrm{max} = \underbrace{\pi \sqrt{\frac{m}{18} + \frac{1}{6m}}}_C \, n^{-\frac{1}{2}}+\left(\frac{m-2}{4}\right)n^{-1}+\cO\left(n^{-\frac{3}{2}}\right) 
\label{xieq}
\end{equation}
as $n\rightarrow\infty$. 
The constant $C$ appearing in our selection of $\xi_\mathrm{max}$ is important for the asymptotic behaviour of $c_n$;
further, we shall see that it lets us define an effective central charge for the theory, or more generally, for an power series of similar growth characteristics.

\subsubsection*{Contribution of the saddle-point}

The integral expression~\eqref{Amgenfunc} is valid for all $\xi>0$, so our discussion up to this point can be viewed simply as selecting the value of $\xi$ that will allow for a proper analysis. 
As sketched above, having decided on $\xi_\mathrm{max}$ we now split the domain of integration $(-\pi,\pi)$ into an interval around the origin and its complement. In this subsection we determine how to split the integral, and approximate the portion of the integral that stays near the origin. 
This approximation gives the dominant asymptotic behaviour of $c_n$, and the next subsection proves the integral over the remaining points is comparatively negligible.

Our arguments rely on bounds coming from Freiman's formula~\cite[p.~119]{postnikov},
\begin{subequations}
\begin{eqnarray}
\ln \left( \prod_{k=1}^{\infty}\frac{1}{1-e^{-ku}} \right)&=&\frac{\pi^{2}}{6u}+\frac{1}{2} \ln \left( \frac{u}{2\pi} \right)+\cO(|u|), \label{freimanformula} \\
\ln \left( \prod_{k=1}^{\infty}\frac{1}{1-e^{-(2k-1)u}} \right) &=& \ln \left( \prod_{k=1}^{\infty}\frac{1}{1-e^{-ku}} \right)-\ln \left( \prod_{k=1}^{\infty}\frac{1}{1-e^{-2ku}} \right)\nonumber = \frac{\pi^{2}}{12u}-\frac{1}{2} \ln 2 +\cO(|u|), \label{freimanformula2}  \\
\ln \left( \prod_{k=1}^{\infty}\frac{1}{1+e^{-(2k-1)u}} \right) &=& \ln \left( \prod_{k=1}^{\infty}\frac{1}{1-e^{-(4k-2)u}} \right)-\ln \left( \prod_{k=1}^{\infty}\frac{1}{1-e^{-(2k-1)u}} \right) =  -\frac{\pi^{2}}{24u}+\cO(|u|), \label{freimannew}
\end{eqnarray}
\label{eq:freimanall}
\end{subequations}
where $u \rightarrow 0$ within any fixed wedge lying strictly in the right half-plane. 
Making the substitution $z=e^{-u}$ in $\ln G(z)$ and expanding with~\eqref{freimanformula} and~\eqref{freimannew} yields, after rewriting $G$ to put it in the correct form, that
\begin{align}
\ln G\left(z\right) &= -2m \ln \left(\prod_{k=1}^\infty \frac{1}{1-z^{6k}}\right) - 2m \ln \left( \prod_{k=1}^{\infty}\frac{1}{1+z^{3(2k-1)}} \right)
+ 2 \ln \left( \prod_{k=1}^{\infty}\frac{1}{1-z^{2km}} \right) + m \ln \left( \prod_{k=1}^{\infty}\frac{1}{1-z^{2k}} \right) \nonumber
\\[+2mm]
&= \frac{C^2}{u} - \left(\frac{m-2}{2}\right) \ln u + \ln\underbrace{\frac{m{\pi}^{\frac{m-2}{2}}}{3^{m}}}_K +\cO(|u|), \label{uexpansion}
\end{align}
where $K$ collects the terms inside the logarithm independent of $u$ and $C$ is the constant introduced in~\eqref{xieq}. 
Next, we use the identity
\begin{equation}
\frac{1}{\xi-i\theta}=\frac{1}{\xi}+\frac{\theta i}{\xi^{2}}-\frac{\theta^{2}}{\xi^{3}}-i \frac{\theta^{3}}{\xi^{4}}\frac{1}{1-\frac{i \theta}{\xi}}, \label{uexpansion2}
\end{equation}
and set $u=\xi_{\mathrm{max}}-i \theta$ in~\eqref{uexpansion} to obtain, making use of \eqref{eq:n_in_terms_of_xi} and \eqref{xieq},
{\small
\begin{align}
&n\xi_{\mathrm{max}}-ni \theta+\ln G\left(e^{-\xi_\mathrm{max}+i\theta}\right) \\
&\quad= n \xi_{\mathrm{max}}-in \theta+ C^{2} \left( \frac{1}{\xi_{\mathrm{max}}}+\frac{\theta i}{\xi_{\mathrm{max}}^{2}}-\frac{\theta^{2}}{\xi_{\mathrm{max}}^{3}}-\frac{i\theta^3}{\xi_{\mathrm{max}}^4}\frac{1}{1-\frac{i \theta}{\xi_{\mathrm{max}}}}\right)- \left(\frac{m-2}{2}\right)\ln \left(\xi_{\mathrm{max}}-i \theta \right)+\ln K+\cO\left( n^{-1/2}+|\theta| \right) \nonumber \\ 
&\quad= Cn^{\frac{1}{2}}+\frac{m-2}{4}-in \theta+Cn^{\frac{1}{2}}-\frac{m-2}{4}+in \theta-\left(\frac{n^{\frac{3}{2}}}{C}\right)\theta^{2}-\left(\frac{m-2}{2}\right) \ln \xi_{\mathrm{max}} \nonumber \\
&\quad\quad -\left(\frac{m-2}{2}\right)\ln \left(1-i\theta\xi_{\mathrm{max}}^{-1} \right)+\ln K+\cO\left( n^{-1/2}+|\theta|+n^{\frac{1}{2}}|\theta| + n^{1/2}\theta^{2}+ n^2|\theta|^3\right) \nonumber \\ 
&\quad= 2Cn^{\frac{1}{2}}-\left(\frac{m-2}{2}\right) \ln \left(Cn^{-\frac{1}{2}} \right)-\left(\frac{m-2}{2}\right) \ln \left(1-i \theta \frac{n^{1/2}}{C}\right)-\left(\frac{n^{\frac{3}{2}}}{C}\right)\theta^{2}+\ln K+\cO\left(n^{-1/2}+|\theta|+n^{\frac{1}{2}}|\theta| + n^{1/2}\theta^{2}+ n^2|\theta|^3\right). \label{gzz-n}
\end{align}
}

\noindent
To apply the saddle-point method, we split $(-\pi,\pi)$ into a vanishingly small interval $(-n^{-\delta},n^{-\delta})$ and its complement. Examining~\eqref{gzz-n}, we want $\delta>2/3$ so that the error terms $|\theta|$, $n^{\frac{1}{2}}|\theta|$, $n^{\frac{1}{2}}\theta^{2}$ and $n^2|\theta|^3$ go to zero as $n\rightarrow\infty$, and $\delta < 3/4$ so that $n^{3/2}\theta^2$ does not decay and we can approximate using a Gaussian integral. Taking any $\delta \in \left(\frac{2}{3},\frac{3}{4}\right)$, we obtain
\begin{align}
&\frac{1}{2\pi} \int_{-n^{-\delta}}^{n^{-\delta}} \exp\left(n\xi_{\mathrm{max}}-ni \theta+\ln G\left(e^{-\xi_{\mathrm{max}}+i\theta}\right) \right) \, \dd \theta \nonumber \\[+2mm]
&\sim \frac{\exp\left(2Cn^{\frac{1}{2}}\right) \, K}{2\pi} \left(Cn^{-\frac{1}{2}} \right)^{\frac{-m+2}{2}}  \int_{-n^{-\delta}}^{n^{-\delta}} \left(1-i\theta \frac{n^{1/2}}{C}\right)^{\frac{-m+2}{2}} \exp\left( -\frac{n^{\frac{3}{2}}}{C}\theta^{2}\right) \dd \theta \nonumber \\[+2mm]
&\sim \frac{K}{2 \pi} \left(Cn^{-\frac{1}{2}} \right)^{\frac{-m+2}{2}} \exp\left(2Cn^{\frac{1}{2}}\right) \int_{-\infty}^{\infty} \exp\left( -\frac{n^{\frac{3}{2}}}{C}\theta^{2}\right) \dd \theta  = \frac{K}{2\sqrt{\pi}}C^{\frac{-m+3}{2}} n^{\frac{m-5}{4}} \exp \left(2Cn^{\frac{1}{2}}\right),
\label{finalAmresult} 
\end{align} 
where we have used the standard estimates
\[\int_{\mathbb{R}} \exp\left( -a\theta^{2}\right) \dd \theta  = \sqrt{\frac{\pi}{a}}\]
for $a>0$ and 
\[\int_{\mathbb{R}\setminus(-b,b)} \exp\left( -a\theta^{2}\right) \dd \theta = \cO\left(a^{-1/2}e^{-ab^2}\right)\]
for $b>0$ and $a\rightarrow\infty$.
It is interesting to note, from the $n^{\frac{m-5}{4}}$ factor in~\eqref{finalAmresult}, that the logarithmic correction term for the $\hat{A}_{m}$ quiver family entropy is negative for $m<5$, vanishes at $m=5$ and thereafter is positive.
A physical interpretation of this behaviour is unclear at the moment.

\subsubsection*{Bounding the tails}
In this paper we only provide computational evidence that~\eqref{finalAmresult} is the dominant contributor to the asymptotics listed in~\cref{tab:saddleasm} for arbitrary $m$ (see ~\cref{fig_univar}).
It remains to prove that the remnant of the contour integral~\eqref{contourdefinition} contributes negligibly.
That is, expressing $c_n$ in terms of the integral~\eqref{Amgenfunc}, and approximating a portion of the integral near the origin using~\eqref{finalAmresult},  it suffices to show that  
\begin{equation}
\int_{(-\pi,\pi) \setminus (-n^{-\delta},n^{-\delta})} \exp\left(n\xi_{\mathrm{max}}-ni \theta+\ln G\left(e^{-\xi_{\mathrm{max}}+i\theta}\right) \right) \, \dd \theta = o\left(n^{\frac{m-5}{4}} \exp \left(2Cn^{\frac{1}{2}}\right)\right). \label{eq:tail}
\end{equation}
We hope to return to a rigorous treatment of~\eqref{eq:tail} in future works.

\subsection{Generalization}

The analysis above can be generalized in many cases. For instance, suppose that
\begin{equation}
G(z)= \prod_{k=1}^{\infty} \frac{\prod_{i=1}^{K}\left(1-z^{r_{i}k}\right)}{\prod_{j=1}^{L}\left(1-z^{s_{j}k}\right)} \label{eq_remark_generalGF}
\end{equation}
for positive integers $K$ and $L$, and vectors of positive integers $\mathbf{r}=(r_{i})_{i=1}^{K}$ and $\mathbf{s}=(s_{j})_{j=1}^{L}$ of lengths $K$ and $L$, respectively.
The equation to find the saddle-point on the positive real axis becomes
\[
0=\phi'(\xi)=n+\sum_{k=1}^{\infty} \left( -\sum_{i=1}^{K} \frac{r_{i}k}{1-e^{\xi r_{i}k}} + \sum_{j=1}^{L} \frac{s_{j}k}{1-e^{\xi s_{j}k}} \right)=n-\frac{\pi^{2}}{6 \xi^{2}} \Delta(\mathbf{r},\mathbf{s})+\cO(\xi^{-1}), 
\]
with solution
\begin{equation}
\xi_{\mathrm{max}}(n)=\pi\sqrt{\frac{\Delta(\mathbf{r},\mathbf{s})}{6}}n^{-\frac{1}{2}}+\cO(n^{-1}),
\end{equation}
where $\Delta(\mathbf{r},\mathbf{s})=\sum_{j=1}^{L}\frac{1}{s_{j}}-\sum_{i=1}^{K}\frac{1}{r_{i}}$. 
It is possible that $\Delta(\mathbf{r},\mathbf{s})\leq0$, in which case $\xi_{\mathrm{max}}$ does not describe a radius that can be used to apply the saddle-point method. When $\Delta(\mathbf{r},\mathbf{s}) > 0$ and $z=1$ is the only singularity where local behaviour of $G$ determines asymptotics, then a slight generalization of the argument above lets us obtain the asymptotic formula
\begin{equation} 
c_n \sim 2^{\frac{3(K-L)-5}{4}}3^{\frac{K-L-1}{4}} \sqrt{ \frac{ \Pi({\mathbf{s}})}{\Pi(\mathbf{r})}} \left[ \Delta(\mathbf{r},\mathbf{s})\right]^{\frac{L-K+1}{4}}n^{\frac{K-L-3}{4}} \exp \left(\sqrt{\frac{2 \Delta(\mathbf{r},\mathbf{s})}{3}}\pi n^{\frac{1}{2}}\right),
\label{eq_remark_generalGF_z=1}
\end{equation}
where $\Pi(\mathbf{x})=\prod_{j=1}^{N}x_{j}$ for a vector $\mathbf{x}=(x_{j})_{j=1}^{N}$.
The special case of $K=0$, $\mathbf{s}=(s)$ for some $s>0$ corresponds to one of a classic result, the Meinardus theorem~\cite{Meinardus1953/54,10.1063/1.530709}.
The more general case of $K=0$ and arbitrary $\mathbf{s}$ has been obtained in~\cite{asymptoticgeneralpartitions}.
\eqref{eq_remark_generalGF_z=1} corroborates both these results. 

This asymptotic expansion is not valid when the behaviour of $G$ near singularities other than $z=1$ contributes to dominant asymptotic behaviour. 
When~\eqref{eq_remark_generalGF_z=1} is valid, it can be proved by bounding the Cauchy integral over points away from the positive real axis, as done in this subsection for the generating function of $\hat{A}_m$. 
Setting $(K,L)=(0,1)$ and $\mathbf{s}=(1)$, we recover the classic Hardy--Ramanujan asymptotics for the generating function for integer partitions,
\begin{equation}
(c_{n})_{\textrm{HR}} \sim \frac{1}{4\sqrt{3}}n^{-1} \exp \left( \sqrt{\frac{2}{3}}\pi n^{\frac{1}{2}} \right).
\end{equation}
If any other singularity $z=e^{i\phi}$, for $\phi \in [0,2\pi)$, is the sole dominant contributor to asymptotic behaviour, it may be possible to modify the above analysis to find this contribution (subject to certain conditions satisfied by the numerator and denominator) by the substitution $z \mapsto ze^{i\phi}$.
In this paper we perform this for the $\hat{A}_{m}$ for $\phi=\pi$.
More generally for this case, after the transformation $z \mapsto -z$ in~\eqref{eq_remark_generalGF}, the generating function takes the form
\begin{equation}
\tilde{G}(z)=\prod_{k=1}^{\infty} \frac{\prod_{i=1}^{K}\left(1-z^{r_{i}k}\right)\prod_{i=1}^{\bar{K}}(1-z^{\bar{r}_{i}(2k-1)})}{\prod_{j=1}^{L}\left(1-z^{s_{j}k}\right)\prod_{j=1}^{\bar{L}}(1-z^{\bar{s}_{j}(2k-1)})}, \label{GF_z=-1_after_transform}
\end{equation}
with input vectors $\mathbf{r},\mathbf{s},\bar{\mathbf{r}},\bar{\mathbf{s}}$, and if $z=1$ is the only singularity where local behaviour of $\tilde{G}$ determines asymptotics, we obtain the asymptotic formula
\begin{equation}
 \tilde{c}_{n} \sim 2^{\frac{3(K-L)-5}{4}+\frac{\bar{K}-\bar{L}}{2}}3^{\frac{K-L-1}{4}} \sqrt{ \frac{ \Pi({\mathbf{s}})}{\Pi(\mathbf{r})}} \left[ \Delta(\mathbf{r},\mathbf{s})+\Delta(2\bar{\mathbf{r}},2\bar{\mathbf{s}})\right]^{\frac{L-K+1}{4}}n^{\frac{K-L-3}{4}} \exp \left(\sqrt{\frac{2(\Delta(\mathbf{r},\mathbf{s})+\Delta(2\bar{\mathbf{r}},2\bar{\mathbf{s}}))}{3}}\pi n^{\frac{1}{2}}\right), \label{eq_remark_generalGF_z=-1}
\end{equation}
whenever $\Delta(\mathbf{r},\mathbf{s})+\Delta(2\bar{\mathbf{r}},2\bar{\mathbf{s}})>0$.
\eqref{eq_remark_generalGF} may be seen as a special case of~\eqref{eq_remark_generalGF_z=-1} with $\bar{\mathbf{r}}=\bar{\mathbf{s}}=0$.
We have corroborated~\eqref{eq_remark_generalGF_z=-1} by computer in several cases, and hope to investigate these asymptotics further in future works.

For the $\hat{A}_{m}$ family, it is interesting to note that for $m=1,2$, i.e. for the $\cN=4$ SYM and $Y^{1,1}$ quivers respectively, using~\eqref{eq_remark_generalGF_z=-1} we are also able to obtain the `sub-dominant'\footnote{Strictly speaking, these quantities will be asymptotic contributions only if the contour integral tails are bounded in the manner described in~\eqref{eq:tail}.} asymptotic contribution from the singularity at $z=1$, 
\begin{equation}
(c_{n})_{\hat{A}_{1}}|_{z=1} \sim \frac{ \sqrt{5}}{18}n^{-1} \exp\left(\frac{ \pi \sqrt{5}}{3}n^{\frac{1}{2}} \right), \;(c_{n})_{\hat{A}_{2}}|_{z=1} \sim \frac{4}{9\sqrt{6}} n^{-\frac{3}{4}} \exp\left(\frac{\pi}{3}n^{\frac{1}{2}} \right).
\end{equation}
In the case of $\cN=4$ SYM, the exponential sector of this contribution from the $z=1$ singularity has been reported in the literature~\cite{murthy2020}. 
However, our calculations and numeric checks show that it is the $z=-1$ singularity that contributes to the leading asymptotics of $\ln |c_{n}|$; see~\cref{tab:saddleasm}.

Given a generating function $G$ and described by integer vectors $\mathbf{r},\mathbf{s}$, it is straightforward to obtain the corresponding generating function $\tilde{G}$ and described by integer vectors $\mathbf{r}_{\mathrm{new}},\mathbf{s}_{\mathrm{new}},\bar{\mathbf{r}},\bar{\mathbf{s}}$ respectively, using the identifications
\begin{equation}
    \mathbf{r}_{\mathrm{new}}=2\mathbf{r}, \; \mathbf{s}_{\mathrm{new}}=2\mathbf{s}, \; \bar{\mathbf{r}}=\mathbf{r}_{\mathrm{even}}\oplus 2\mathbf{r}_{\mathrm{odd}}\oplus \mathbf{s}_{\mathrm{odd}}, \; \bar{\mathbf{s}}=\mathbf{s}_{\mathrm{even}}\oplus 2
    \mathbf{s}_{\mathrm{odd}}\oplus \mathbf{r}_{\mathrm{odd}}. \label{eq:identifications}  
\end{equation}
Here we denote by $\mathbf{x}_{\mathrm{even}}$ (respectively, $\mathbf{x}_{\mathrm{odd}}$) the vector composed out of the even (respectively, odd) entries of a vector $\mathbf{x}$ -- the ordering is irrelevant for this purpose -- and the symbol $\oplus$ denotes a concatenation, i.e. direct sum of entries.
Using~\eqref{eq:identifications}, it is straightforward to derive that the condition on$~\tilde{G}$, $\Delta(\mathbf{r}_{\mathrm{new}},\mathbf{s}_{\mathrm{new}})+\Delta(2\bar{\mathbf{r}},2\bar{\mathbf{s}})>0$ corresponds to the condition $\Delta(\mathbf{r},\mathbf{s})>\frac{3}{4}\Delta(\mathbf{r}_{\mathrm{odd}},\mathbf{s}_{\mathrm{odd}})>$  on $G$.
Hence we may summarize the conditions for obtaining, using our method, an asymptotic contribution to the growth of the coefficients of $G$ from the singularities at $z=\pm 1$ as follows:
\begin{subequations}
\label{conditions:z}
    \begin{align}
        z=1: \;\;\; \Delta(\mathbf{r},\mathbf{s}) &> 0, \label{condition:z=1} \\
        z=-1: \;\;\; \Delta(\mathbf{r},\mathbf{s})-\frac{3}{4}\Delta(\mathbf{r}_{\mathrm{odd}},\mathbf{s}_{\mathrm{odd}}) &> 0. \label{condition:z=-1}
    \end{align}
\end{subequations}
For quivers outside the $\hat{A}_{m}$ family, neither obtained asymptotic contribution from $z=\pm 1$ necessarily need be the true, i.e. dominant asymptotics.
For example, for the $Y^{2,2}$ quiver, only~\eqref{condition:z=-1} is satisfied, giving from~\eqref{eq_remark_generalGF_z=-1} the asymptotic contribution
\begin{equation}
    (c_{n})_{Y^{2,2}}|_{z=-1} \sim (-1)^{n} \frac{16\sqrt{3}}{27} 2^{\frac{1}{4}}n^{\frac{1}{4}} \exp \left( \frac{\pi}{3\sqrt{2}} n^{\frac{1}{2}} \right) \label{univariateasymptoticsY22z=-1}
\end{equation}
for the generating function~\eqref{gf_Ypq_q_ne_0}.
However numeric checks indicate this is not the true asymptotic behaviour -- the exponential sector of the growth is apparently of the Hardy--Ramanujan form, but is different for two distinct subsequences, indicating that other singularities must be responsible. 
For other $Y^{p,q}$ quivers, the values of $x$ and $y$ appearing in~\eqref{gf_Ypq_q_ne_0} for $q \ne 0$ are generally not rational; we leave an analytical investigation of their asymptotic behaviour to future work.
Two important families with rational exponents in the generating functions are given by $q=p$ and $q=0$, of which we have already investigated~$q=p=1$.
For $Y^{p,p}$ quivers with $p \ge 3$, neither~\eqref{condition:z=1} nor~\eqref{condition:z=-1} is satisfied, but from several examples we have checked (some are plotted in~\cref{fig_Y_asymptotics}) the exponential sector appears to grow in Hardy--Ramanujan fashion for $p>3$, in general differently along multiple subsequences.
See~\cref{tab:Ypp} for some explicit conjectures on the growth.

\subsection{Polynomially-growing asymptotics}
\label{sec:polyasm}
We now show that the $dP3$, $Y^{3,3},$ and $Y^{2,0}$ quivers exhibit qualitatively different asymptotic behaviour than $\hat{A}_m$ -- their asymptotics appear to grow polynomially.
Based on numeric checks, we conjecture this to be true for all $Y^{p\ge 2,0}$ quivers.
From the Jacobi triple product identity (see~\cref{appendix:jacobitripleproductandmacdonaldidentities}) we obtain
\[\sum_{k \in \mathbb{Z}} (-1)^{k}x^{k^{2}}= \prod_{k=1}^{\infty} \frac{1-x^{k}}{1+x^{k}},\]
so that
\begin{equation}
\prod_{k=1}^{\infty} \frac{(1-x^{k})^{2s}}{(1-x^{2k})^{s}}=\prod_{k=1}^{\infty} \left(\frac{1-x^{k}}{1+x^{k}}\right)^{s}= \sum_{n \ge 0} (-1)^{n}r_{s}(n)x^{n}, \label{jacobissquaresGF}
\end{equation}
where $r_{s}(n)$ is the number of ways of representing the natural number $n$ as the sum of $s$ squares of integers,
\begin{equation}
r_{s}(n)= \left|\left\{ (n_{1},\ldots,n_{s}) \in \mathbb{Z}^{s}: \sum_{i=1}^{s} n_{i}^{2}=n \right\}\right|.
\end{equation}
As enumerators of a subset of $\mathbb{Z}^{s}$ of length scale $n$, the $r_{s}(n)$ grow at most as $\cO\left(n^{s}\right)$.\footnote{For tighter bounds on the growth, see e.g.~\cite{hardy_squares} and the discussion in~\cite{456863}.}.
Hence a series that can be composed out of a finite product of functions of the form~\eqref{jacobissquaresGF} will also have coefficients that grow at most polynomially.

Explicit forms for the $r_{s}(n)$ are known for several values of $s$.
For example, defining
\begin{equation}
D_{i}(n,k)=\left\{d \in \mathbb{Z}^{+}: d \mid n, \; d \equiv i \pmod k \right\}, \; 
S_{i}(n,k)= \left\{d \in \mathbb{Z}^{+}: d \mid n, \;  \frac{n}{d} \equiv i \pmod k \right\}, 
\end{equation}
for positive integers $n$ and $k$, it is known~\cite{jacobi1829fundamenta,hardy2008introduction}\footnote{See also e.g.~\cite{ChanJacobiSixSquares,bhargava1988simple}.} that
\begin{align}
r_{2}(n) &=  \sum_{d \in D_{1}(n,2)}(-1)^{ \frac{d-1}{2}}, \label{formofr2n} \\
r_{4}(n) &= 8 \sum_{\substack{d \mid n, 4 \nmid d}} d, \label{formofr4n} \\
r_{6}(n) &=  4 \left( \sum_{d \in D_{3}(n,4)}d^{2} - \sum_{d \in D_{1}(n,4)}d^{2}  \right) + 16 \left( \sum_{d\in S_{1}(n,4)}d^{2}-  \sum_{d\in S_{3}(n,4)}d^{2} \right). \label{formofr6n}
\end{align}
 
The generating function for $dP3$ directly corresponds to the case $s=6$ in \eqref{jacobissquaresGF}, giving 
\begin{equation}
(c_{n})_{dP3} \sim 4(-1)^{n} \left( \sum_{d \in D_{3}(n,4)}d^{2} - \sum_{d \in D_{1}(n,4)}d^{2}  \right) + 16(-1)^{n} \left( \sum_{d\in S_{1}(n,4)}d^{2}-  \sum_{d \in S_{3}(n,4)}d^{2} \right).
\end{equation}

Similarly, from~\eqref{gf_Yp0} we see that the $Y^{2,0}$ generating function directly corresponds to $s=4$ in~\eqref{jacobissquaresGF}, and after a monomial substitution $z^{3} \mapsto z$, the generating function for $Y^{3,3}$ can be factored as
\begin{equation}
G(z)= \prod_{k=1}^{\infty} \frac{(1-z^{k})^{12}}{(1-z^{4k})^{2}(1-z^{2k})^{2}}= \left[ \sum_{n\ge 0}(-1)^{n}r_{6}(n)z^{n} \right]\left[ \sum_{n\ge 0}(-1)^{n}r_{2}(n)z^{2n} \right].
\end{equation}
Hence, the $Y^{2,0}$ and $Y^{3,3}$ quivers also exhibit at most polynomial growth, which we have checked numerically (see~\cref{tab:polyasm}).
We express the growth in $\Theta$ notation: for two functions $f(n)$ and $g(n) \ge 0$ over non-negative integers $n$, $f(n)=\Theta(g(n))$ when there exist constants $C_{1},C_{2}$ and some $n_{0}$ such that $C_{1}g(n) \le |f(n)| \le C_{2}g(n)$ for all $n>n_{0}$.
In general, we conjecture from numeric checks that the $Y^{p \ge 3,0}$ generating functions show polynomial growth, $\Theta(n^{2p-3})$; we have verified this up to $p=8$; but a proof of this assertion is not obvious at the moment.
See~\cref{fig_polynomial_asymptotics} for a numeric check of some of these asymptotics.

\subsection{Bivariate asymptotics}
\label{subsection:bivariate}

The univariate generating functions studied above  may contain limited physical information, as one fugacity dependency is eliminated. 
One way to circumvent this potential issue is to consider a different degeneracy of fugacities in the index. 
A full bivariate analysis of~\eqref{indexasgeneratingfunction} is more involved, so to complement our results above we consider a few examples of \textit{main diagonals}
\begin{equation}
    \cI^{(1,1)}_{\infty}(pq)=\sum_{n}c_{n,n}p^{n}q^{n}. \label{maindiagonalindex}
\end{equation}
Let us obtain a more explicit expression for \eqref{maindiagonalindex}. For positive integers $V$ we introduce the constants $(d_{V}(k))_{k \ge 0}$ which are generated by the infinite product
\begin{equation}
\prod_{k=1}^{\infty} \left(1-x^{k} \right)^{V}= \sum_{k=0}^{\infty} d_{V}(k)x^{k}. \label{dvk}
\end{equation}
The $d_{V}(k)$ are related to coloured distinct integer partitions (see \cref{appendix:jacobitripleproductandmacdonaldidentities} for details).
In this $d$-representation~\eqref{deflargeNindex} is
\begin{equation}
    \cI_{\infty}(p,q)=\frac{\sum_{k,l=0}^{\infty}d_{V}(k)d_{V}(l)p^{k}q^{l}}{\prod_{k=1}^{\infty} \det M((pq)^{k/2})}. \label{supconfindexindrep}
\end{equation}
As the denominator of \eqref{supconfindexindrep} is already a function of $pq$, we only need to trim out a subsequence out of the numerator to get a $d$-representation of~\eqref{maindiagonalindex}. 
With the variable substitution $pq=z^{m}$ for a suitable $m$, we hence obtain the generating function
\begin{equation}
    G(z)= \cI_{\infty}^{(1,1)}(z^{m})=\frac{\sum_{k=0}^{\infty}d_{V}^{2}(k)z^{mk}}{\prod_{k=1}^{\infty}\det M\left(z^{mk/2} \right)}. \label{GFbivariatemaindiagonal}
\end{equation}
Note that generating functions of this form have positive coefficients, i.e. in a physical sense, the enumerated BPS states at fixed $R$-charge always have a bosonic excess.
In this paper we consider only $V=1,3$, which correspond to the $\cN=4$ SYM quiver and the $\hat{A}_{3}$ quiver respectively. 
$d_{1}(k)$ and $d_{3}(k)$ have known expressions -- see \eqref{jacobi2} and \eqref{V=3nonsquaredmain} respectively in \cref{appendix:jacobitripleproductandmacdonaldidentities}.
Similarly, $d_{1}^{2}(k)$ is known (see \eqref{jacobi3}) and we can construct $d_{3}^{2}(k)$ (see \eqref{V=3squaredmain}). 
Hence, we obtain the following two generating functions.
\begin{table}[H]
\begin{center}
\begin{tabular}{|>{$}c<{$}|>{$}c<{$}|>{$}c<{$}|}
\hline
\textrm{Quiver} & \; \textrm{Variable substitution} \; & \; G(z) \;
\\ \hline \hline && \\[-1em]
\cN=4 \textrm{ SYM} & pq=z^{3} & \prod_{k=1}^{\infty} \dfrac{(1-z^{9k})}{(1-z^{3k}+z^{6k})(1-z^{k})^3} \\
\hline && \\[-1em] && \\[-1em]
\hat{A}_{3} & pq=z^{3} & \left[1+ 8 \sum_{k=1}^{\infty} \dfrac{kz^{3k}(1-3z^{3k})}{1-z^{6k}} \right] \prod_{k=1}^{\infty} \dfrac{ \left(1-z^{3k}\right)\left(1+z^{3k}\right)^{2}}{(1-z^{3k})^{2}(1-z^{k})^{3}} \\
\hline
\end{tabular}
\caption{Table of bivariate on-shell generating functions for $\cN=4$ SYM and $\hat{A}_{3}$.}
\label{tablebivariateGFs}
\end{center}
\end{table}

In the $\cN=4$ SYM case, algebraic manipulation lets us write the bivariate generating function in the univariate form
\begin{equation}
G(z)=\prod_{k=1}^{\infty} \frac{(1-z^{9k})^{2}(1-z^{6k})}{(1-z^{18k})(1-z^{3k})(1-z^{k})^3}. \label{Gzbivariatecloveralternate}
\end{equation}
The structure of this generating function heuristically suggests that the dominant contribution to the asymptotics ought to be from $z=1$ (as we verify numerically; see~\cref{fig_bivar}).
From~\eqref{eq_remark_generalGF_z=1} this is
\begin{equation}
(c_{n})_{\hat{A}_{1}} \sim \frac{2^{\frac{1}{4}}}{24} n^{-\frac{5}{4}}\exp \left(\sqrt{2}\pi n^{\frac{1}{2}}\right). \label{bivariateasymptoticsclover}
\end{equation}
The $\hat{A}_{3}$ case is more involved. 
In this case, delineating the two constituent parts of the generating function as follows,
\begin{equation}
G(z)=\underbrace{\left[1+ 8 \sum_{k=1}^{\infty} \frac{kz^{3k}(1-3z^{3k})}{1-z^{6k}} \right]}_{=H(z)} \underbrace{ \prod_{k=1}^{\infty} \frac{ \left(1-z^{6k}\right)^{2}}{(1-z^{3k})^{3}(1-z^{k})^{3}}}_{=G_{0}(z)}, \label{modifiedbivariateGFA3}
\end{equation}
we proceed along the lines of our analysis in~\cref{subsec:asymototics_review} and attempt to obtain an asymptotic formula, treating $H(z)$ as a `perturbation' on $G_{0}(z)$. 
The structure of $G_{0}(z)$ suggests that $z=1$ will be the site of the dominant contribution.
From \eqref{phixi}, taking derivatives with respect to $\phi$ gives the following condition for finding the saddle point,
\begin{equation}
\phi'(\xi)=n+\frac{\dd}{\dd \xi} \ln \left(H(e^{-\xi})\right)-\frac{11\pi{^2}}{18\xi^{2}}+ \frac{2}{\xi}+\cO\left(1\right)=0, \label{phixiderbivariate1}
\end{equation}
where if we disregarded the presence of $H(z)$, the singularity at $z=1$ would have a saddle point near it solely due to $G_{0}(z)$, and we could directly output the asymptotics from~\eqref{eq_remark_generalGF}. 
However, we need to obtain an expansion of $\frac{\dd}{\dd \xi} \ln \left(H(e^{-\xi})\right)$ in small $\xi$ to find the order of its contribution to the saddle point.
Since
\begin{eqnarray}
H\left(e^{-\xi}\right) &=& 1+8 \sum_{k=1}^{\infty} \frac{k\left(e^{3\xi k}-3\right)}{e^{6 \xi k}-1} \nonumber \\
&=& 1+ \frac{8}{9\xi^{2}} \underbrace{ \int_{0}^{\infty} \dd y \, \frac{y\left(e^{y}-3\right)}{e^{2y}-1}}_{=0} +\frac{4}{\xi}\left[ \frac{y\left(e^{3y}-3 \right)}{e^{6y}-1} \right]_{0}^{\infty}+\cO\left(1 \right) = \cO\left(\xi^{-1}\right), \label{contributionH}
\end{eqnarray}
$\frac{\dd}{\dd \xi} \ln \left(H(e^{-\xi})\right)$ is also of order $\cO\left(\xi^{-1}\right)$ and is irrelevant for evaluating the saddle point.
Next, we need to check the contribution of $\ln H\left(e^{-\xi_{\mathrm{max}}+i\theta}\right)$ when expanded around $\xi_{\mathrm{max}}$, to evaluate its contribution to \eqref{gzz-n}. Let $f(w)=\frac{w(e^{w}-3)}{e^{2w}-1}$ and $\Gamma(u)$ be the line $\{xu:x \ge 0\}$.
Then for $u=\xi_{\mathrm{max}}-i\theta$, we can use contour integration to instead evaluate over a vanishingly small circular arc.
The remainder term corresponding to this integral vanishes in the Euler-Maclaurin formula:
\begin{equation}
 H\left(e^{-\xi_{\mathrm{max}+i\theta}}\right)= 1+\frac{8}{9u^{2}} \int_{\Gamma(u)} \dd w \, f(w)+ \frac{4}{3u}\left(f(\infty)-f(0^{+}) \right)+\cO(1)=1+\frac{4}{3u}+\cO(1).
\label{contributionHmain}
\end{equation}
The leading remainder term is of order $n^{\frac{1}{2}}$. 
Integrating over the interval $\left(-n^{-\delta},n^{-\delta}\right)$ as in \eqref{finalAmresult}, we see that it suffices to take $\delta \in \left( \frac{1}{2}, \frac{3}{4} \right)$ to ignore all contributions from terms containing $\theta$, leaving only the leading order contribution from $\xi_{\mathrm{max}}$.
The exact $\cO\left(n^{\frac{1}{2}}\right)$ dependency may be obtained from \eqref{phixiderbivariate1} to be $\left(\frac{4}{3}\cdot \frac{3 \sqrt{2}}{\sqrt{11}\pi} \right)n^{\frac{1}{2}}= \frac{4 \sqrt{2}}{\sqrt{11}\pi}n^{\frac{1}{2}}$, and applying this multiplicative factor to the result we obtain for the `unperturbed' generating function $G_{0}(z)$ from~\eqref{eq_remark_generalGF}, we obtain
\begin{equation}
(c_{n})_{\hat{A}_{3}} \sim \frac{11^{\frac{3}{4}}2^{\frac{1}{4}}}{72 \pi} n^{-\frac{5}{4}}\exp \left( \frac{\pi\sqrt{22}}{3} n^{\frac{1}{2}}\right). \label{bivariateasymptoticsA3}
\end{equation}
We have computationally verified this formula (see~\cref{fig_bivar}).

\subsection{Effective central charge}
\label{subsec:effective_central_charge}

Recently a quantity called the effective central charge has been proposed and studied (see~\cite{ceffgukov,adams2025crmeffresurgencestokes,harichurn2025ctexteffsurgerymodularity} and references therein) in the context of $d=3$ theories.
For a general power series $f(z)=\sum_{n \ge 0}c_{n}z^{n}$, the effective central charge may be defined, when it exists, as
\begin{equation}
c_{\mathrm{eff}}:=\lim_{n \rightarrow \infty} \frac{3}{2} \frac{(\ln |c_{n}|)^{2}}{\pi^{2}n}. \label{c_eff_defn}
\end{equation}
This limit may be different across subsequences, in which case we would speak of limit points.
In physical contexts the power series is typically a supersymmetric partition function, and $c_{\mathrm{eff}}$ characterizes the growth of BPS states in such theories, and is related e.g. to modular invariants in the $d=3$ case.
This definition is motivated by the $n^{\frac{1}{2}}$ growth law of the logarithm of BPS state counts\footnote{The logarithm of non-BPS state counts for such indices typically grows as $n^{\frac{d-1}{d}}$~\cite{Kutasov_2001}.} at leading order, as observed e.g. in some $d=3$~\cite{ceffgukov} and $d=4$~\cite{Sen_2010} superconformal indices, and also in the $d=4$ toric quiver indices in our work.

The properties of $c_{\mathrm{eff}}$ are also of general mathematical interest in the context of the asymptotic formulae we have obtained or conjectured.
For the general formula~\eqref{eq_remark_generalGF_z=-1}, that (conditionally) captures the asymptotics for the generating function~\eqref{GF_z=-1_after_transform}, we have the simple expression
\begin{equation}
c_{\mathrm{eff}}=\Delta(\mathbf{r},\mathbf{s})+\Delta(2\bar{\mathbf{r}},2\bar{\mathbf{s}}) \label{c_eff_general}
\end{equation}
in terms of the tuplets of integers describing the constituents of the infinite product.
Of course, this appears to only be valid when the asymptotics are indeed due to the $z=1$ singularity, and the restriction $\Delta(\mathbf{r},\mathbf{s})+\Delta(2\bar{\mathbf{r}},2\bar{\mathbf{s}})>0$ is satisfied.
A look at the derivation of~\eqref{finalAmresult}, in particular at the expression for finding the saddle point~\eqref{xieq}, shows that $c_{\mathrm{eff}}$ in such cases originates from the Euler--Maclaurin approximations involved in that procedure. 

Specializing to the $d=4$ superconformal index of toric quivers, it is evident that $c_{\mathrm{eff}}$ can be expressed in terms of the $R$-charges of the quiver, and through~\eqref{amaximizationmain}, in terms of its $a$ and $c$ central charges.
We remind the reader that these charges are expressed through fermionic traces, and that they are equal due to the vanishing of $\tr R$.
Further, for infinite quiver families, we may study how $c_{\mathrm{eff}}$ changes with quiver size.
Let us now briefly look at these aspects for toric quivers, in particular for the $\hat{A}_{m}$ family, whose asymptotics we have comprehensively described.
From the on-shell asymptotics listed in~\cref{tab:saddleasm}, we find
\begin{equation}
c_{\mathrm{eff}}[\hat{A}_{m}]=\frac{m}{3}+\frac{1}{m}.
\end{equation}
It is unclear at the moment how this number is related to the $R$-charges of the $U$, $V$ and $Y$ chiral multiplets; recall that there are $m$ of each kind, with on-shell $R$-charge $\frac{2}{3}$ each, and hence from~\eqref{amaximizationmain}, the central charges are $a=c=\frac{m}{4}$. 
In the large quiver size limit, we see that $c_{\mathrm{eff}}[\hat{A}_{m}] \sim \frac{m}{3}$, and it attains the value of $\frac{28}{15}$ when the logarithmic correction to the entropy changes sign, i.e. at $m=5$.
This corresponds to $a=c=\frac{5}{4}$.
For the $Y^{p,p}$ quivers, we have $a=c=\frac{p}{2}$.
For $p=2$ values in~\cref{tab:Ypp} let us conjecture
\begin{equation}
c_{\mathrm{eff}}[Y^{2,2}]=\begin{cases}
\frac{27}{128}, & n \equiv 3 \pmod 4, \\
\frac{75}{128}, & n \equiv 1 \pmod 2,
\end{cases}
\end{equation}
corresponding to $a=c=1$.
It would be interesting to obtain $c_{\mathrm{eff}}$ for general $Y^{p,q}$.
In the case of the bivariate main-diagonal generating functions in~\cref{tablebivariateGFs}, we obtain
\begin{equation}
c_{\mathrm{eff},\mathrm{biv}}[\hat{A}_{1}]=3, \; c_{\mathrm{eff},\mathrm{biv}}[\hat{A}_{3}]= \frac{11}{3}.
\end{equation}
These generating functions are different from the original index, and it is unclear what interpretation the effective central charge may have in this case.
For the quivers demonstrating polynomial growth (detailed in~\cref{sec:polyasm}), $c_{\mathrm{eff}}$ manifestly vanishes.

\section{Giant graviton expansion}
\label{section:matrixmodelandgiantgrav}

Having looked at factorization and asymptotics of the large $N$ superconformal index, we now briefly discuss the matrix model underlying the index, and the giant graviton expansion associated with it. 
The giant graviton expansion is the decomposition of the finite $N$ matrix model into a sum whose terms appear to admit a physical interpretation in the context of duality.
The sum is typically presented in a normalized form,
\begin{equation}
\frac{\cZ_{N}(\cQ)}{\cZ_{\infty}(\cQ)}=\sum_{j \in J}G^{(j)}_{N}(\cQ), \label{giantgravexpansion_canonical_form}
\end{equation}
for a partition function $\cZ_{N}(\cQ)$ dependent on a set of fugacities, i.e. couplings $\cQ=\{q_{i}\}_{i \in I}$, with large $N$ limit $\cZ_{\infty}(\cQ):= \lim_{N \rightarrow \infty} \cZ_{N}(\cQ)$.
Here $I$ and $J$ are at most countable index sets, with $J$ typically countable, and the $G^{(j)}_{N}(\cQ)$ are the giant graviton expansion parameters at level $j$.

There are multiple approaches and underlying principles towards developing such an expansion; we base ours on the approach of~\cite{murthy_giantgravitons}, which utilizes the Hubbard--Stratonovich transform and the Borodin--Okounkov determinantal formula to develop the expansion for \textit{scalar} couplings, and which is further extended in~\cite{Liu_2023,Ezroura:2024wmp} in the aspect of explicit calculation of the expansion terms.
For consistency, we will closely follow the notation and steps of~\cite{murthy_giantgravitons,Liu_2023}.

Finite $N$ giant graviton expansions have been studied in the context of certain toric quiver gauge theories in general, with the focus being largely on the $\frac{1}{16}-$BPS index of $\cN=4$ SYM and its reduced-fugacity specializations; see ~\cite{murthy_giantgravitons,Fujiwara:2023bdc,Ezroura:2024wmp,hatsuda2025deformedschurindicesmacdonald,Eleftheriou_2024,Lee2024,Gautason:2024nru} and references therein.
Explicit formulae for the terms of such expansions have been determined only in some particular cases.
We do not foray into an examination of the existence, regime of validity, and uniqueness\footnote{See e.g.~\cite{Lee2024} for a homological explanation of the non-uniqueness of the $\frac{1}{2}$-BPS expansion; we thank Ji Hoon Lee for bringing this work to our attention.} of giant graviton expansion, which are interesting problems in their own regard requiring more involved mathematical constructions; we postpone this to future work.
The main results we obtain in this section are the expansion~\eqref{eq:giant_grav_expansion_multi_model} for the \textit{matrix}-coupling model~\eqref{eq:mainproblemmultimatrix}, and the corresponding generating function~\eqref{general_Gmz_V} for the giant graviton terms.
These generalize the respective scalar-coupling results,~\eqref{giant_graviton_scalar_coupling_unnormalized} and~\eqref{general_Gmz_V=1}, for the model~\eqref{eq:mainmodel}, which were developed in~\cite{murthy_giantgravitons,Liu_2023}. 
We conclude with a look at the specialization of this expansion for the univariate index for two infinite families of quivers we have considered, the $\hat{A}_{m}$ and the $Y^{p,p}$; we computationally find some explicit forms in~\cref{appendix:misc} for a few examples from these quiver families, verifying that~\eqref{general_Gmz_V} does indeed generate an iterative correction up
to the first few terms of the expansion.

\subsection{Review of the matrix model}

The index~\eqref{defindextoricquiver} is a multi-variable generalization of the matrix integral over the normalized Haar measure on $\mathrm{U}(N)$
\begin{equation}
     \cZ_{N}(\mathbf{g})= \int_{\mathrm{U}(N)} \dd U \, \exp \left( \sum_{k=1}^{\infty} \frac{1}{k}g_{k} \tr U^{k} \tr U^{-k} \right), \label{eq:mainmodel}
\end{equation}
for a set of coupling constants $\mathbf{g}=\left\{g_{k} \right\}_{k\ge 1}$.
The $g_{k}$ are assumed appropriate for convergence and well-definedness of this integral.
In physical contexts, the coupling constants are generally of the restricted form $g_{k}=g(t_{i}^{k})_{i \in I}$ for some parameters $t_{i}$ indexed by $I$. 
At large $N$, the integral evaluates into the infinite product
\begin{equation}
  \cZ_{\infty}(\mathbf{g}):=\lim_{N \rightarrow \infty}\cZ_{N}(\mathbf{g})=\prod_{k=1}^{\infty} \frac{1}{1-g_{k}}. \label{eq:generalindexatlargeN}
\end{equation}
A standard way of deriving~\eqref{eq:generalindexatlargeN} is the Coulomb gas formalism -- see e.g.~\cite{Dolan_2009,shirazall,gadde,hagedorn,holocheck} for reviews on the derivation.
Essentially, this approach begins by recasting the matrix model in terms of contour integrals over the parameters characterizing the maximal tori of the respective gauge groups, as we have briefly discussed in~\cref{sec:review}.
Another is deploying the Hubbard--Stratonovich (HS) transformation~\cite{stratonovich,hubbard} (see also \cite{_lvarez_Gaum__2006}) -- which we discuss shortly -- on a random partition description of a related matrix model, the generalized Gross--Witten--Wadia (GWW) model~\cite{gwworiginal1,wadia1980cp,wadia2012study}, but we do not take up that perspective in this paper beyond an application of the Cauchy identity.
A third approach, also not considered in this paper, but which has proven especially useful in investigating the existence of and convergence to the large $N$ limit, as used in~\cite{murthy_giantgravitons} for the scalar coupling case, is converting the model directly into a sum over partitions utilizing the Frobenius formula for characters of $\mathrm{U}(N)$~\cite{fultonharris}.
We intend to explore these alternative approaches in future works.

The matrix model~\eqref{eq:mainmodel} may be generalized to
\begin{equation}
        \cZ_{N}(\fM)=\int_{\bigtimes_{i=1}^{V}\mathrm{U}(N)} \prod_{i=1}^{V}\dd U_{i} \, \exp\left(\sum_{k=1}^{\infty} \frac{1}{k} \sum_{i,j=1}^{V}(\sM_{k})_{ij} \tr U_{i}^{k} \tr U_{j}^{-k} \right), \label{eq:mainproblemmultimatrix}
    \end{equation}
for a set of $V\times V$ matrix couplings $\fM=\left\{\sM_{k}\right\}_{k \ge 1}$ assumed diagonalizable, with appropriate entries to ensure convergence.
At large $N$ this integral evaluates into the infinite product~\cite{Benvenuti_2007,Pasukonis_2013,Kimura_2021}\footnote{~\cite{Kimura_2021} considers a large $N$ treatment of a different index, the chiral ring index, which involves a very similar matrix model.},
\begin{equation}
\cZ_{\infty}(\fM)= \prod_{k=1}^{\infty} \frac{1}{\det\left(I- \sM_{k}\right)}, \label{eq:coulombgasfinalunitary}
\end{equation}
which may also be obtained either by the Coloumb gas or random partition approaches as generalizations of the scalar coupling case.
Note that~\eqref{eq:coulombgasfinalunitary} encodes just the eigenvalue data of the matrix couplings; in general we cannot find a transformation in $\mathrm{GL}(V,\mathbb{C})$ that simultaneously diagonalizes the $\sM_{k}$ while keeping the product measure invariant.
However, this difficulty becomes irrelevant in the large $N$ limit.

In the context of our work, the couplings $\sM_{k}$ are identified in terms of the large $N$ index matrix \eqref{mainmatrix} as 
\begin{equation}
    \sM_{k}=I- \frac{M((pq)^{k/2})}{(1-p^{k})(1-q^{k})}. \label{eq:connectionwithsupindex}
\end{equation}

In~\eqref{eq:mainproblemmultimatrix}, if the role of the unitary group is reprised by other gauge groups from infinite families ranked by $N$, the structure of the large $N$ limit changes. 
Without going into details, we have seen that, for example, if each copy of $\mathrm{U}(N)$ is replaced by $\mathrm{SO}(2N)$, a similar analysis can be performed, and the large $N$ index of the scalar coupling model~\eqref{eq:mainmodel} appears to be
\begin{equation}
\cZ_{\mathrm{SO}(2N \rightarrow \infty)}(\mathbf{g})= \prod_{k=1}^{\infty} \frac{\exp \left( \frac{g_{2k}}{2k(1-2g_{2k})} \right)}{\sqrt{1-2g_{k}}}, \label{eq:coulombgasfinalSO_even}
\end{equation}
and we hope to investigate the generalized matrix-coupling version in future works.
See e.g.~\cite{Sei:2023fjk} for a random partition description of this matrix model over the classical Lie groups; it would be interesting to corroborate our observations with results therein.

\subsection{Review of the scalar coupling expansion}

Let us briefly review the derivation of the scalar-coupling giant graviton expansion in~\cite{murthy_giantgravitons}.
One of the two building-blocks central to this setup is the Hubbard--Stratonovich (HS) transformation,
\begin{equation}
\exp \left(f ab \right)= \int_{\mathbb{C}} \frac{\dd t \wedge \dd \bar{t}}{2\pi i f} \exp \left(-\frac{t\bar{t}}{f}+t a+\bar{t} b \right), \label{eq:gaussiantransform}
\end{equation}
where $a,b\in\mathbb{C}$, and $\dd t$, $\dd \bar{t}$ are the canonical holomorphic (respectively, anti-holomorphic) differentials on $\mathbb{C}$.
In physical terms, this may be viewed as a transformation relating a field theory with one field, and linear in matter content, to a quantum-mechanical theory with quadratic matter content.
A deeper understanding of its role and interpretation in a gauge-duality context appears unclear for the moment, but essentially, we can switch from a double-trace to a single-trace theory with the price of adding some degrees of freedom.
The generalized GWW partition function is
\begin{equation}
\tilde{\cZ}_{N}(\mathbf{t}^{+},\mathbf{t}^{-})=\int_{\mathrm{U}(N)} \dd U \exp \left( \sum_{k=1}^{\infty} \frac{1}{k} t^{+}_{k}\tr U^{k}+t^{-}_{k}\tr U^{-k} \right), ~\label{gen_gww_scalar}
\end{equation}
where $\mathbf{t}^{+}=\left\{t^{+}_{k}\right\}_{k \ge 1}, \mathbf{t}^{-}=\left\{t^{-}_{k}\right\}_{k \ge 1}$ are sets of coupling constants.
Unless otherwise noted, for the remainder of this review, we shall assume that these are complex conjugate, i.e. $\overline{t^{+}_{k}}=t^{-}_{k}$ for all $k$; this means that that~\eqref{gen_gww_scalar} is a real number.

Employing the HS transformation,~\eqref{eq:mainmodel} can be expressed in terms of~\eqref{gen_gww_scalar}
as follows,
\begin{equation}
\cZ_{N}(\mathbf{g})= \prod_{k=1}^{\infty} \left( \frac{1}{ g_{k}} \int_{\mathbb{C}} \frac{\dd t^{+}_{k} \wedge \dd t^{-}_{k}}{2\pi i k} e^{-\frac{t^{+}_{k}t^{-}_{k}}{k g_{k}}} \right) \tilde{\cZ}_{N}(\mathbf{t}^{+},\mathbf{t}^{-}). \label{eq:HStransform_sc_coup}
\end{equation}
With the compact notation
\begin{equation}
\left\langle f(\mathbf{t}^{+},\mathbf{t}^{-};\mathbf{g}) \right\rangle_{\mathbf{g}}:=\prod_{k=1}^{\infty} \left( \frac{1}{g_{k}} \int_{\mathbb{C}} \frac{\dd t^{+}_{k} \wedge \dd t^{-}_{k}}{2\pi i k}  e^{-\frac{t^{+}_{k}t^{-}_{k}}{k g_{k}}} \right) f(\mathbf{t}^{+},\mathbf{t}^{-};\mathbf{g}) \label{general_transform_scalar}
\end{equation}
for a function $f$ depending $\mathbf{g}, \mathbf{t}^{+},\mathbf{t}^{-}$, which following~\cite{Liu_2023} we shall call the $\left\langle \cdots \right\rangle_{\mathbf{g}}$-average of $f$, the relation between the two partition functions can be simply written as $\cZ_{N}(\mathbf{g})= \left\langle \tilde{\cZ}_{N}(\mathbf{t}^{+},\mathbf{t}^{-}) \right\rangle_{\mathbf{g}}$.

The generalized GWW partition function~\eqref{gen_gww_scalar} can be recast into a random partition form, employing the language of symmetric polynomials~\cite{borodinokounkov1999fredholm} and using the Miwa variable parametrization $\tr X^{k}=t^{+}_{k}, \tr Y^{k}=t^{-}_{k}$, where $X,Y$ are countably infinite diagonal matrices, as follows,
\begin{equation}
\tilde{\cZ}_{N}(\mathbf{t}^{+},\mathbf{t}^{-})= \sum_{\ell(\lambda) \le N}s_{\lambda}(X)s_{\lambda}(Y). \label{gen_gww_random_partition}
\end{equation}
Here $s_{\lambda}(Z)$ is the Schur polynomial indexed by the partition $\lambda$ in the entries $\{Z_{i}\}_{i \ge 1}$ of a diagonal matrix $Z$.
Employing these notations, the large $N$ limit can be re-expressed using the Cauchy identity\footnote{This may be viewed as the exponentiated free energy at large $N$ for the ungapped phase. Essentially this is the $N \rightarrow \infty$ limit of Gessel's theorem, see~\cite{borodinokounkov1999fredholm} and references therein for details. See also the strong Szeg\"o theorem~\cite{1958PhT....11Q..38G}.} and the plethystic exponential~\eqref{defplethystic} as
\begin{equation}
\tilde{\cZ}_{\infty}(\mathbf{t}^{+},\mathbf{t}^{-})= \sum_{\lambda}s_{\lambda}(X)s_{\lambda}(Y)= \prod_{i,j \ge 1} \frac{1}{1-X_{i}Y_{j}}= \exp \left( \sum_{k=1}^{\infty} \frac{1}{k}t^{+}_{k}t^{-}_{k} \right). \label{cauchy_identity}
\end{equation}
Note that~\eqref{cauchy_identity} holds true even if the $t^{\pm}_{k}$ are not complex conjugate.  
Also note that the generalized GWW model has an intricate phase space structure~\cite{Kimura_2021,Kimura_2021_next,Kimura_2024} at large $N$ in the fugacities $\bm{t}^{\pm}$; there are additional corrective terms to the free energy in the so-called \textit{bulk} region of the phase space, but these do not appear relevant to our computations.

The second central building-block is a Fredholm determinantal expansion due to Borodin and Okounkov~\cite{borodinokounkov1999fredholm}.
We refer to~\cite{Okounkov2001-kf,murthy_giantgravitons} for details on the particular presentation of the Borodin--Okounkov formula used, which utilizes the so-called free fermion formalism. 
Also note that there are alternate ways~\cite{Chen:2024cvf,Eniceicu2024} to arrive at the giant graviton expansion using instantons, but we postpone a discussion of this approach to future work.
In the language of the formulation in~\cite{murthy_giantgravitons}, the determinantal expansion is given by
\begin{equation}
\frac{\tilde{\cZ}_{N}(\mathbf{t}^{+},\mathbf{t}^{-})}{\tilde{\cZ}_{\infty}(\mathbf{t}^{+},\mathbf{t}^{-})} =\sum_{m=0}^{\infty}(-1)^{m} \sum_{\substack{N<r_{1}< \dots <r_{m} \\ r_{1},\ldots,r_{m} \in \mathbb{Z}+\frac{1}{2}}} \det \left(\tilde{K}(r_{p},r_{q};\mathbf{t}^{+},\mathbf{t}^{-})\right)_{p,q=1}^{m}. \label{giant_graviton_scalar}
\end{equation}
The function $\tilde{K}$ is defined through the formal series coefficients of the following generating function, 
\begin{equation}
\frac{J(z;\mathbf{t}^{+},\mathbf{t}^{-})}{J(w;\mathbf{t}^{+},\mathbf{t}^{-})}\frac{\sqrt{zw}}{z-w}= \sum_{r,s \in \mathbb{Z}+\frac{1}{2}}\tilde{K}(r,s;\mathbf{t}^{+},\mathbf{t}^{-})z^{r}w^{-s}, \; J(z;\mathbf{t}^{+},\mathbf{t}^{-})= \exp \left( \sum_{k=1}^{\infty} \frac{1}{k}\left(t^{+}_{k}z^{k}-t^{-}_{k}z^{-k} \right) \right). \label{ktildeGF}
\end{equation}
The generating function $J$ may be regarded as a generalization of that of the Bessel functions~\cite{Okounkov2001-kf}.
Similar to~\eqref{cauchy_identity}, note that~\eqref{ktildeGF}~also holds true when the $t^{\pm}_{k}$ are not complex conjugate.

Plugging in~\eqref{giant_graviton_scalar} into~\eqref{eq:HStransform_sc_coup}, the expansion
\begin{equation}
\cZ_{N}(\mathbf{g})=\sum_{m=0}^{\infty}(-1)^{m} \sum_{\substack{N<r_{1}< \dots <r_{m} \\ r_{1},\ldots,r_{m} \in \mathbb{Z}+\frac{1}{2}}}\left\langle \tilde{Z}_{\infty}(\mathbf{t}^{+},\mathbf{t}^{-}) \det \left(\tilde{K}(r_{p},r_{q};\mathbf{t}^{+},\mathbf{t}^{-})\right)_{p,q=1}^{m} \right\rangle_{\mathbf{g}}, \label{giant_graviton_scalar_coupling_unnormalized}   
\end{equation}
is obtained which, after normalization by $\cZ_{\infty}(\mathbf{g})$, can be written in the following form,
\begin{equation}
\frac{\cZ_{N}(\mathbf{g})}{\cZ_{\infty}(\mathbf{g})}=\sum_{m=0}^{\infty} G^{(m)}_{N}(\mathbf{g}), \label{contrib_expansion_normalized_scalar}
\end{equation}
where at level $m$,
\begin{equation}
G^{(m)}_{N}(\mathbf{g})= \frac{(-1)^{m}}{\cZ_{\infty}(\mathbf{g})} \sum_{\substack{N<r_{1}< \dots <r_{m} \\ r_{1},\ldots,r_{m} \in \mathbb{Z}+\frac{1}{2}}}\left\langle \tilde{Z}_{\infty}(\mathbf{t}^{+},\mathbf{t}^{-}) \det \left(\tilde{K}(r_{p},r_{q};\mathbf{t}^{+},\mathbf{t}^{-})\right)_{p,q=1}^{m} \right\rangle_{\mathbf{g}} \label{giant_graviton_scalar_coupling}
\end{equation}
may be given the interpretation~\cite{murthy_giantgravitons} of contributions from $m$ giant gravitons in the dual theory.
Note that $G^{(0)}_{N}(\mathbf{g})=1$, and the terms $\cZ_{\infty}(\mathbf{g})G^{(m)}(\mathbf{g})$ for $m>0$ may be viewed as corrections to $\cZ_{\infty}(\mathbf{g})$ to output the finite $N$ index $\cZ_{N}(\mathbf{g})$.
It is interesting to note that the giant graviton contributions oscillate in sign with level $m$; see~\cite{lee2024bulkthimblesdualtrace,Eleftheriou_2024} for an explanation of this behaviour for the $\frac{1}{2}$-BPS specialization from the gravity side.\footnote{We thank Ji Hoon Lee for bringing these works to our attention.}

At this juncture we note a interesting property that has been conjectured~\cite{gaiotto2021giantgravitonexpansion} from observation for the specific form of coefficients $g_{k}=f(t^{k},u^{k},v^{k};p^{k},q^{k})$, where
\begin{equation}
f(t,u,v;p,q)=1-\frac{(1-t)(1-u)(1-v)}{(1-p)(1-q)}. \label{firstletter_N=4SYM_generic}
\end{equation}
This, sans the gauge group character data, is the general form of the single-letter $\cN=4$ SYM index, which we specialize to $t=u=v=(pq)^{\frac{1}{3}}$ in our work.
There appears to be an `inversion' symmetry for the matrix model for this form of $f$, which interchanges the roles of the fugacities $p,q$ in the denominator with two of the three fugacities $t,u,v$ in the numerator:
\begin{equation}
\frac{\cZ_{N}(t,u,v;p,q)}{\cZ_{\infty}(t,u,v;p,q)}=\sum_{m=0}^{\infty}t^{mN}\cZ_{m}(t^{-1},p,q;u,v), \label{conj:gaiotto_et_al}
\end{equation}
i.e. in the language of giant graviton expansion,
\begin{equation}
G_{N}^{(m)}(t,u,v;p,q)=t^{mN}\cZ_{m}(t^{-1},p,q;u,v). 
\end{equation}
\eqref{conj:gaiotto_et_al} can be analytically proven using e.g. the Frobenius formula~\cite{murthy_giantgravitons} for the special case of the $\frac{1}{2}-$BPS index, for which two fugacities each in the numerator and denominator are switched off, e.g. $g_{k}=t^{k}$, but remains conjectural in general. 
We do not pursue this further in this paper, but it would be interesting to explore possible generalizations, in particular for matrix couplings, of this property.
Importantly, the functional form of $\cZ$ on the right-hand-side of~\eqref{conj:gaiotto_et_al} is the analytic continuation of that in the regime $|t|<1$, and not the expression obtained by evaluating the contour integrals resulting from the maximal toral decomposition of the matrix model.
See~\cite{Ezroura:2024wmp} for detailed discussion on the analytic properties of the index determined by~\eqref{firstletter_N=4SYM_generic}, when the fugacities lie outside the unit circle.

For the first, trivial mode $m=0$, it is straightforward to show that $G_{N}^{(0)}(\mathbf{g})=1$. 
The contribution from the first non-trivial mode, $m=1$ is also relatively simple to analyze, for the determinant reduces to a sum over one index,
\begin{equation}
G^{(1)}_{N}(\mathbf{g}):=-\frac{1}{\cZ_{\infty}(\mathbf{g})} \sum_{\substack{N<r \\ r \in \mathbb{Z}+\frac{1}{2}}}\left\langle \tilde{\cZ}_{\infty}(\mathbf{t}^{+},\mathbf{t}^{-}) \tilde{K}(r,r;\mathbf{t}^{+},\mathbf{t}^{-}) \right\rangle_{\mathbf{g}}.  
\end{equation}
To compute $G^{(1)}_{N}(\mathbf{g})$, a generating function in the index $N$ may be constructed (see~\cite{murthy_giantgravitons} for details),
\begin{equation}
\tilde{K}(\zeta;\mathbf{g}):= \sum_{N \in \mathbb{Z}} G^{(1)}_{N}(\mathbf{g}) \zeta^{-N} = -\frac{\zeta}{(1-\zeta)^{2}}\left( \sum_{k=1}^{\infty} \frac{1}{k}(-2+\zeta^{k}+\zeta^{-k}) \frac{g_{k}}{1-g_{k}} \right), \label{gs_1st_order_giantgrav_main_scalar}  
\end{equation}
which is especially useful for computation in the following sense: if the $g_{k}$ are of the particular form $g_{k}=g(z^{k})$ in some fugacity variable $z$ (as happens for the univariate index we consider in this paper), and the $\gamma_{k}:=\frac{g_{k}}{1-g_{k}}$ have the power series expansion
\begin{equation}
\gamma_{k}(z)=\sum_{n}c_{n}z^{kn}, \label{gamma_k}
\end{equation}
then~\eqref{gs_1st_order_giantgrav_main_scalar} assumes the infinite product form
\begin{equation}
\tilde{K}(\zeta;z)=\frac{1}{(1-\zeta)(1-\zeta^{-1})}\prod_{n} \left[ \frac{(1-z^{n})^{2}}{(1-\zeta z^{n})(1-\zeta^{-1}z^{n})} \right]^{c_{n}}.
\label{gs_1st_order_giantgrav_quiver_productform_scalar}
\end{equation}
The form~\eqref{gs_1st_order_giantgrav_quiver_productform_scalar}, which may be readily derived using the plethystic exponential, is understood to be evaluated in the regime $|z|<|\zeta|<1$, with the infinite product over $n$ performed first to obtain a power series in $z$, which has as coefficients Laurent polynomials in $\zeta$.
Note that it is typically assumed $g(z)=\cO(z)$, i.e. so that $z=0$ corresponds to switching off the coupling constants in~\eqref{eq:mainmodel}.
This means that the $\gamma_{k}$ are also $\cO(z)$, and hence the indexing in $n$ is understood to be over positive integers.
Generalizing the steps in this calculation, an explicit form can eventually be obtained for the $G^{(m)}_{N}(\mathbf{g})$ for all $m \ge 1$~\cite{Liu_2023}, in terms of the coefficients of a multi-variable generating function in an alphabet of complex variables $\{z_{l},w_{l}: 1 \le l \le m\}$,
\begin{equation}
G^{(m)}_{N}(\mathbf{g})=\frac{(-1)^{m}}{m!} \left[ \prod_{l=1}^{m} \frac{w_{l}/z_{l}}{1-w_{l}/z_{l}} \det \left( \frac{1}{1-w_{l}/z_{l'}} \right)_{l,l'=1}^{m} \exp \left( -\sum_{k=1}^{\infty} \frac{1}{k}\alpha_{k}\alpha_{-k} \gamma_{k} \right) \right]_{\prod_{l=1}^{m}z^{N}_{l}w^{-N}_{l}}, \label{general_Gmz_V=1}
\end{equation}
where $\alpha_{k}=\sum_{l=1}^{m}(z_{l}^{k}-w_{l}^{k})$ for all non-zero $k \in \mathbb{Z}$, and $\gamma_{k}=\frac{g_{k}}{1-g_{k}}$; we refer the reader to~\cite{Liu_2023} for details.
We also note that for the specific case of the $\frac{1}{2}-$BPS index,~\cite{Liu_2023} report obtaining different expressions for the $G^{(m)}_{N}(\mathbf{g})$ by computation compared to the analytical result.
This discrepancy is further analyzed in~\cite{Eniceicu:2023uvd,Ezroura:2024wmp}; it indicates that the expansion may not be uniquely defined, but also that the different approaches may be related to one another through analytic properties of the index.

It was proven~\cite{Liu_2023} that if the $\gamma_{k}$ are of the form~\eqref{gamma_k}, with leading order behaviour $\gamma_{k}=c_{\alpha}z^{\alpha k}$, i.e. the first non-zero coefficient is $c_{\alpha}$, that of the giant graviton expansion coefficients is at least $\cO(z^{\alpha m(N+m)})$, i.e.
\begin{equation}
G^{(m)}_{N}(z) = c_{N,m}z^{\alpha( m(N+m)+\beta_{N,m})}+\ldots, \label{G_scalar_leading_order_behaviour_general}
\end{equation}
for non-zero $c_{N,m}$ and non-negative $\beta_{N,m}$. 
In several cases, it has been observed that this bound is saturated, i.e. $\beta_{m,N}=0$.
We shall observe a matrix-coupling analogue of this leading-order behaviour in the quiver examples we study.

\subsection{Matrix-coupling generalization}

We now develop an analogous notion of giant graviton expansion for the matrix-coupling model~\eqref{eq:mainproblemmultimatrix}.
The natural generalization of the HS transformation~\eqref{eq:gaussiantransform} in this case is
\begin{equation}
\exp \left(\mathbf{a}^{T}\sF \mathbf{b} \right)= \frac{1}{\det \sF} \prod_{j=1}^{V}\int_{\mathbb{C}} \frac{\dd t_{j}\, \wedge \dd \bar{t}_{j}}{2\pi i} \exp\left(-\bar{\mathbf{t}}^{T}\Lambda^{-1}\mathbf{t}+\mathbf{a}^{T}\bm{\tau}+\bm{\sigma}^{T} \mathbf{b} \right), \label{eq:generalizationofgaussiannew}    
\end{equation}
for a $V \times V$ matrix $\sF$ diagonalizing as $\sF=P \Lambda P^{-1}$, coupling constants $\mathbf{a},\mathbf{b} \in \mathbb{C}^{V}$, holomorphic (respectively, anti-holomorphic) differentials $\left\{\dd t_{j}\right\}_{1\le j \le V}$, $\left\{\dd \bar{t}_{j}\right\}_{1 \le j \le V}$ on $V$ copies of $\mathbb{C}$, and the transformed indices $\bm{\tau}=P\mathbf{t}, \, \bm{\sigma}=(P^{-1})^{T}\bar{\mathbf{t}}$ for the coordinates $\mathbf{t}=(t_{j})_{i \le j \le V}$ on $\mathbb{C}^{V}$. 
Note that in general $\bm{\tau}$ and $\bm{\sigma}$ are not complex conjugate unless $(P^{-1})^{T}=\bar{P}$; we shall soon see that this particular case actually arises in our work in specific quiver examples.

Using~\eqref{eq:generalizationofgaussiannew}, we are able to rewrite~\eqref{eq:mainproblemmultimatrix} in the following way.
Suppose $\fM=\left\{\sM_{k}\right\}_{k\ge 1}$ is a set of $V \times V$ matrix couplings which diagonalize as $\sM_{k}=P_{k}\Lambda_{k}P_{k}^{-1}$, for all $k$.
Let $\mathbf{t}^{+}=\left\{t^{+}_{jk}\right\}_{j \le 1 \le V,k \ge 1}$, $\mathbf{t}^{-}=\left\{t^{-}_{jk}\right\}_{j \le 1 \le V,k \ge 1}$ be doubly-indexed sets of coupling constants.
Hereon in this section, unless otherwise mentioned, we shall adopt the convention that the variable $j$ runs from $1$ to $V$, and the variable $k$ over $\mathbb{Z}^{+}$.
Likewise, unless otherwise mentioned, we shall assume complex conjugacy, $\overline{t^{+}_{jk}}=t^{-}_{jk}$.
We introduce the shorthand vector notations
\begin{equation}
\mathbf{t}^{\pm}_{j}=(t^{\pm}_{jk})_{k \ge 1}, \; t^{\pm}_{k}=(\mathbf{t}^{\pm}_{jk})_{1 \le j \le V}. \label{t_shorthand} 
\end{equation}
Employing these conventions, we define the generalized GWW partition function
\begin{equation}
\tilde{\cZ}_{N}(\mathbf{t}^{+},\mathbf{t}^{-};\fM)=\int_{\bigtimes_{i=1}^{V}\mathrm{U}(N)} \prod_{i=1}^{V}\dd U_{i} \exp \left( \sum_{k=1}^{\infty} \frac{1}{k} \sum_{i=1}^{V} \left(\tau_{ik} \tr U_{i}^{k} + \sigma_{ik} \tr U_{i}^{-k}\right) \right), ~\label{gen_gww}
\end{equation}
where we introduce the doubly-indexed parameters $\bm{\tau}_{k}=P_{k}\mathbf{t}^{+}_{k}=(\tau_{jk})_{1 \le j \le V, k \ge 1}, \bm{\sigma}_{k}=\left(P_{k}^{-1} \right)^{T}\mathbf{t}_{k}^{-}=(\sigma_{jk})_{1 \le j \le V,k \ge 1}$.
Similar to~\eqref{t_shorthand}, we will use the notation
\begin{equation}
\mathbf{x}_{j}=(\mathbf{x}_{jk})_{k \ge 1}, \;  \mathbf{x}_{k}=(\mathbf{x}_{jk})_{1 \le j \le V},\label{tau_sigma_shorthand} 
\end{equation}
for $\mathbf{x}=\bm{\tau},\bm{\sigma}$.
It is useful to note, for later use, the `invariance of scalar product' identities, which follow from definition,
\begin{equation}
\bm{\sigma}_{k}^{T}\bm{\tau}_{k}=(\mathbf{t}_{k}^{-})^{T}\mathbf{t}_{k}^{+}. \label{scalar_product_tausigma}
\end{equation}

Notice that in~\eqref{gen_gww}, at each level $k$ we have a $\fS_{V} \times \mathbb{C}^{V}$-fold redundancy in ordering and scaling the eigenvectors constituting the columns of the $P_{k}$ (here $\fS_{V}$ is the permutation group on $V$ elements).
We shall see that precisely fixing this redundancy is irrelevant to computing the giant graviton expansion.
For scalar couplings~\eqref{gen_gww} reduces to~\eqref{gen_gww_scalar} up 
to a $\mathbb{C}$-fold scaling redundancy, and it can be seen in the general formula~\eqref{general_Gmz_V=1}, which our result will reduce to for $V=1$, that this is irrelevant.
In general~\eqref{gen_gww} exhibits dependency on the internal structure of the entries of $\fM$, but we shall see that in special cases and limits, this dependency is simplified or eliminated.
Also note that $\bm{\tau}$ and $\bm{\sigma}$ are not necessarily complex conjugate, unless $(P_{k}^{-1})^{T}=\overline{P}_{k}$; this is a special case we shall encounter later in this section.

As a generalization of the scalar product scenario, it is straightforward to derive the relation 
\begin{equation}
\cZ_{N}(\fM)= \prod_{k=1}^{\infty} \left( \frac{1}{\det \sM_{k}} \prod_{j=1}^{V} \int_{\mathbb{C}} \frac{\dd t^{+}_{jk} \wedge \dd t^{-}_{jk}}{2\pi i k}  e^{-\frac{1}{k} \left(\mathbf{t}_{k}^{+}\right)^{T}\Lambda_{k}^{-1}\mathbf{t}_{k}^{-}} \right) \tilde{\cZ}_{N}(\mathbf{t}^{+},\mathbf{t}^{-};\fM). \label{eq:HStransform_gen}
\end{equation}
which, introducing the notation
\begin{equation}
\left\langle f(\mathbf{t}^{+},\mathbf{t}^{-};\fM) \right\rangle_{\fM}:=\prod_{k=1}^{\infty} \left( \frac{1}{\det \sM_{k}} \prod_{j=1}^{V} \int_{\mathbb{C}} \frac{\dd t^{+}_{jk} \wedge \dd t^{-}_{jk}}{2\pi i k}  e^{-\frac{1}{k} \left(\mathbf{t}_{k}^{+}\right)^{T}\Lambda_{k}^{-1}\mathbf{t}_{k}^{-}} \right)f(\mathbf{t}^{+},\mathbf{t}^{-};\fM), \label{general_transform}
\end{equation}
for a function $f$ depending on $\mathbf{t}^{+},\mathbf{t}^{-},\fM$, is compactly written as $\cZ_{N}(\fM)=\left\langle \tilde{\cZ}_{N}(\mathbf{t}^{+},\mathbf{t}^{-};\fM) \right\rangle_{\fM}$.
We may call the right-hand-side of~\eqref{general_transform} the $\left\langle \cdots \right\rangle_{\fM}$-average of $f$.

As noted in~\cite{murthy_giantgravitons} for the scalar case, the $\langle \cdots \rangle_{\fM}$-average vanishes for terms in the series expansion of $f$ that contain unequal powers of the conjugate variables $t^{+}_{jk}$ and $t^{-}_{jk}$,
i.e. ignoring questions of convergence, if $f$ is formally expanded as
\begin{equation}
f(\mathbf{t}^{+},\mathbf{t}^{-};\fM)=\sum_{\mathbf{a},\mathbf{b}} C_{\mathbf{a}\mathbf{b}} \prod_{k=1}^{\infty} \prod_{j=1}^{V}(t^{+}_{jk})^{a_{jk}}(t^{-}_{jk})^{b_{jk}}, \label{formal_expansion_only_equal_powers}
\end{equation}
where $\mathbf{a}=\left\{a_{jk}\right\}_{1 \le j \le V,k \ge 1}$, $\mathbf{b}=\left\{b_{jk}\right\}_{1 \le j \le V, k \ge 1}$ are multisets of integers, then $C_{\mathbf{a}\mathbf{b}}=0$ whenever $\mathbf{a}\ne\mathbf{b}$, and $\left\langle f(\mathbf{t}^{+},\mathbf{t}^{-};\fM) \right\rangle_{\fM}$ is a function solely of the $C_{\mathbf{a}\mathbf{a}}$.

Note that~\eqref{gen_gww} factorizes into $V$ copies as $\tilde{\cZ}_{N}(\mathbf{t}^{+},\mathbf{t}^{-};\fM)=\prod_{j=1}^{V}\tilde{Z}_{N,j}(\mathbf{t}^{+},\mathbf{t}^{-};\fM)$, where
\begin{equation}
\tilde{\cZ}_{N,j}(\mathbf{t}^{+},\mathbf{t}^{-};\fM)= \int_{\mathrm{U}(N)} \dd U \exp \left( \sum_{k=1}^{\infty} \frac{1}{k}  \left(\tau_{jk} \tr U^{k} + \sigma_{jk} \tr U^{-k}\right) \right),
\end{equation}
and hence we may treat each copy separately.
By the Cauchy identity~\eqref{cauchy_identity}, at the $j$-th level we have
\begin{equation}
\tilde{\cZ}_{\infty,j}(t^{+},t^{-};\fM):=\lim_{N \rightarrow \infty} \tilde{Z}_{N,j}(t^{+},t^{-};\fM)=\exp \left( \sum_{k=1}^{\infty} \frac{1}{k}\tau_{jk}\sigma_{jk}  \right), \label{Ztilde_jth}
\end{equation}
and then by the Borodin--Okounkov determinantal formula, similar to~\eqref{giant_graviton_scalar} we have
\begin{equation}
 \frac{\tilde{\cZ}_{N,j}(\mathbf{t}^{+},\mathbf{t}^{-};\fM)}{\tilde{\cZ}_{\infty,j}(\mathbf{t}^{+},\mathbf{t}^{-};\fM)} =\sum_{m=0}^{\infty}(-1)^{m} \sum_{\substack{N<r_{1}< \dots <r_{m} \\ r_{1},\ldots,r_{m} \in \mathbb{Z}+\frac{1}{2}}} \det \left(\tilde{K}(r_{p},r_{q};\bm{\tau}_{j},\bm{\sigma}_{j})\right)_{p,q=1}^{m}, \label{giant_graviton_jth}
\end{equation}
where we remind the reader of the notation~\eqref{tau_sigma_shorthand}, and that~$\tilde{K}$ is defined by~\eqref{ktildeGF}.

Using~\eqref{scalar_product_tausigma} and~\eqref{Ztilde_jth}, we obtain the large $N$ limit of~\eqref{eq:mainproblemmultimatrix},
\begin{equation}
\tilde{\cZ}_{\infty}(\mathbf{t}^{+},\mathbf{t}^{-};\fM)=\exp \left( \sum_{k=1}^{\infty} \frac{1}{k} \bm{\sigma}^{T}_{k} \bm{\tau}_{k}\right)=\exp \left( \sum_{k=1}^{\infty} \frac{1}{k} (\mathbf{t}^{-})^{T}_{k} \mathbf{t}^{+}_{k}\right). \label{eq_Zinf}
\end{equation}
We see that in the large $N$ limit, the generalized GWW partition function depends solely on the size $V$ of the quiver, and is insensitive to its internal structure, $\tilde{\cZ}_{\infty}(\mathbf{t}^{+},\mathbf{t}^{-};\fM) \equiv \tilde{\cZ}_{\infty}(\mathbf{t}^{+},\mathbf{t}^{-};V)$.
The information about the internal structure in the eventual expansion comes from the $\det \sM_{k}$, the $\det \tilde{K}$ and the exponential term containing the inverse eigenvalues $\Lambda^{-1}_{k}$.

We are now ready to develop the giant graviton expansion.
Taking a product over all $j$ in~\eqref{giant_graviton_jth}, we obtain 
\begin{equation}
\cZ_{N}(\fM)= \left\langle  \tilde{\cZ}_{\infty}(\mathbf{t}^{+},\mathbf{t}^{-};V) \prod_{j=1}^{V}\left( \sum_{m_{j}=0}^{\infty}(-1)^{m_{j}} \sum_{\substack{N<r_{1}< \dots <r_{m_{j}} \\ r_{1},\ldots,r_{m} \in \mathbb{Z}+\frac{1}{2}}} \det \left(\tilde{K}(r_{p},r_{q};\bm{\tau}_{j},\bm{\sigma}_{j})\right)_{p,q=1}^{m_{j}} \right) \right\rangle_{\fM}.\label{eq:giant_grav_expansion_multi_model}
\end{equation}
 which we have written in compact product form for brevity.
 When fully expanded,~\eqref{eq:giant_grav_expansion_multi_model} is a sum over $V$-tuplets $(m_{1},\ldots,m_{V}) \in \mathbb{N}^{V}$.
The first term of the expansion is given by $\left\langle \tilde{\cZ}_{\infty}(t^{+},t^{-};V) \right\rangle_{\fM}$, and it is straightforward to see from~\eqref{eq_Zinf} that it is~\eqref{eq:coulombgasfinalunitary}.
The successive terms may be viewed as corrections to~\eqref{eq:coulombgasfinalunitary} to output $\cZ_{N}(\fM)$.
Analogous to the scalar case~\eqref{contrib_expansion_normalized_scalar}, after normalization we define the expansion
\begin{equation}
\frac{\cZ_{N}(\fM)}{\cZ_{\infty}(\fM)}:=\sum_{m_{1},\ldots,m_{V}=0}^{\infty}G^{(m_{1},\ldots,m_{V})}_{N}(\fM), \label{result:giant_grav_expansion}
\end{equation}
over $V$-tuplets $\mathbf{m}:=(m_{1},\ldots,m_{V}) \in \mathbb{N}^{V}$, where, denoting $|\mathbf{m}|=\sum_{j=1}^{V}m_{j}$,
\begin{equation}
G^{(\mathbf{m})}_{N}(\fM)= \frac{(-1)^{|\mathbf{m}|}}{\cZ_{\infty}(\fM)\prod_{j=1}^{V}m_{j}!} \left\langle  \tilde{\cZ}_{\infty}(\mathbf{t}^{+},\mathbf{t}^{-};V) \left( \prod_{j=1}^{V} \sum_{\substack{N<r_{j1}, \dots ,r_{jm_{j}} \\ r_{j1},\ldots,r_{jm_{j}} \in \mathbb{Z}+\frac{1}{2}}} \det \left(\tilde{K}(r_{jp},r_{jq};\bm{\tau}_{j},\bm{\sigma}_{j})\right)_{p,q=1}^{m_{j}} \right) \right\rangle_{\fM}, \label{contrib_expansion_normalized} 
\end{equation}
where, employing column and row symmetries in the $\det \tilde{K}$, it is convenient to unrestrict the ordering of the dummy variables in the sums inside the $\langle \cdots \rangle_{\fM}$-average for ease of later computation~\cite{Liu_2023}, leading to a pre-factor of $1/(\prod_{j=1}^{V}m_{j}!)$, and to introduce the labelling subscript $_{j}$ in these variables for clarity.
As noted, we have $G^{(\mathbf{0})}_{N}(\fM)=1$.
Also note that the sign of the expansion parameters depends upon $|\mathbf{m}|$; it would be interesting to better understand this behaviour in the gravity dual setup. 

When the couplings $\fM$ describe a physical theory, we intuitively expect that the giant graviton expansion terms may characterize physically degenerate modes. 
A natural signature of such degeneracy would be the invariance of the $G^{(\mathbf{m})}_{N}(\fM)$ under the action of the permutation group $\fS_{V}$ or some of its subgroups on the entries of $\mathbf{m}$. 
Such invariance would arise from symmetries in the coupling constants $\sM_{k}$.
The strictest case is that the $G^{(\mathbf{m})}_{N}(\fM)$ are invariant under the action of $\fS_{V}$, i.e. the expansion is expressed as a sum purely over integer partitions $\lambda$,
\begin{equation}
\frac{\cZ_{N}(\fM)}{\cZ_{\infty}(\fM)}:=\sum_{\substack{\lambda \\ \ell(\lambda) \le V}} (-1)^{|\lambda|}\binom{V}{\mathbf{r}}G^{(\lambda)}_{N}(\fM), \label{conj:giantgrav_partition}
\end{equation}
for class representatives $G^{(\lambda)}_{N}(\fM)$ of correction.
Here $\lambda$ is a partition of length $\ell(\lambda)$ and size $|\lambda|$, and the degeneracy is expressed through the multinomial coefficients $\binom{V}{\mathbf{r}}= \frac{V!}{\prod_{i \ge 1}r_{i}!}$, where $\mathbf{r}=(r_{i})_{i \in \mathbb{Z}^{+}}$ is the frequency representation of $\lambda$ (so that $|\lambda|=\sum_{i \ge 1}ir_{i}$).
It turns out this expectation is unrealistic; we shall now see, by constructing an explicit formula for the expansion parameters for $|\mathbf{m}| \ge 1$, that the symmetry subgroup of $\fS_{V}$ involved is typically much reduced.

\subsubsection*{First-order correction}

At the first order, the total correction to the index has a relatively simple expression in terms of the generating function for the $\tilde{K}$ in terms of $V$ modes,
\begin{equation}
\sum_{|\mathbf{m}|=1}G^{(\mathbf{\mathbf{m}})}_{N}(\fM)= -\frac{1}{\cZ_{\infty}(\fM)} \sum_{j=1}^{V} \sum_{\substack{N<r_{j} \\ r_{j} \in \mathbb{Z}+\frac{1}{2}}} \left\langle  \tilde{Z}_{\infty}(t^{+},t^{-};\fM) \tilde{K}(r,r;\bm{\tau}_{j},\bm{\sigma}_{j}) \right\rangle_{\fM}.\label{contrib_expansion_normalized_firstorder}
\end{equation}
Generalizing the calculations performed for the scalar case in~\cite{murthy_giantgravitons}, and keeping the discussion around~\eqref{formal_expansion_only_equal_powers} in mind, we obtain, using the generating function~\eqref{ktildeGF}, the correction as follows.
Let $\mathbf{1}_{j}$ denote the mode with $1$ at the $j$-th entry and $0$ otherwise.
Then we have the generating function for this mode,
\begin{eqnarray}
\hat{K}^{(\mathbf{1}_{j})}(\zeta;\fM) &:=& \sum_{N \in \mathbb{Z}}G_{N}^{(\mathbf{1}_{j})}(\fM)\zeta^{-N}  \nonumber \\ &=& -\frac{1}{\cZ_{\infty}(\fM)} \frac{\zeta}{(1-\zeta)^{2}} \left\langle  \tilde{Z}_{\infty}(t^{+},t^{-};\fM)\frac{J(z;\bm{\tau}_{j},\bm{\sigma}_{j})}{J(w;\bm{\tau}_{j},\bm{\sigma}_{j})} \right\rangle_{\fM} \nonumber \\ &=& -\frac{\zeta}{(1-\zeta)^{2}}\exp \left( \sum_{k=1}^{\infty} \frac{1}{k}(-2+\zeta^{k}+\zeta^{-k}) ((\sM_{k}^{-1}-I)^{-1})_{jj} \right),\label{gs_1st_order_giantgrav_main}
\end{eqnarray}
and the generating function for the full correction is simply $\sum_{j=1}^{V}\hat{K}^{(\mathbf{1}_{j})}(\zeta;\fM)$.
This formula generalizes~\eqref{gs_1st_order_giantgrav_main_scalar}, and reduces to it for $V=1$.
We provide some details on the involved calculations in~\cref{appendix:misc}, in the context of the general $|\mathbf{m}| \ge 1$ case, which we soon discuss.

For the quivers we have considered in this paper, for which the couplings are given by~\eqref{eq:connectionwithsupindex}, we have
\begin{equation}
(\sM_{k}^{-1}-I)^{-1}=\sM_{k}(1-\sM_{k})^{-1}=(1-p^{k})(1-q^{k})M^{-1}\left((pq)^{k/2} \right)-I, \label{specialization_bivariate_Mk_giantgrav}
\end{equation}
and~\eqref{gs_1st_order_giantgrav_main} becomes
\begin{equation}
\hat{K}^{(\mathbf{1}_j)}(\zeta;p,q)=-\frac{\zeta}{(1-\zeta)^{2}}\exp \left( \sum_{k=1}^{\infty} \frac{1}{k}(-2+\zeta^{k}+\zeta^{-k}) \left((1-p^{k})(1-q^{k})(M^{-1}((pq)^{k/2}))_{jj} -1\right) \right). \label{gs_1st_order_giantgrav_quiver}
\end{equation}
Following~\cite{murthy_giantgravitons}, we may write~\eqref{gs_1st_order_giantgrav_quiver} in a form more amenable for calculations.
Anticipating a look at some quiver examples, let us consider the univariate regime $p=q=t=z^{3}$, adopting the parametrization most often used in~\cref{sec:analysis}, for which all power series in $z$ in our setup have terms with integer exponents. 
Defining\footnote{This particular definition is in anticipation of its eventual use in the matrix-coupling case later in the section.}
\begin{equation}
\hat{f}_{jj'}(z):=(1-z^{3})^{2}(M^{-1}(z^{3}))_{jj'} -\delta_{jj'}=\sum_{n}c_{n}^{(j,j')}z^{n} \label{f_jjprime}
\end{equation}
for $1\le j,j' \le V$, we obtain, from~\eqref{gs_1st_order_giantgrav_quiver} and utilizing the plethystic exponential, the infinite product form\footnote{Note that this generating function is invariant under the transformation $\zeta \mapsto \zeta^{-1}$; this implies $G_{N}^{\mathbf{1}_{j}}(\zeta;z)=G_{-N}^{\mathbf{1}_{j}}(\zeta;z)$.
Index corrections with negative gauge group rank can be mathematically defined despite not having any obvious physical meaning.}
\begin{equation}
\hat{K}^{(\mathbf{1}_{j})}(\zeta;z)= \frac{1}{(1-\zeta)(1-\zeta^{-1})}\prod_{n} \left[ \frac{(1-z^{n})^{2}}{(1-\zeta z^{n})(1-\zeta^{-1}z^{n})} \right]^{c^{(j,j)}_{n}}.\label{gs_1st_order_giantgrav_quiver_productform}
\end{equation}
As with the scalar case, the form~\eqref{gs_1st_order_giantgrav_quiver_productform} is understood to be evaluated in the regime $|z|<|\zeta|<1$, with the infinite product over $z$ performed first to obtain a Laurent series in $\zeta$.
Note\footnote{Strictly speaking, we avoid the singularity due to $t^{2} \cM(t^{-1})$ by imposing the off-shell conditions, as then the $R$-charges cannot exceed $2$.} from~\eqref{mainmatrix} that as $M(0)=I$, $\hat{f}_{jj'}=\cO(z)$ as $z \rightarrow 0$ for all $j,j'$, and hence the dummy index $n$ in~\eqref{f_jjprime} and~\eqref{gs_1st_order_giantgrav_quiver_productform} is understood to be a positive integer.

Let us remark on the dependence of the leading order behaviour of the $G^{(\mathbf{1}_{j})}_{N}(z)$ on the coefficients of the expansion~\eqref{f_jjprime}; for brevity we drop the mode superscript $^{(j,j)}$.
If the first non-zero coefficient $c_{\alpha}$ is negative, then a check of~\eqref{gs_1st_order_giantgrav_quiver_productform} shows that the leading-order term of $G^{(\mathbf{1}_{j})}_{N}(z)$ alternates in sign with $N$, and has conventional binomial coefficients modulo sign,
\begin{equation}
G^{(\mathbf{1}_{j})}_{N}(z)=(-1)^{N} \binom{|c_{\alpha}|}{N+1}z^{\alpha(N+1)}, \; 1 \le N \le |c_{\alpha}|-1. \label{1st_order_correction_behaviour_negative}
\end{equation}
The behaviour for $N \ge c_{\alpha}$ depends on the sign of the next non-zero coefficients; it continues to be of such an oscillatory-in-sign binomial form unless there exists a positive $c_{n}$.
In this case, after $N$ exceeds a certain value, the coefficient of the leading order term always remains negative, and proportional to generalized binomial coefficients.
Making this more precise in the context of examples we will study, suppose the first positive coefficient is $c_{\alpha'}$, then we will have
\begin{equation}
G^{(\mathbf{1}_{j})}_{N}(z)=-\binom{c_{\alpha'}+N}{c_{\alpha'}-1} z^{\alpha'(N+1)}, \; N \ge \operatorname{max}\left\{|c_{\alpha<\alpha'}|\right\}, \label{1st_order_correction_behaviour_positive}
\end{equation}
where we understand that if there are no negative $c_{n}$, $\operatorname{max}(|c_{\alpha<\alpha'}|)=1$.
In a mathematical sense,~\eqref{1st_order_correction_behaviour_negative} and~\eqref{1st_order_correction_behaviour_positive} may be said to correspond to `fermionic' and 'bosonic' behaviour respectively, though a physical interpretation, if any is not immediately apparent.
Note that the leading order behaviour in either case is identical to that in the scalar-coupling regime,~\eqref{G_matrix_leading_order_behaviour_general}, for $m=1$.
Similar but more involved relations, which we only briefly discuss soon, may be obtained for the leading order behaviour in the more general case of $|\mathbf{m}|>1$.

\subsubsection*{General order correction}

Generalizing the steps in~\cite{Ezroura:2024wmp} and the above arguments, we may write down an explicit formula for $G^{(\mathbf{m})}_{N}(\fM)$ for arbitrary $\mathbf{m}=(m_{j})_{j=1}^{V}$, the matrix-coupling generalization of~\eqref{general_Gmz_V=1}.
Utilizing an alphabet of complex variables $\{z_{jl},w_{jl}: 1 \le j \le V, \, 1 \le l \le m_{j} \}$, we obtain
\begin{multline}
G^{(\mathbf{m})}_{N}(\fM)=\frac{(-1)^{|\mathbf{m}|}}{\prod_{j=1}^{V}m_{j}!} \left[ \prod_{j=1}^{V} \left( \prod_{l=1}^{m_{j}} \frac{w_{jl}/z_{jl}}{1-w_{jl}/z_{jl}} \det \left( \frac{1}{1-w_{jl}/z_{jl'}} \right)_{l,l'=1}^{m_{j}} \right) \right. \\ \left. \exp \left( -\sum_{k=1}^{\infty} \frac{1}{k}\sum_{j,j'=1}^{V}\alpha_{jk} \alpha_{j',-k} ((\sM_{k}^{-1}-I)^{-1})_{jj'}\right) \right]_{\prod_{j=1}^{V}\prod_{l=1}^{m_{j}}z^{N}_{jl}w^{-N}_{jl}}, \label{general_Gmz_V}
\end{multline}
where $\alpha_{jk}=\sum_{l=1}^{m_{j}}(z^{k}_{jl}-w^{k}_{jl})$.
For working with indices considered in this paper,~\eqref{specialization_bivariate_Mk_giantgrav} holds, and we substitute $((\sM_{k}^{-1}-I)^{-1})_{jj'}$ by $(1-p^{k})(1-q^{k})(M^{-1}((pq)^{k/2}))_{jj'} -\delta_{jj'}$, and eventually by $(1-z^{3k})^{2}(M^{-1}(z^{3k}))_{jj'} -\delta_{jj'}$ for the univariate case. 
We provide details on the calculations involved in~\cref{appendix:misc}.
At the first order, i.e. $\mathbf{m}=\mathbf{1}_{j}$,~\eqref{general_Gmz_V} reduces to~\eqref{gs_1st_order_giantgrav_main}.

Similar to the product forms~\eqref{gs_1st_order_giantgrav_quiver_productform_scalar} or~\eqref{gs_1st_order_giantgrav_quiver_productform}, employing the plethystic exponential we may rewrite~\eqref{general_Gmz_V} for the univariate case,
\begin{multline}
G^{(\mathbf{m})}_{N}(z)=\frac{(-1)^{|\mathbf{m}|}}{\prod_{j=1}^{V}m_{j}!} \left[ \prod_{j=1}^{V} \left( \prod_{l=1}^{m_{j}} \frac{w_{jl}/z_{jl}}{1-w_{jl}/z_{jl}} \det \left( \frac{1}{1-w_{jl}/z_{jl'}} \right)_{l,l'=1}^{m_{j}} \right) \right. \\
\left. \prod_{n} \prod_{j,j'=1}^{V} \left[  \prod_{l=1}^{m_{j}}\prod_{l'=1}^{m_{j'}} \frac{\left(1-\frac{z_{jl}}{z_{j'l'}}z^{n} \right) \left(1-\frac{w_{jl}}{w_{j'l'}}z^{n} \right)}{\left(1-\frac{z_{jl}}{w_{j'l'}}z^{n} \right) \left(1-\frac{w_{jl}}{z_{j'l'}}z^{n} \right)} \right]^{c_{n}^{(j,j')}} \right]_{\prod_{j=1}^{V}\prod_{l=1}^{m_{j}}z^{N}_{jl}w^{-N}_{jl}}, \label{general_Gmz_V_productform}
\end{multline}
where the Laurent series expansion~\eqref{f_jjprime} defines the coefficients $c^{(j,j')}_{n}$.
As we observed following~\eqref{gs_1st_order_giantgrav_quiver_productform}, the infinite sum and product over $n$ are taken over the positive integers. 
For computation purposes, the products over $z$ in~\eqref{general_Gmz_V_productform} must be evaluated first, as a Taylor series in $z$ up to a suitable order, and then the full product is to be expanded in the $z_{jl}$ and $w_{jl}$ as a formal series; for convergence of the series in $z$, it suffices to assume $|z|<|w_{jl}/z_{j'l'}|<1$, and $|z|<\left|z_{jl}/z_{j'l'} \right|,\left|w_{jl}/w_{j'l'}\right|$ for all $j,j',l,l'$.

We find from our computational checks (as detailed in~\cref{appendix:misc}) that for some quivers of small size considered in our paper, the leading-order behaviour appears to be
\begin{equation}
G^{(\mathbf{m})}_{N}(z)=c_{N,\mathbf{m}}z^{ \sum_{j=1}^{V}\alpha_{j} m_{j}(N+m_{j})}, \label{G_matrix_leading_order_behaviour_general}
\end{equation}
where the $\alpha_{j}$ appear to depend on the leading order behaviour of the matrix coefficients $M^{-1}_{jj'}(z^{3})$ and the $m_{j'}$, for all $j'$ in general; this pattern is consistent with~\eqref{G_scalar_leading_order_behaviour_general}, but it remains to be seen if it holds true in more general cases. 
We leave an analytic check and refinement of this observation, in particular a precise description of the $\alpha_{j}$ and the $c_{N,\mathbf{m}}$, to later work.

\subsection{Symmetries and quiver examples}

The physical interpretation of the generic giant graviton expansion~\eqref{result:giant_grav_expansion} we have developed for the $V \times V$-coupling matrix model~\eqref{eq:mainproblemmultimatrix}, with the exact form~\eqref{general_Gmz_V} or equivalently~\eqref{general_Gmz_V_productform} for the univariate case, is not clear to us at the moment.
It is plausible that when the matrix model arises from a suitable SCFT, each term in the expansion corresponds, in some sense, to a contribution from a $V$-tuplet of giant gravitons in the dual gravity theory.
As we noted following~\eqref{contrib_expansion_normalized}, we may have degeneracies originating from some inherent symmetries, and the question naturally arises how one may most conveniently organize these tuplets according to their degeneracy structure. 
From~\eqref{general_Gmz_V}, it is clear that such degeneracy would be encoded in the matrices $(\sM_{k}-1)^{-1}$.
In the bivariate specialization~\eqref{specialization_bivariate_Mk_giantgrav}, the matrix $(1-p)(1-q)M^{-1}((pq)^{1/2})-I$, which by definition~\eqref{mainmatrix} is related to the inverse of the weighted adjacency matrix $\cM((pq)^{1/2})$, manifestly plays a fundamental role, and symmetries in this matrix will directly impose degeneracies in the expansion parameters.
Let us now have a closer look at this relation for a few specializations.

One computational hurdle in the most general case is that the matrices $\sM_{k}$ do not simultaneously diagonalize, and except for the leading order term of the expansion, the others all appear to retain `memory' of the matrix couplings rather than just encoding their diagonal, i.e. eigenvalue data.
The $\tilde{K}$ are defined independently of the quiver parameters, but it does not appear obvious if the $\tilde{K}(r_{p},r_{q};\bm{\tau}_{j},\bm{\sigma}_{j})$ can be succinctly expressed in terms of the $\tilde{K}(r_{p},r_{q};\mathbf{t}^{+}_{j},\mathbf{t}^{-}_{j})$ in the general case.
In addition, for large quiver size, the weighted adjacency matrices (and their derivatives) we consider are \textit{band} matrices, i.e. the non-zero entries are concentrated near the diagonal.
The inverse of a band matrix is not band unless the matrix is purely diagonal.

However, for many of the quivers considered in this paper, the large $N$ index matrix~\eqref{mainmatrix} exhibits several symmetries which through~\eqref{eq:connectionwithsupindex} make the analysis of the giant graviton expansion more tractable.
Looking back at~\cref{section:factorizations}, note that when we switch to on-shell $R$-charges, $M(t)$ is always circulant for the $Y^{p,p}$ and the $\hat{A}_{m}$ families; further, it is always symmetric in the latter case.
In addition, it is circulant for the $Y^{1,0}$, $Y^{2,0}$ and $dP3$ quivers (but not for $Y^{p \ge3,0}$).

The eigenvectors of circulant matrices form, up to column interchange symmetry, the matrix of a discrete Fourier transform (DFT), and are independent of the eigenvalues. 
Specifically, at level $k$ we have $\tau_{k}=Pt^{+}_{k}, \sigma_{k}=\bar{P}t^{-}_{k}$, where the DFT matrix $P$ is canonically given by $P_{ij}=\frac{1}{\sqrt{V}} \omega^{(i-1)(j-1)}$ for $\omega=\exp(2\pi i/V)$.
It is unitary and symmetric, $P^{-1}=P^{\dag}=\bar{P}$; in addition, $P^{2}=J$, whose entries are either $0$ or $1$ as follows: $J_{ij}=\delta_{i+j,2}+\delta_{i+j,V+2}$, and $P^{4}=1$. 
Hence for these quivers, the $\sM_{k}$ are circulant, and simultaneously diagonalizable by the DFT matrix.
Since the inverses of circulant matrices are also circulant, from~\eqref{general_Gmz_V} we see that the $G^{(\mathbf{m})}_{N}(\fM)$ are identical at least up to a cyclic permutation of the entries of $\mathbf{m}$.

Let us now look at the correction terms~\eqref{general_Gmz_V} for some examples from the $\hat{A}_{m}$ and $Y^{p,p}$ infinite families on-shell.
For simplicity we will look at the univariate index with variable substitution $p=q=t=z^{3}$, in accordance with the analysis in~\cref{sec:analysis}, so that we may utilize~\eqref{general_Gmz_V_productform} directly.
We begin by determining $M^{-1}(z^{3})_{jj'}$ for these families.

For the $\hat{A}_{m}$ family, the case $m=1$ has been studied in~\cite{murthy_giantgravitons}.
From computer algebra we observe the general form
\begin{equation}
(M^{-1}(z^{3}))_{jj'}[\hat{A}_{m}]=\frac{(z^{2|j-j'|}+z^{2m-2|j-j'|})}{(1-z^{2m})(1-z^{2}
)(1-z^{4})}.  \label{Minverse_ahat_ij}
\end{equation}
We have computationally checked this result for several values of $m$.
Literature and OEIS checks~\cite{OEIS_A115139,Lang01112000} in fact indicate that
\begin{equation}
(M^{-1}(z^{3}))_{jj'}=\frac{1}{\det M(z^{3})}\frac{(z^{2|j-j'|}+z^{2m-2|j-j'|})}{(1+z^{2m})}\cC^{m}(z)S_{m-1}\left(\frac{\cD(z)}{\cC(z)}\right), \label{Minverse_Ahat_ij_singleletter}
\end{equation}
where $\cD(z)$ and $\cC(z)$ are the diagonal and off-diagonal on-shell entries of $M$ respectively, 
\begin{equation}
\cD(z)=1-z^{6}-z^{2}+z^{4}, \; \cC(z)=z^{4}-z^{2}, \label{entries_chebyshev}
\end{equation}
the polynomials $S_{n}(x)$ are scaled Chebyshev polynomials of the second kind, $S_{n}(x)=U_{n}\left(\frac{x}{2} \right)$, and $\det M(z^{3})$ is given by~\eqref{detMAn} on-shell.

For the $Y^{p,p}$ family, we are unable to obtain similarly well-described expressions in $z$; we find that, in the simplest form involving polynomials irreducible over the integers,
\begin{equation}
(M^{-1}(z^{3}))_{jj'}[Y_{p,p}]=\frac{P_{a_{p}(j-j')}(z)}{(1-z^{4p})^{2}(1-z^{2b_{p}})}, \label{Minverse_Ypp_11}
\end{equation}
where the sequence $(b_{p})_{p \ge 1}$ is simply $3$ when $p \equiv 0 \pmod 3$ and $1$ otherwise, and for the specific case of the diagonal entries, $P_{a_{p}(0)}(z)$ is a polynomial of degree $a_{p}(0)=8p-6+2b_{p}$ with non-trivial roots clustered fairly symmetrically around the origin, apart from $p=0$, when it is trivial, and $p=1$, when it is $1+z^{4}$ -- identical to the equivalent $\hat{A}_{2}$ quiver in~\eqref{Minverse_ahat_ij}. 
For $p \ge 2$ and $p \not\equiv 0 \pmod 3$ we observe that $P_{a_{p}(0)}(z)=1-z^{2}+\cO(z^{3})$; see~\cref{appendix:misc} for the first few explicit forms.

In~\cref{appendix:misc} we list out some explicit computed expressions for the expansion terms using~\eqref{general_Gmz_V_productform}, and demonstrate the robustness of the giant graviton expansion for some small quivers from the $\hat{A}_{m}$ and $Y^{p,p}$ families, and small gauge group rank.
We find that the giant graviton expansion indeed iteratively corrects the large $N$ index to output the finite $N$ index.
In addition we find that in at least some cases, for a fixed $\mathbf{m}$ and scaling by $z^{\alpha N}$ for some $\alpha$ dependent on the quiver family, the corrections appear to converge to a limiting functional form.

\section{Discussion and future work}
\label{section:conclusions}

In this paper we study the asymptotics of the superconformal index in a few classes of quivers, in both the univariate and main-diagonal bivariate cases.
Apart from isolated examples and the $Y^{p,0}$ family, for which the degeneracy grows sub-exponentially, all the other cases we study display characteristic Hardy--Ramanujan growth in degeneracy, and this is encouraging from the perspective of investigating gauge-gravity duality and for possibly related black holes.

In our work we switched off flavour symmetries; it would be insightful to investigate the flavoured superconformal index in likewise fashion, as has been explored previously in the literature, e.g. the $Y^{p,q}$ quivers have been briefly discussed in~\cite{gadde} for the finite $N$ case; switching off these fugacities amounts to setting them to $1$.
It would be interesting to study the cycle structures in these cases, as it would be to look deeper into the relation between the cycles of the quiver superpotential and those arising in the factorization of the large $N$ index.
For example, for the $dP3$ quiver, the relationship between~\eqref{superpotentialdP3} and~\eqref{detMdP3} is not immediately obvious. 

From the CFT perspective, the appearance of the central charges (equal in our case for $d=4$ toric quivers) in the index can be investigated in more depth, especially the role of the effective central charge in characterizing them.
From both gauge and gravity perspectives, an important possible expansion of study is the analysis of the index at finite $N$, especially its growth.
The logarithmic contribution to the finite $N$ index has been explicitly calculated in the case of $\cN=4$ SYM~\cite{Cabo-Bizet2019,lezcano2020paper,Gonz_lez_Lezcano_2021} using Bethe ansatz techniques, and its gravity dual is well-studied. 
Similarly, the superconformal index for some $d=3$ theories also has been investigated~\cite{Bobev_2023}.
We hope to investigate a finite $N$ computation of the index for more generic supersymmetric theories in the future.
Once the dual theory is better-described for such cases, we could perhaps proceed with computing the index from the gravity side, as has been recently done in the case of $\frac{1}{2}-$BPS states in $\cN=4$ SYM~\cite{Eleftheriou_2024,Lee2024,eleftheriou2025localizationwallcrossinggiantgraviton,Gautason:2024nru}.\footnote{We thank the authors of~\cite{Gautason:2024nru} for bringing this work to our attention.}

The dual theory may be useful in analyzing the giant graviton expansion: as we have seen, the expansion parameters are relatively simple for sufficiently `uncomplicated' quivers such as the $\hat{A}_{m}$ and the $Y^{p,p}$, and it would be interesting to see if similar patterns hold true for other quivers.
The symmetry of the quiver appears to play a significant role. 
For infinite families in particular, it would be interesting to explore what connections expansions of the form~\eqref{conj:giantgrav_partition}, which appear to hold true for at least some such families, have to the Schur polynomials and their generalizations, and what implications such large-size limits may have in a physical context.
We have also noticed a link between the expansion for other $d=4$ indices, e.g. the $\frac{1}{2}-$ and $\frac{1}{8}-$BPS, i.e. Schur index~\cite{kucharski2025quiversbpsstates3d} for $\cN=4$, and generating functions of moduli spaces of symmetric quivers~\cite{Efimov_2012}. 
This may also be related via the knots-quivers correspondence~\cite{Kucharski_2019} to $d=3$ correlation functions of knot polynomials, the coloured link invariants, hence presenting an interesting prospect of study of triality.

From an analytic number theory perspective, one interesting line of research is to study the asymptotics of generic bivariate and multivariate generating functions using the current available techniques~\cite{Pemantle_Wilson_Melczer_2024}. 
It would be interesting to eventually attempt to classify quiver gauge theories for example according to their asymptotic index behaviour. 
Our work applies classic saddle-point techniques to a fairly constrained class of generating functions expressible as infinite products. 
It remains to investigate the full bivariate asymptotic analysis of generating functions in two or more fugacities; eventually, we can perhaps extend these techniques to a wider class of generating functions, treated by saddle-point techniques and supplemented by e.g. the Mellin transform or resurgence techniques.
In particular, it would be really nice to undertake a deeper study of the dominant singularities in such generating functions. 
For a generating function expressible as an infinite product, it may perhaps make sense to speak of a `hierarchy' of singularities in the sense constructing an `expansion' of the large $N$ index. 

With regard to gauge symmetry, we could consider variants of the index with $\mathrm{U}(N)$ replaced by other gauge groups, e.g. $\mathrm{SO}(N)$ or $\mathrm{Sp}(N)$.
This could be done more generally with the matrix model, as we briefly touched upon in~\cref{section:matrixmodelandgiantgrav}.
The asymptotic behaviour of the giant graviton expansion at large rank or quiver size would also be interesting to look at.
Finally, the inverse of the weighted adjacency matrix plays an important role in developing the expansion, and it would be interesting to look into implications thereof.

\subsection*{Acknowledgements}

We are especially grateful to Stephen Melczer for his contributions in the initial stages of the project.
We also acknowledge useful discussions with Aditya Bawane, Taro Kimura, Alfredo Gonz\'alez Lezcano, Sameer Murthy and Piotr Su\l{}kowski.
SP is supported by OPUS grant no.~2022/47/B/ST2/03313 “Quantum geometry and BPS states” and NCN Sonata Bis 9 grant no.~2019/34/E/ST2/00123
funded by the National Science Centre, Poland; SP also thanks the Galileo Galilei Institute of Theoretical Physics and the \'Ecole de Physique des Houches for hospitality, and the
INFN Italy and the Universit\'e Grenoble-Alpes for partial support during the completion of the work.
ZQ is supported by the Natural Sciences and Engineering Research Council of Canada / Conseil de recherches en sciences naturelles et en g\'enie du Canada (NSERC/CRSNG), reference no. / n\textdegree \,de r\'ef\'ererence PGSD-568936-2022.

\subsection*{Data availability}
Code supporting the findings of our work is available upon request.

\appendix

\section{Infinite series identities}
\label{appendix:jacobitripleproductandmacdonaldidentities}

In this Appendix we survey some of the infinite series identities used in this paper.
\subsubsection*{Jacobi triple product}
For suitable $t$ and $w$ to ensure convergence, the Jacobi triple product~\cite{jacobi1829fundamenta} is given by
\begin{equation}
\prod_{k=1}^{\infty} \left(1-t^{2k}\right)\left(1+t^{2k-1}w^{2}\right)\left(1+\frac{t^{2k-1}}{w^{2}} \right)=\sum_{k\in \mathbb{Z}}t^{k^{2}}w^{2k}. \label{jacobitripleproduct} 
\end{equation}
Many useful results can be obtained from algebraic manipulation or variable reparametrization in the Jacobi triple product identity, including the following that are used in our paper.

\begin{subequations}
\begin{itemize}
    \item[(a)] Setting $t=-x$ and $w=1$ gives 
    \begin{equation}
    \prod_{k=1}^{\infty} \frac{1-x^{k}}{1+x^{k}}=\sum_{k \in \mathbb{Z}} (-1)^{k}x^{k^{2}}. \label{jacobi1}  
    \end{equation}
    \item[(b)] Setting $t=x^{\frac{3}{2}}$ and $w=ix^{-\frac{1}{4}}$ yields Euler's pentagonal number theorem, a sum over terms with generalized pentagonal number powers,
    \begin{equation}
    \prod_{k=1}^{\infty} \left(1-x^{k}\right)=\sum_{k\in \mathbb{Z}}(-1)^{k}x^{\frac{k(3k-1)}{2}}. \label{jacobi2}  
    \end{equation} 
    \item[(c)] Setting $t=x^{\frac{3}{2}}$ and $w=x^{-\frac{1}{4}}$ flips the negative coefficients in the right-hand-side of \eqref{jacobi2}, giving
    \begin{equation}
    \prod_{k=1}^{\infty} \frac{1-x^{3k}}{1-x^{k}+x^{2k}}=\sum_{k\in \mathbb{Z}}x^{\frac{k(3k-1)}{2}}. \label{jacobi3}
    \end{equation}
    \item[(d)] Setting $t=x^{\frac{1}{2}}$ and $w=x^{\frac{1}{4}}$ gives a sum over triangular number powers,
    \begin{equation}
    \prod_{k=1}^{\infty} \left(1-x^{k}\right)\left(1+x^{k}\right)^{2}=\sum_{k=0}^{\infty}x^{\frac{k(k+1)}{2}}. \label{jacobi4} 
\end{equation}
\end{itemize}    
\end{subequations}
Additional details can be found in many standard texts on combinatorial analysis, such as~\cite{macmahon}.

\subsubsection*{Macdonald identities}
More generally, the Macdonald identities~\cite{macdonald} give explicit series expansions for certain generating functions of $V$-coloured distinct partitions summed under parity,
\begin{equation}
\prod_{k=1}^{\infty} \left(1-x^{k} \right)^{V}=\sum_{k=0}^{\infty} \left(q_{V,\mathrm{even}}(k)-q_{V,\mathrm{odd}}(k) \right)x^{k} = \sum_{k=0}^{\infty} d_{V}(k)x^{k}, \label{partitionfunctionVcoloursevenoddparitymain}
\end{equation}
where $V$ is a positive integer and $q_{V}(k)=q_{V,\mathrm{even}}(k)+q_{V,\mathrm{odd}}(k)$ is the number of distinct partitions in $V$ colours (where no two entries are of the same colour).  The quantity
$q_{V}(k)$ can be split up into the number of such partitions with even entries, $q_{V,\mathrm{even}}(k)$, and odd entries, $q_{V,\mathrm{odd}}(k)$ respectively, and $d_{V}(k)$ is the difference of these two numbers. Note that in \eqref{partitionfunctionVcoloursevenoddparitymain} we set $q_{V,\mathrm{even}}(0)=1$ and $q_{V,\mathrm{odd}}(0)=0$.

Macdonald found expressions for $d_{V}(k)$ in the dimension $V=\dim \mathbf{g}$ of the semi-simple Lie algebras $\mathbf{g}$. So far, exact expressions are known for $V=1,3,8,10,14,15,21,24,26,28,35,36,\ldots$~(see \cite{hannekrasov} for further discussion).
Note that \eqref{jacobi2} and~\eqref{jacobi3} involve the generating functions of $d_{1}(k)$ and  $d_{1}^{2}(k)$ respectively.

Note that some observations may be readily made about the $d_{V}(k)$ from a series expansion of~\eqref{partitionfunctionVcoloursevenoddparitymain} obtained using the Jacobi triple product even if their exact functional forms are not known or relevant.
For example, from the expansion~\cite{WINQUIST196956}
\begin{equation}
\prod_{k=1}^{\infty}(1-x^{k})^{8}=\sum_{\substack{i \ge 0 \\ z \in \mathbb{Z}}} (2i+1)^{2}(6j+1)x^{i(i+1)+j(3j+1)}-\sum_{\substack{i \ge 0 \\ z \in \mathbb{Z}}} (2i+2)^{2}(6j+2)x^{(i+2)^{2}+j(3j+2)},
\end{equation}
a straightforward check modulo $4$ shows that $d_{8}(k)=0$ when $k \equiv 3 \pmod 4$. 

\subsubsection*{Nekrasov--Okounkov hook-length formula}
The coefficients $d_{V}(k)$ in \eqref{partitionfunctionVcoloursevenoddparitymain} also have an elegant expression in terms of the hook-length of integer partitions. 
Let $\cP(\lambda)$ be the set of all partitions of $\lambda \in \mathbb{N}$ with $\cP(0)= \emptyset$ by convention. The Nekrasov--Okounkov formula~\cite{nekrasovokounkovmain} (see also~\cite{hannekrasov}) states that, for any $V \in \mathbb{C}$,
\begin{equation}
\prod_{k=1}^{\infty}\left(1-x^k\right)^{V}= \sum_{n=0}^{\infty} x^{n} \sum_{P \in \cP(n)} \prod_{\nu \in P} \left(1-\frac{V+1}{h_{\nu}^2}\right), \label{nekokouformulamain}
\end{equation}
where $h_{\nu}$ is the `hook-length' of any box $\nu$ in the Young diagram of $P \in \cP(\lambda)$. A definition of hook-length may be found in~\cite{fultonharris}.
As an example of the utility of~\eqref{nekokouformulamain}, one may obtain the generating function for $d_{3}(k)$, 
\begin{equation}
\prod_{k=1}^{\infty}\left(1-x^{k}\right)^{3}=\sum_{k=0}^{\infty}(-1)^{k}(2k+1)x^{\frac{k(k+1)}{2}}, \label{V=3nonsquaredmain}
\end{equation}
from the observation that the only partitions lacking boxes with hook-length $2$ are the `staircase' partitions, which correspond to triangular numbers~\cite{hannekrasov}.
Note that it is also possible to obtain~\eqref{V=3nonsquaredmain} from the Jacobi triple product~\cite{jacobi1829fundamenta}.

The form of $d_{3}(k)$ being known,~\eqref{jacobi4} now lets us obtain a generating function for $d_{3}^{2}(k)$. 
Denoting by $S(x)$ the infinite series on the right-hand side of \eqref{jacobi4}, 
\begin{align}
\sum_{k=0}^{\infty} (2k+1)^{2}x^{\frac{k(k+1)}{2}} &= S(x)+8xS'(x)= S(x)+8xS(x) \sum_{k=1}^{\infty} \frac{\frac{\dd}{\dd x} \left[(1-x^k)(1+x^k)^{2}\right]}{(1-x^k)(1+x^k)^{2}} \nonumber \\
&= \left[1+ 8 \sum_{k=1}^{\infty} \frac{kx^{k}(1-3x^{k})}{1-x^{2k}} \right] \prod_{k=1}^{\infty} \left(1-x^{k}\right)\left(1+x^{k}\right)^{2}. \label{V=3squaredmain}
\end{align}

\section{Supplementary material}
\label{appendix:misc}

In this Appendix we document additional computational evidence and analytic details for some of the conjectures and results presented in this paper. 

\begin{table}[H]
\begin{center}
\begin{tabular}{|>{$}c<{$}|>{$}c<{$}|>{$}c<{$}|>{$}c<{$}|}
\hline
n & \; \frac{1}{\pi n^{\frac{1}{2}}}\ln |c_{n}| \; & \; \frac{1}{\pi (n+2)^{\frac{1}{2}}}\ln |c_{n+2}| \; & \; \frac{1}{\pi (n+3)^{\frac{1}{2}}}\ln |c_{n+3}| \;
\\ \hline \hline && \\[-1.4em]
4000 & \; 0.617223 \; & \; 0.617225 \; & \; 0.373656 \;
\\ \hline && \\[-1.4em]
8000 & \; 0.619713 \; & \; 0.619714 \; & \; 0.373991 \;
\\ \hline && \\[-1.4em]
12000 & \; 0.620724 \; & \; 0.620724 \; & \; 0.374047 \;
\\ \hline && \\[-1.4em]
16000 & \; 0.621292 \; & \; 0.621292 \; & \; 0.374045 \; \\
\hline
\end{tabular}
\caption{Numerical support for our conjecture~\cref{tab:Ypp} of the exponential sector of the $Y^{2,2}$ quiver, correct up to six digits. We observed a very gradual decrease in the value $3$ modulo $4$ from $n\approx 14000$ onwards, but presume this may be due to limitations of numerical calculation.}
\label{tab:Y22}
\end{center}
\end{table}

We now present some computations of the giant graviton index corrections for some quivers.
The finite $N$ index in all the examples presented in this Appendix is computed by interpreting the contour integrals in~\eqref{quiverfullindex} as a combinatorial problem of counting constant terms in the maximal torus alphabet expansion, and the giant graviton terms are evaluated using~\eqref{general_Gmz_V_productform}, employing the procedure outlined following it.
For brevity, we use Fraktur font notation $\mathfrak{m}$ for the degenerate identical modes $(0, \ldots, 0,m,0,\ldots,0)$, in each case; hence, $\fone$, $\ftwo$ and so forth. 
We present all series and corrections up to $\cO(z^{15})$ for $N=1$, and $\cO(z^{12})$ for $N=2$.
Constrained by considerable computation time with increasing $N$ and quiver size $V$, we present second and higher order corrections only for $N=1$.
We hope to come back to these computations to make further observations of these trends.

Our observations on giant graviton corrections for the $\hat{A}_{2}$ quiver are presented below.

\begin{subequations}
\begin{align}
N=1: \;  \cZ_{N}[\hat{A}_{2}] &= 1+2z^{2}-4z^{3}+5z^{4}-4z^{5}+2z^{6}-5z^{8}+20z^{9}-20z^{10} \nonumber \\ &\;\; +36z^{12} -68z^{13}+72z^{14}-32z^{15}, \label{begin_results} \\
\cZ_{\infty}[\hat{A}_{2}] &=1+2z^{2}-4z^{3}\textcolor{red}{+7z^{4}-8z^{5}+16z^{6}-28z^{7}+39z^{8}-48z^{9}+80z^{10}} \nonumber \\ &\;\; \textcolor{red}{-124z^{11}+163z^{12}-208z^{13}+312z^{14}-452z^{15}}, \\
\cZ_{\infty}[\hat{A}_{2}](1+2G^{(\fone)}_{N}[\hat{A}_{2}]) &=1+2z^{2}-4z^{3}+5z^{4}-4z^{5}+2z^{6}\textcolor{red}{-19z^{8}+40z^{9}-78z^{10}}\nonumber \\ &\;\; \textcolor{red}{+112z^{11}-153z^{12}+184z^{13}-258z^{14}+372z^{15}}, \\ 
\cZ_{\infty}[\hat{A}_{2}](1+2G^{(\fone)}_{N}[\hat{A}_{2}]+G^{((1,1))}_{N}[\hat{A}_{2}]) &=1+2z^{2}-4z^{3}+5z^{4}-4z^{5}+2z^{6}-5z^{8}+20z^{9}-20z^{10} \nonumber \\ &\;\; \textcolor{red}{+34z^{12}-64z^{13}+56z^{14}+24z^{15}}, \\
\cZ_{\infty}[\hat{A}_{2}](1+2G^{(\fone)}_{N}[\hat{A}_{2}]+G^{((1,1))}_{N}[\hat{A}_{2}]+2G^{(\ftwo)}_{N}[\hat{A}_{2}]) &= 1+2z^{2}-4z^{3}+5z^{4}-4z^{5}+2z^{6}-5z^{8}+20z^{9}-20z^{10} \nonumber \\ &\;\; +36z^{12} -68z^{13}+72z^{14}-32z^{15}.
\end{align}
\end{subequations}

\begin{subequations}
\begin{align}
N=2: \;  \cZ_{N}[\hat{A}_{2}] &= 1+2z^{2}-4z^{3}+7z^{4}-8z^{5}+14z^{6}-24z^{7}+25z^{8}-20z^{9}+14z^{10}+8z^{11}-47z^{12}, \\
\cZ_{\infty}[\hat{A}_{2}] &=1+2z^{2}-4z^{3}+7z^{4}-8z^{5}\textcolor{red}{+16z^{6}-28z^{7}+39z^{8}-48z^{9}+80z^{10}-124z^{11}+163z^{12}}, \\
\cZ_{\infty}[\hat{A}_{2}](1+2G^{(\fone)}_{N}[\hat{A}_{2}]) &=1+2z^{2}-4z^{3}+5z^{4}-8z^{5}+14z^{6}-24z^{7}+25z^{8}-20z^{9}+14z^{10}+8z^{11}\textcolor{red}{-77z^{12}}. 
\end{align}
\end{subequations}

We observe similar trends for the $\hat{A}_{3}$ quiver.
For $N=1$, we find that a correction term from the $(1,1,1)$ mode must be added to fully correct at least up to $\cO(z^{15})$.
This appears consistent with our general observation~\eqref{G_matrix_leading_order_behaviour_general}: we find $\alpha_{1,2,3}=2$ in all these cases, since the $(1,1,1)$ and $(2,0,0)$ modes start contributing at $\cO(z^{12})$.
\begin{subequations}
\begin{align}
N=1: \;  \cZ_{N}[\hat{A}_{3}] &= 1+3z^{2}-6z^{3}+6z^{4}-12z^{5}+15z^{6}-12z^{7}+16z^{9}-27z^{10} \nonumber \\ &\;\; +54z^{11}-36z^{12}-36z^{13}+123z^{14}-216z^{15}, \\
\cZ_{\infty}[\hat{A}_{3}] &=1+2z^{2}-6z^{3}\textcolor{red}{+9z^{4}-18z^{5}+33z^{6}-54z^{7}+84z^{8}-134z^{9}+207z^{10}} \nonumber \\ &\;\; \textcolor{red}{-312z^{11}+456z^{12}-666z^{13}+969z^{14}-1380z^{15}}, \\
\cZ_{\infty}[\hat{A}_{3}](1+3G^{(\fone)}_{N}[\hat{A}_{3}]) &=1+3z^{2}-6z^{3}+6z^{4}-12z^{5}+15z^{6}-12z^{7}\textcolor{red}{-9z^{8}+40z^{9}-108z^{10} } \nonumber \\ &\;\; \textcolor{red}{+240z^{11}-450z^{12}+726z^{13}-1188z^{14}+1866z^{15}}, \\
\cZ_{\infty}[\hat{A}_{3}](1+3G^{(\fone)}_{N}[\hat{A}_{3}]+3G^{((1,1,0))}_{N}[\hat{A}_{3}]) &=1+3z^{2}-6z^{3}+6z^{4}-12z^{5}+15z^{6}-12z^{7}+16z^{9}-27z^{10} \nonumber \\ &\;\; +54z^{11}\textcolor{red}{-18z^{12}-96z^{13}+300z^{14}-618z^{15}}, \\
\cZ_{\infty}[\hat{A}_{3}](1+3G^{(\fone)}_{N}[\hat{A}_{3}]+3G^{((1,1,0))}_{N}[\hat{A}_{3}]+3G^{(\ftwo)}_{N}[\hat{A}_{3}]) &=1+3z^{2}-6z^{3}+6z^{4}-12z^{5}+15z^{6}-12z^{7}+16z^{9}-27z^{10} \nonumber \\ &\;\; +54z^{11}\textcolor{red}{-15z^{12}-102z^{13}+321z^{14}-690z^{15}},
\\
\cZ_{\infty}[\hat{A}_{3}](1+3G^{(\fone)}_{N}[\hat{A}_{3}]+3G^{((1,1,0))}_{N}[\hat{A}_{3}]+3G^{(\ftwo)}_{N}[\hat{A}_{3}] & \nonumber \\ +G^{((1,1,1))}_{N}[\hat{A}_{3}]) &= 1+3z^{2}-6z^{3}+6z^{4}-12z^{5}+15z^{6}-12z^{7}+16z^{9}-27z^{10} \nonumber \\ &\;\; +54z^{11}-36z^{12}-36z^{13}+123z^{14}-216z^{15}. 
\end{align}
\end{subequations}

\begin{subequations}
\begin{align}
N=2: \;  \cZ_{N}[\hat{A}_{3}] &= 1+3z^{2}-6z^{3}+9z^{4}-18z^{5}+30z^{6}-48z^{7}+66z^{8}-92z^{9}+108z^{10}-114z^{11}+96z^{12}, \\
\cZ_{\infty}[\hat{A}_{3}] &=1+2z^{2}-6z^{3}+9z^{4}-18z^{5}\textcolor{red}{+33z^{6}-54z^{7}+84z^{8}-134z^{9}+207z^{10}-312z^{11}+456z^{12}}, \\
\cZ_{\infty}[\hat{A}_{3}](1+3G^{(\fone)}_{N}[\hat{A}_{3}]) &=1+3z^{2}-6z^{3}+9z^{4}-18z^{5}+30z^{6}-48z^{7}+66z^{8}-92z^{9}+108z^{10}-114z^{11}\textcolor{red}{+84z^{12}}. 
\end{align}
\end{subequations}

Here are our observations on the corrections to the index for the $Y^{2,2}$ quiver.
Corresponding to~\eqref{G_matrix_leading_order_behaviour_general}, we find $\alpha_{1,2,3,4}=2,3$ or $4$ in various cases.
At the second order, the correction agrees up to $\cO(z^{11})$ upon the addition of the doubly-degenerate $(1,0,1,0)$ mode, which contributes at $\cO(z^{8})$; the four-fold degenerate $(1,1,0,0)$ mode contributes from $\cO(z^{12})$, however as we saw with the $\hat{A}_{3}$ quiver, a full correction up to $\cO(z^{15})$ requires the addition of the four-fold degenerate $(1,1,1,0)$ mode.

\begin{subequations}
\begin{align}
N=1: \;  \cZ_{N}[Y^{2,2}] &= 1-8z^{3}+2z^{4}+16z^{6}-8z^{7}+5z^{8}+48z^{9}-12z^{10} 
\nonumber \\  
&\;\; +16z^{11}-130z^{12}+48z^{13}-40z^{14}-360z^{15}, \\
\cZ_{\infty}[Y^{2,2}] &=1-8z^{3}+2z^{4}\textcolor{red}{+20z^{6}-16z^{7}+7z^{8}+40z^{10}} \nonumber \\
&\;\; \textcolor{red}{-56z^{11}-56z^{12}+140z^{14}-48z^{15}}, \\
\cZ_{\infty}[Y^{2,2}](1+4G^{(\fone)}_{N}[Y^{2,2}]) &= 1-8z^{3}+2z^{4}+16z^{6}-8z^{7}\textcolor{red}{+3z^{8}+48z^{9}-32z^{10}} \nonumber \\
&\;\; \textcolor{red}{+32z^{11}-296z^{12}+240 z^{13}-400 z^{14}+560 z^{15}}, \\
\cZ_{\infty}[Y^{2,2}](1+4G^{(\fone)}_{N}[Y^{2,2}]+2G^{((1,0,1,0))}_{N}[Y^{2,2}]) &= 1-8z^{3}+2z^{4}+16z^{6}-8z^{7}+5z^{8}+48z^{9}-12z^{10}  \nonumber \\
&\;\;+16z^{11}\textcolor{red}{-198z^{12}+96z^{13}-176z^{14}-88z^{15}}, \\
\cZ_{\infty}[Y^{2,2}](1+4G^{(\fone)}_{N}[Y^{2,2}]+2G^{((1,0,1,0))}_{N}[Y^{2,2}] \nonumber \\ +4G^{((1,1,0,0))}_{N}[Y^{2,2}]) &= 1-8z^{3}+2z^{4}+16z^{6}-8z^{7}+5z^{8}+48z^{9}-12z^{10}  \nonumber \\
&\;\;+16z^{11}\textcolor{red}{-94z^{12}+16z^{13}+88z^{14}-904z^{15}}, \nonumber \\
\cZ_{\infty}[Y^{2,2}](1+4G^{(\fone)}_{N}[Y^{2,2}]+2G^{((1,0,1,0))}_{N}[Y^{2,2}] \nonumber \\ +4G^{((1,1,0,0))}_{N}[Y^{2,2}]+4G^{(1,1,1,0)}_{N}[Y^{2,2}]) &=  1-8z^{3}+2z^{4}+16z^{6}-8z^{7}+5z^{8}+48z^{9}-12z^{10} 
\nonumber \\  
&\;\; +16z^{11}-130z^{12}+48z^{13}-40z^{14}-360z^{15}.
\end{align}
\end{subequations}

\begin{subequations}
\begin{align}
N=2: \;  \cZ_{N}[Y^{2,2}] &= 1-8z^{3}+2z^{4}+20z^{6}-16z^{7}+7z^{8}+36z^{10}-48z^{11}-74z^{12}, \\
\cZ_{\infty}[Y^{2,2}] &=1-8z^{3}+2z^{4}+20z^{6}-16z^{7}+7z^{8}\textcolor{red}{+40z^{10}-56z^{11}-56z^{12}}, \\
\cZ_{\infty}[Y^{2,2}](1+4G^{(\fone)}_{N}[Y^{2,2}]) &=1-8z^{3}+2z^{4}+20z^{6}-16z^{7}+7z^{8}+36z^{10}-48z^{11}\textcolor{red}{-76z^{12}}. \label{end_results}
\end{align}
\end{subequations}

We note down the explicit forms of a few first-order expansion terms per unit mode for the $\hat{A}_{m}$ and $Y^{p,p}$ quivers (and additionally, report corroboration with~\cite{murthy_giantgravitons} for $m=1$).
We only show up to five terms for brevity; more accurate expressions, when relevant, are utilized in obtaining the results~\eqref{begin_results} through~\eqref{end_results}.

\begin{table}[H]
\begin{center}
\begin{tabular}{|>{$}c<{$}|>{$}c<{$}|>{$}c<{$}|>{$}c<{$}|}
\hline
N & \; G^{(\fone)}_{N}(z)[\hat{A}_{2}] \; & \; G^{(\fone)}_{N}(z)[\hat{A}_{3}]\; & \; G^{(\fone)}_{N}(z)[\hat{A}_{4}] \;
\\ \hline \hline && \\[-1.4em]
1 & \; -z^{4}+2z^{5}-5z^{6}+6z^{7}-4z^{8} \; & \; -z^{4}+2z^{5}-3z^{6}+2z^{7}-z^{8} \; & \; -z^{4}+3z^{5}-3z^{6}+2z^{7}+z^{8} \;
\\ \hline && \\[-1.4em]
2 & \; -z^{6}+2z^{7}-5z^{8}+6z^{9}-8z^{10}\; & \; -z^{6}+2z^{7}-3z^{8}+2z^{9}-3z^{10} \; & \; -z^{6}+2z^{7}-3z^{8}+2z^{9}-z^{10} \;
\\ \hline && \\[-1.4em]
3 & \; -z^{8}+2z^{9}-5z^{10}+6z^{11}-8z^{12} \; & \; -z^{8}+2z^{9}-3z^{10}+2z^{11}-3z^{12} \; & \; -z^{8}+2z^{9}-3z^{10}+2z^{11}-z^{12} \;
\\ \hline && \\[-1.4em]
4 & \; -z^{10}+2z^{11}-5z^{12}+6z^{13}-8z^{14} \; & \; -z^{10}+2z^{11}-3z^{12}+2z^{13}-3z^{14} \; & \; -z^{10}+2z^{11}-3z^{12}+2z^{13}-z^{14} \; \\
\hline
\end{tabular}
\caption{The form of the first-order correction per unit mode for the $\hat{A}_{m}$ quivers.}
\label{tab:G1Ahat}
\end{center}
\end{table}

\begin{table}[H]
\begin{center}
\begin{tabular}{|>{$}c<{$}|>{$}c<{$}|>{$}c<{$}|>{$}c<{$}|}
\hline
N & \; G^{(\fone)}_{N}(z)[Y^{2,2}] \; & \; G^{(\fone)}_{N}(z)[Y^{3,3}]\; & \; G^{(\fone)}_{N}(z)[Y^{4,4}] \;
\\ \hline \hline && \\[-1.4em]
1 & \; -z^{6}+2z^{7}-z^{8}+4z^{9}+10z^{11} \; & \; -z^{6}+6z^{9}-z^{12}+6z^{15}+12z^{18} \; & \; -z^{6}+4z^{9}+2z^{11}-4z^{15}-z^{16}\;
\\ \hline && \\[-1.4em]
2 & \; -z^{10}+2z^{11}-5z^{12}+6z^{13}-6z^{14}\; & \; -5z^{12}+18z^{15}-13z^{18}+18z^21+63z^{24} \; & \; -4z^{12}-z^{14}+10z^{15}+6z^{17}+2z^{19}\;
\\ \hline && \\[-1.4em]
3 & \;-z^{14}+2z^{15}-5z^{16}+6z^{17}-16z^{18} \; & \; -15z^{18}+50z^{21}-68z^{24}+98z^{27}+160z^{30} \; & \; -10z^{18}-4z^{20}+24z^{21}-z^{22}+18z^{23} \;
\\ \hline && \\[-1.4em]
4 & \;-z^{18}+2z^{19}-5z^{20}+6z^{21}-16z^{22} \; & \; -35z^{24}+120z^{27}-226z^{30}+416z^{33}+3z^{36} \; & \; -20z^{24}-10z^{26}+50z^{27}-4z^{28}+44z^{29} \; \\
\hline
\end{tabular}
\caption{The form of the first-order expansion term per unit mode for the $Y^{p,p}$ quivers.
Note that $p=1$ corresponds to $m=2$ in~\cref{tab:G1Ahat}.}
\label{tab:G1Ypp}
\end{center}
\end{table}
The higher-order expansion terms per unit mode we have used in this Appendix for $N=1$, shown up to $\cO(z^{15})$, are
\begin{subequations}
\begin{align}
G_{N}^{((1,1))}(z)[\hat{A}_{2}] &= 14z^{8}-20z^{9}+30z^{10}-16z^{11}-51z^{12}+156z^{13}-242z^{14}+200z^{15}, \\
G_{N}^{(\ftwo)}(z)[\hat{A}_{2}] &= z^{12}-2z^{13}+6z^{14}-20z^{15}, \\
G_{N}^{((1,1,0))}(z)[\hat{A}_{3}] &= 3z^{8}-8z^{9}+18z^{10}-20z^{11}+15z^{12}+20z^{13}-74z^{14}+132z^{15}, \\
G_{N}^{(\ftwo)}(z)[\hat{A}_{3}] &= z^{12}-2z^{13}+4z^{14}-12z^{15}, \\
G_{N}^{((1,1,1))}(z)[\hat{A}_{3}] &=-21z^{12}+66z^{13}-135z^{14}+150z^{15}, \\
G_{N}^{((1,0,1,0))}(z)[Y^{2,2}] &= z^{8}+10z^{10}+47z^{12}+8z^{13}+72z^{14}+68z^{15}, \\
G_{N}^{((1,1,0,0))}(z)[Y^{2,2}] &= 26z^{12}-20z^{13}+66z^{14}+4z^{15}, \\
G_{N}^{((1,1,1,0))}(z)[Y^{2,2}] &= -9z^{12}+8z^{13}-32z^{14}+64z^{15}. 
\end{align}
\end{subequations}

The first few polynomials $P_{a_{p}(0)}(z)$ appearing in~\eqref{Minverse_Ypp_11} are
\begin{multline}
1+z^4,\; 1-z^2+z^4+2 z^6+z^8-z^{10}+z^{12},\; 1+3 z^6+10 z^{12}+3
   z^{18}+z^{24}, \; 1-z^2+3 z^6-2 z^8-z^{10}+5 z^{12} \\-2 z^{14}+5 z^{16}-z^{18}-2
   z^{20}+3 z^{22}-z^{26}+z^{28}, \; 1-z^2+3 z^6-3 z^8+z^{10}+4 z^{12}-5 z^{14}+3
   z^{16}+4 z^{18}+3 z^{20} \\ -5 z^{22}+ 4 z^{24}+z^{26} -3 z^{28}+3
   z^{30}-z^{34}+z^{36}, \; 1+2 z^6+3 z^{12}+4 z^{18}+16 z^{24}+4 z^{30}+3 z^{36}+2
   z^{42}+z^{48}.
\end{multline}

We conclude this Appendix with a sketch of the derivation of~\eqref{general_Gmz_V} from~\eqref{contrib_expansion_normalized} by constructing a generating function that generalizes the ones considered in~\cite{murthy_giantgravitons} and~\cite{Liu_2023}.
Our arguments are a straightforward extension of those in~\cite{Liu_2023}, to which we refer for more details.

In the complex variable alphabet $\mathbf{z}=\{z_{jl}: 1 \le j \le V, \, 1 \le l \le m_{j} \}$, $\mathbf{w}=\{w_{jl}: 1 \le j \le V, \, 1 \le l \le m_{j} \}$, consider the generating function
\begin{equation}
F(\mathbf{z},\mathbf{w})= \left\langle \tilde{\cZ}_{\infty}(\mathbf{t}^{+},\mathbf{t}^{-};V)  \prod_{j=1}^{V} \sum_{\substack{r_{j1},\ldots,r_{jm_{j}}, \\ s_{j1},\ldots,s_{jm_{j}} \in \mathbb{Z}+\frac{1}{2}}} \prod_{l=1}^{m_{j}} \tilde{K}(r_{jl},s_{jl};\bm{\tau}_{j},\bm{\sigma}_{j})z_{jl}^{r_{jl}}w_{jl}^{-s_{jl}} \right\rangle_{\fM}, \label{Fzw_orig}
\end{equation}
from which the giant graviton expansion terms~\eqref{contrib_expansion_normalized} may be extracted as
\begin{align}
G_{N}^{(\mathbf{m})}(\fM) &= \frac{(-1)^{|\mathbf{m}|}}{\cZ_{\infty}(\fM) \prod_{j=1}^{V}m_{j}!} \left[  \sum_{\substack{\sigma_{1} \in \fS_{m_{1}},\ldots, \\ \sigma_{V} \in \fS_{m_{V}}}} (-1)^{\sum_{j=1}^{V} |\sigma_{j}|} \left(\prod_{j=1}^{V}\prod_{l=1}^{m_{j}} \frac{1}{1-w_{j,\sigma^{-1}_{j}(l)}/z_{jl}} \right) F(\mathbf{z},\mathbf{w}) \right]_{\prod_{j=1}^{V}\prod_{l=1}^{m_{j}}(w_{jl/}z_{jl})^{-N-\frac{1}{2}}} \nonumber \\
&=\frac{(-1)^{|\mathbf{m}|}}{\cZ_{\infty}(\fM)  \prod_{j=1}^{V}m_{j}!} \left[ \left( \prod_{j=1}^{V} \prod_{l=1}^{m_{j}} \det \left( \frac{1}{1-w_{jl}/z_{jl'}} \right)_{l,l'=1}^{m_{j}} \right)  F(\mathbf{z},\mathbf{w}) \right]_{\prod_{j=1}^{V}\prod_{l=1}^{m_{j}}(w_{jl/}z_{jl})^{-N-\frac{1}{2}}}. \label{GmN_extraction}
\end{align}
The above expression is obtained by carefully expanding out the determinants involving the $\tilde{K}$ in~\eqref{contrib_expansion_normalized}, and effecting the dummy variable substitutions $s_{jl} \mapsto s_{j,\sigma_{j}(l)}$ in~\eqref{Fzw_orig}.
Note that we assume operating in the regime $|w_{jl}/z_{jl'}|<1$, so that the determinant term is well-defined.
Since the $\langle \cdots \rangle_{\fM}$-average is commutative with the coefficient extraction operations, it suffices to first evaluate $F(\mathbf{z},\mathbf{w})$.

We now expand $F(\mathbf{z},\mathbf{w})$ using the generating function~\eqref{ktildeGF} (through which the $\tilde{K}$ are defined), as
\begin{align}
F(\mathbf{z},\mathbf{w}) &= \left\langle  \tilde{\cZ}_{\infty}(\mathbf{t}^{+},\mathbf{t}^{-};V)  \prod_{j=1}^{V} \prod_{l=1}^{m_{j}} \frac{\sqrt{z_{jl}w_{jl}}}{z_{jl}-w_{jl}} \frac{J(z_{jl};\bm{\tau}_{j},\bm{\sigma}_{j})}{J(w_{jl};\bm{\tau}_{j},\bm{\sigma}_{j})} \right\rangle_{\fM} \nonumber \\
&= \left( \prod_{j=1}^{V} \prod_{l=1}^{m_{j}} \frac{\sqrt{w_{jl}/z_{jl}}}{1-w_{jl}/z_{jl}} \right) \left\langle  \tilde{\cZ}_{\infty}(\mathbf{t}^{+},\mathbf{t}^{-};V) \exp \left( \sum_{k=1}^{\infty} \frac{1}{k} \left( \sum_{j=1}^{V} \left(\tau_{jk}\alpha_{jk}-\sigma_{jk}\alpha_{-jk} \right) \right) \right) \right\rangle_{\fM}, \label{F_expanded}
\end{align}
where $\alpha_{jk}=\sum_{l=1}^{m_{j}}(z^{k}_{jl}-w^{k}_{jl})$.
Noting that $\tilde{\cZ}_{\infty}$ is defined through~\eqref{eq_Zinf}, we now work out the $\langle \cdots \rangle_{\fM}$-average.
Using the definition~\eqref{general_transform}, we may decompose the average, which for brevity we denote just by $\langle \cdots \rangle_{\fM}$, into an infinite product of integrals, indexed by $k$ and $j$, as follows, 
\begin{align}
\langle \cdots \rangle_{\fM} &= \left( \prod_{k=1}^{\infty} \frac{1}{\det \sM_{k}} \right) \prod_{k=1}^{\infty} \left( \left( \prod_{j=1}^{V} \int_{\mathbb{C}} \frac{\dd t^{+}_{jk} \wedge \dd t^{-}_{jk}}{2\pi i k} \right)  e^{-\frac{1}{k}  \left(\mathbf{t}_{k}^{+}\right)^{T}(\Lambda_{k}^{-1}-I)\mathbf{t}_{k}^{-}}\exp \left( \frac{1}{k} \sum_{j=1}^{V} \left( \tau_{jk}\alpha_{jk}-\sigma_{jk}\alpha_{-jk} \right) \right) \right).
\end{align}
We remind the reader of the diagonalizations $\sM_{k}=P_{k}\Lambda_{k} P_{k}^{-1}$; hence we also have $(\sM_{k}^{-1}-I)^{-1}=P_{k}(\Lambda_{k}^{-1}-I)^{-1} P_{k}^{-1}$.
This observation, along with the definitions of the $\bm{\tau}_{k}$ and $\bm{\sigma}_{k}$, allows us to deploy the generalized HS transformation~\eqref{eq:generalizationofgaussiannew} at each level $k$, with the identification $\sF=(\sM_{k}^{-1}-I)^{-1}$ in each case, and respective $\mathbf{a},\mathbf{b}$. 
We get
\begin{align}
\langle \cdots \rangle_{\fM} &= \left( \prod_{k=1}^{\infty} \frac{1}{\det \sM_{k}} \right) \left( \prod_{k=1}^{\infty} \det (\sM_{k}^{-1}-I)^{-1} \right) \exp \left( -\sum_{k=1}^{\infty} \frac{1}{k}\sum_{j,j'=1}^{V}\alpha_{jk} \alpha_{j',-k} ((\sM_{k}^{-1}-I)^{-1})_{jj'}\right) \nonumber  \\ &= \cZ_{\infty}(\fM) \exp \left( -\sum_{k=1}^{\infty} \frac{1}{k}\sum_{j,j'=1}^{V}\alpha_{jk} \alpha_{j',-k} ((\sM_{k}^{-1}-I)^{-1})_{jj'}\right). \label{incalc_HS} 
\end{align}
Plugging in~\eqref{incalc_HS} in~\eqref{F_expanded}, we obtain
\begin{equation}
F(\mathbf{z},\mathbf{w})=\cZ_{\infty}(\fM) \left( \prod_{j=1}^{V} \prod_{l=1}^{m_{j}} \frac{\sqrt{w_{jl}/z_{jl}}}{1-w_{jl}/z_{jl}} \right) \exp \left( -\sum_{k=1}^{\infty} \frac{1}{k}\sum_{j,j'=1}^{V}\alpha_{jk} \alpha_{j',-k} ((\sM_{k}^{-1}-I)^{-1})_{jj'}\right), \label{F_finalform}
\end{equation}
which when plugged into~\eqref{GmN_extraction} yields~\eqref{general_Gmz_V}, after including a multiplicative factor of $\prod_{j=1}^{V}\prod_{l=1}^{m_{j}} \sqrt{w_{jl}/z_{jl}}$.

\bibliographystyle{utphys}
\bibliography{refsAQSI}

\end{document}